\documentstyle[preprint,aps]{revtex}
\tighten
\draft
\begin{document}
\widetext
\input epsf
\preprint{CLNS 97/1464, HUTP-97/A004, NUB 3153}
\bigskip
\bigskip
\title{Couplings In Asymmetric Orbifolds and Grand Unified String Models}
\medskip
\author{Zurab Kakushadze$^{1,2}$\footnote{E-mail: 
zurab@string.harvard.edu}, 
Gary Shiu$^3$\footnote{E-mail: shiu@mail.lns.cornell.edu}, 
S.-H. Henry Tye$^3$\footnote{E-mail: tye@mail.lns.cornell.edu}}

\bigskip
\address{$^1$Lyman Laboratory of Physics, Harvard University, Cambridge, 
MA 02138\\
$^2$Department of Physics, Northeastern University, Boston, MA 02115\\
$^3$Newman Laboratory of Nuclear Studies, Cornell University,
Ithaca, NY 14853-5001}
\date{April 15, 1997}

\bigskip
\medskip
\maketitle

\begin{abstract}

{}Using the bosonic supercurrent (or covariant lattice) formalism, 
we review how to compute 
scattering amplitudes in asymmetric orbifold string models. This method is
particularly useful for calculating scattering of multiple 
asymmetrically twisted string states, where the twisted states are 
rewritten as ordinary momentum states. We show how to reconstruct some
of the $3$-family grand unified string models in this formalism, and identify 
the quantum numbers of the massless states in 
their spectra. 
The discrete symmetries of these models are rather intricate. 
The superpotentials for the $3$-family $E_6$ model
and a closely related $SO(10)$ model are discussed in some detail.
The forms of the superpotentials of the two $3$-family $SU(6)$ models (with 
asymptotically-free hidden sectors $SU(3)$ and $SU(2)\otimes SU(2)$) are 
also presented.

\end{abstract}
\narrowtext

\section{Introduction}

{}A phenomenologically interesting string model must contain the
standard model of strong and electroweak interactions in its low energy 
effective field theory limit. Many of such realizations are constructed 
as orbifolds, both symmetric \cite{DHVW} and asymmetric \cite{NSV}, of
the $4$-dimensional heterotic string.
Recently, $3$-family grand unified string models were constructed via
asymmetric orbifolds \cite{kt}. Analyses of these models would require
the determination of the couplings of states in their spectra. 
The main goal of the present work is to provide a prescription for calculating
their correlation functions and scattering amplitudes. In the process,
the quantum numbers of the massless states
are found and the selection rules of their couplings are determined.

{}The prescription for calculating any correlation function 
in orbifold conformal field theory is given in Ref \cite{DFMS}. However, 
the actual calculations can be quite non-trivial \cite{LMN}, when the 
couplings of twisted string states are involved. The problem becomes even 
more difficult in asymmetric orbifold models, where one typically encounters 
some ambiguities which are not easily resolved. 

{}In contrast to symmetric orbifolds, where the original lattice of the
$6$ compactified dimensions ({\em e.g.}, the compactification radii)
can be arbitrary, consistency of asymmetric orbifolds imposes strong 
constraints on the allowed lattices, typically with enhanced 
(discrete or local) symmetry. This enhanced symmetry allows us to treat 
twists as shifts in the momentum lattices, in the so-called bosonic 
supercurrent (or covariant lattice) formalism \cite{bsc,sw}.
Twisted states in an asymmetric orbifold now become ordinary momentum states 
in this bosonic supercurrent formalism; 
so their quantum numbers are straightforward to identify and 
calculating their correlation functions becomes relatively easy.

{}In this paper, we shall first review this bosonic supercurrent 
formalism, and discuss briefly its rules for model-building. 
This formalism is a generalization of the free fermionic string model 
construction \cite{KLT}. So their rules for model-building are quite similar.
Next we discuss the prescription for calculating 
the correlation functions, in particular for the twist states. All the 
couplings can be determined from the correlation functions, or the 
scattering amplitudes. For clarification, we shall apply this approach 
to the original ${\bf Z}_3$ symmetric (at special radii) and 
asymmetric models \cite{DHVW,NSV}. Then we shall discuss the
construction of higher-level string models. Our goal 
is to apply this formalism to determine the couplings in the $3$-family 
grand unified models constructed recently. After we explain 
the prescription for calculating the couplings, the form of the 
superpotential for a number of these models will be presented. Their 
phenomenological implications will be discussed elsewhere.

{}Within the framework of conformal (free) field theory and orbifolds,
we find only one $3$-family $E_6$ model with a non-abelian hidden 
sector \cite{kt}.
Here, we give its construction in the bosonic supercurrent formalism, 
where we see that its discrete symmetry and the charge assignments in the
spectrum are quite non-trivial. There are two orbifold constructions of this 
model, where some of the twisted 
states in one orbifold appear as untwisted states in the other orbifold. 
The supercurrents in these two constructions are rather 
different in the bosonic supercurrent formalism; however, the form of the 
superpotential (at least at tree-level) turns out to be the same.
The same procedure to obtain the forms of superpotentials is 
also applied to the two interesting $3$-family $SU(6)$ models, one with 
an $SU(3)$, and the other with an $SU(2) \otimes SU(2)$, asymptotically-free 
hidden sector.

{}Without much work, we can obtain the superpotential for the
$3$-family grand unified models (or standard-like models) that can be 
obtained by giving the $E_6$ adjoint Higgs an appropriate vacuum 
expectation value. We illustrate this with the $3$-family $SO(10)$ model
that can be obtained this way, and which was constructed as an orbifold 
before. 

{}The plan is as follows: some preliminary discussions are given in
section II, where the basic idea, the advantage, and the issues of 
the bosonic supercurrent formalism are reviewed. 
The orbifold rules for level-$1$ models are discussed in section III 
and that for higher-level models are discussed in section IV. Here,  
the prescription for calculating correlation functions is also discussed.
Since the rules for model-building are given in the light-cone gauge, 
we explain how to calculate the scatterings in the covariant gauge.
Section V contains a discussion of the original ${\bf Z}_3$ 
symmetric (at special radii) and asymmetric models. The construction of 
the $3$-family $E_6$ grand unified model in this formalism is discussed in 
Section VI, where some terms in its superpotential is also presented. 
The $SO(10)$ model 
is discussed in Section VII. Section VIII discusses the $SU(6)$ models. 
The quantum numbers of the massless spectra of these models and the form 
of the relevant supercurrents are given in the tables. 
Section IX contains some concluding remarks.

\section{Preliminaries}

{}To be specific, let us consider $4$-dimensional heterotic string 
models within the conformal field theory (CFT) framework, where free 
fields are used. Consider such a model with a 
Lorentzian lattice $\Gamma^{6,22}$. An orbifold is realized via modding out 
this lattice by a point group $P$, ${\em i.e.}$, a group of discrete 
rotations, 
or twists. We shall restrict ourselves to Abelian twists only.
Let $X ({\overline z})$ be one of the right-moving complex chiral bosons 
in the $6$ compactified dimensions in $\Gamma^{6,22}$. 
In terms of 2 real bosons, $X=(X_1 +iX_2)/{\sqrt 2}$. For a ${\bf Z}_N$
twist, in the neighborhood of a twist field located at the origin, 
$X ({\overline z})$ undergoes a phase rotation 
\begin{eqnarray}\label{monod}
 \partial X ({\overline z}e^{-2\pi i})=\exp(-2\pi i k/N) \partial X
({\overline z})~,
\end{eqnarray}
which is called the monodromy of X. (Note that $k$ is an integer.)
The basic twist field $\sigma ({\overline z})$ has conformal weight 
$h=k(1-k/N)/{2N}$. It 
twists $X$ by $\exp(-2\pi i k/N)$ and its 
complex conjugate ${\overline X}$ by  $\exp(2\pi i k/N)$, ${\em i.e.}$, 
their operator product expansions (OPEs) are \cite{DFMS}
\begin{eqnarray}\label{tau}
 i \partial X ({\overline z}) \sigma ({\overline w}) &=& ({\overline z} -
{\overline w})^{-(1-k/N)} \tau({\overline w}) + ...  ~,\nonumber \\
 i \partial {\overline X} ({\overline z}) \sigma ({\overline w}) &=&
({\overline z} -{\overline w})^{-k/N} \tau ' ({\overline w}) + ... ~,
\end{eqnarray}
where $\tau $ and $\tau '$ are excited twist fields. 

{}Suppose we can rewrite the $i \partial X$ as exponentials of a pair
of boson fields $\phi _1$ and $\phi _2$, 
\begin{equation}
 i \partial X ({\overline z}) = \exp (i e \cdot \phi ({\overline z})) + ...~. 
\end{equation}
Here $e$ is a $2$-dimensional vector and the conformal weight $h=1$ condition
requires $e^2 =2$. Then, the phase rotation of $\partial X$ in 
Eq. (\ref{monod}) becomes a shift in $\phi$
\begin{equation}
 	\phi ({\overline z}e^{-2\pi i})= \phi ({\overline z}) - 2\pi u ~,
\end{equation}
where $e\cdot u=k/N$; that is, a twist on $\partial X$ becomes a shift in 
$\phi$.
To recover the correct OPEs of $\partial X$ and $\partial {\overline X}$, 
it turns out that the $\partial X$ must be written as a linear 
combination of such exponential terms. The proper monodromy condition of
$i \partial X$ then
implies that $\phi$ must be compactified in some lattice.
$\phi$ provides a particularly useful basis if the supercurrent, 
as well as the twist fields such as $\sigma$ and $\tau$, can also be 
written in terms of ordinary momentum states. This is the basic idea of the 
bosonic supercurrent formalism \cite{bsc}.

{}Let us see more explicitly how this conversion of twists to shifts can be 
realized, and the advantages of this approach.
In this paper, we are mostly interested in ${\bf Z}_2$ and ${\bf Z}_3$
twists.
For $N=3$ in Eq. (\ref{monod}), the lattice must have a ${\bf Z}_3$ symmetry, 
so the $U(1)^2$ symmetry enhances to a $SU(3)$ symmetry. 
Here, $\partial X_1$ and 
$\partial X_2$ are the Cartan generators and the six root generators are 
given by 
$e^{{\pm}ie_\alpha \cdot X({\overline z})}c({\pm} e_\alpha)$, $\alpha =1,2,3$,
where the 2-dimensional vectors $e_1$ and $e_2$ are the simple roots of 
$SU(3)$ and we define $e_3 =-e_1 -e_2$. 
Note that $e_\alpha \cdot e_\beta = 2$ for 
$\alpha = \beta$, and $-1$ otherwise.
For convenience, we shall not always explicitly display
the cocycle operator $c({\pm} e_\alpha )$: its presence is understood.

{}In the standard orbifold formalism, the supercurrent for the right-movers
can be written as
\begin{equation}
 T_F={i\over 2}\psi \partial X + {\mbox {h.c.}} + ...~,
\end{equation}
where $\psi$ is a world-sheet complex fermion.
Since each $e^{{\pm}ie_\alpha \cdot X}$ has conformal weight 1, same as the 
$i \partial X$,
we may choose to rewrite the supercurrent in terms of the $SU(3)$ roots.
The choice is constrained by the condition that they must obey the same 
OPEs as $i \partial X$ and $i \partial {\overline X}$.
For any Lie algebra, one can always 
find such a subalgebra by rotating the Cartan generators to the root system 
of the algebra. The new $U(1)$ generators will always be a linear combination
of the root generators. For $SU(3)$, the choice is unique,
\begin{equation}\label{simple3}
 i \partial X = {1 \over {\sqrt 3}} \sum_\alpha e^{-ie_\alpha \cdot \phi}~,~~~
 i \partial {\overline X} = {1 \over {\sqrt 3}} \sum_\alpha e^{ie_\alpha 
\cdot \phi}~.
\end{equation}
Now, there are three twist fields $\sigma _\alpha$ in the ${\bf Z}_3$ twisted 
sector. In this basis, it is easy to check that,
\begin{eqnarray}
 i \partial X ({\overline z}) \sigma _\alpha ({\overline w}) = ({\overline z} -
{\overline w})^{-2/3} \tau_\alpha ({\overline w}) + ... ~,\\ \nonumber
 i \partial {\overline X} ({\overline z}) \sigma _\alpha ({\overline w}) =
({\overline z} -{\overline w})^{-1/3} \tau_\alpha ' ({\overline w}) + ... ~,
\end{eqnarray}
and these OPEs agree with Eq. (\ref{tau}) for $N=3$, where
\begin{eqnarray}
 \sigma_\alpha =  e^{ie_\alpha \phi /3}~, ~~~
 \tau_\alpha = {1 \over {\sqrt 3}} e^{-2ie_\alpha \phi /3} ~,~~~
 \tau_\alpha' = {1\over \sqrt{3}} \sum_{\beta \not=\alpha}  
 e^{i(e_\alpha/3+e_\beta ) \phi }~.
\end{eqnarray}
Note that $\tau_\alpha'$ is a linear combination of the corresponding vertex 
operators for {\em two} states with the same highest weight. 

{}The usefulness of the bosonic supercurrent formalism is now clear. 
The untwisted states lie in the original lattice while the twist states
lie in the shifted lattice. Since they are all expressed in terms of 
ordinary momentum states, their normalizations and degeneracies, as well 
as their OPEs with each other, are easy to calculate. 

{}For a ${\bf Z}_2$ twist, it is sometimes convenient to decompose $X$ into 
2 real 
bosons, $X_1$ and $X_2$.
If the compactification radii of the bosons are 1, we can use the 
fermion-boson equivalence in CFT to rewrite each $X_i$ as a
complex fermion. So, for orbifolds with only ${\bf Z}_2$ twists, the 
internal parts of the supercurrent can be rewritten entirely into world-sheet
fermions. This is the free fermionic string model construction \cite{KLT}.
In the next section, we shall give the 
rules for model-building using this  bosonic supercurrent formalism.
Since some of the $3$-family grand unified models can be constructed with 
the bosonic supercurrent formalism, it is natural to use this formalism 
to calculate the couplings in these models.

{}In the examples with only level-$1$ current algebras, we shall use the 
above supercurrent with (\ref{simple3}) involving the $SU(3)$ lattice. 
For the $3$-family grand unified models, the supercurrents are more 
complicated. Consistency with the transformation of specific twists to 
shifts in a given orbifold model essentially fixes the supercurrent. 
As mentioned earlier, 
there are two equivalent constructions of the $3$-family $E_6$ model.
In one construction, namely, the $E1$ model, the supercurrent involves a 
$E_6$ lattice and its $SU(3)^2$ sub-lattice; while, in the other construction 
of the same model, namely, the $E2$ model, the supercurrent involves the 
appropriate $SU(2)^4$, $SU(6)$ and $SU(3)^2$ sub-lattices of $E_6$.

\section{Model-Building Rules}

{}In this section we give the rules for constructing Abelian asymmetric 
orbifolds using the bosonic supercurrent formalism.
The rules that we present here are less general than the ones in 
Ref \cite{kt} because we confine our attention to a more limited class 
of orbifolds. Nevertheless, some of the three-family grand unified models 
found recently \cite{kt} can be constructed within the present
framework. Moreover, this formalism is particularly useful in 
computing couplings and identifying quantum numbers 
(both local and discrete) of the physical states. 
Throughout the paper, we consider only heterotic strings compactified to 
four space-time dimensions. 

\subsection{Framework}

{}In this subsection we set up the framework for the remainder of this 
section. In the light -cone gauge which we adopt, 
we have the following world-sheet degrees of freedom: one right-moving 
complex boson $X^0$ (which along with its left-moving counterpart 
corresponds to two transverse space-time coordinates); three 
right-moving complex bosons $X^{1,2,3}$ (corresponding to six 
internal coordinates); four right-moving complex fermions $\psi^a$, 
$a=0,1,2,3$ ($\psi^0$ is the world-sheet superpartner of $(X^0)^\dagger$, 
whereas $\psi^{1,2,3}$ are the world-sheet superpartners of 
$(X^{1,2,3})^\dagger$); two left-moving real bosons 
${\cal X}^\mu_L$, $\mu=1,2$ (these are the left-moving counterparts 
of the two real bosons ${\cal X}^\mu_R$ corresponding to 
$X^0$ via $X^0=({\cal X}^1_R+i{\cal X}^2_R)/\sqrt{2}$); $22$ 
left-moving real bosons $\varphi^A$, $A=1,...,22$ 
(corresponding to twenty-two internal coordinates). 
Before orbifolding, the corresponding string model has $N=4$ 
space-time supersymmetry and the internal momenta span an even 
self-dual Lorentzian lattice $\Gamma^{6,22}$.
The underlying conformal field theory of the internal degrees of
freedom is given by $G_L \otimes G_R$ 
where $G_L$ and $G_R$ are the left- and right-moving level-$1$ Kac-Moody 
algebras with central charges $c_L=22$ and $c_R=9$. 
The right-moving Kac-Moody algebra consists of two factors, {\em i.e.},
$G_R={\cal G}_R \otimes SO(6)_1$ where ${\cal G}_R$ is a level $1$ 
Kac-Moody algebra (with central charge $6$) corresponding to the 
right-moving part of the lattice $\Gamma^{6,22}$ and $SO(6)_1$ (with central
charge $3$) comes from the right-moving fermions.
After orbifolding the 
underlying conformal field theory becomes
$(G^{\prime}_L \otimes {\cal C}_L)\otimes (G^{\prime}_R \otimes {\cal C}_R)$.
Here $G^{\prime}_L$ and $G^{\prime}_R$ are left- and right-moving Kac-Moody 
algebras, and ${\cal C}_L$ and ${\cal C}_R$ are certain cosets that arise in  
the breakings $G_L\equiv {\cal G}_L \supset {\cal G}^{\prime}_L$ and 
${\cal G}_R \supset {\cal G}^{\prime}_R$, 
respectively. (Note that since we are considering Abelian orbifolds, the 
$SO(6)$
subgroup of $G_R$ breaks to a level-1 subgroup, and the corresponding coset 
is trivial.) 
In this section, we restrict ourselves to the cases where
after orbifolding, the left- and right-moving Kac-Moody algebras are
realized at level $1$ and the cosets are trivial. The rules for higher level 
models will be discussed in the next section.

{}It is convenient to organize the string states into sectors labeled by 
the monodromies of the string degrees of freedom. Consider the 
right-movers
\begin{eqnarray}
 \psi^a ({\overline z}e^{-2\pi i}) &=& \exp(-2\pi i V^a_i) \psi^a 
(\overline z)~,\nonumber\\
 \partial X^a ({\overline z}e^{-2\pi i}) &=& \exp(-2\pi i T^a_i) \partial X^a 
({\overline z})~.
\end{eqnarray}
Here we note that 
$T^0_i$ must always be zero as ${\cal X}^\mu (ze^{2\pi i},
{\overline z}e^{-2\pi i})={\cal X}^\mu (z,{\overline z})$ since 
${\cal X}^\mu (z,{\overline z})={\cal X}^\mu_L (z)+{\cal X}^\mu_R 
({\overline z})$ correspond to space-time coordinates. Let us
define $s_i \equiv V^0_i$. The sectors with 
$s_i \in {\bf Z}$ give rise to the space-time bosons, whereas the sectors 
with $s_i \in {\bf Z}+1/2$ give rise to space-time fermions. 
The monodromy of the supercurrent
\begin{equation}
 T_F={i\over 2}\sum_{a=0}^{3} \psi^a \partial X^a +{\mbox {h.c.}}~.
\end{equation}
is given by $s_i$, {\em i.e.}, 
$T_F ({\overline z}e^{-2\pi i})=\exp(-2\pi is_i) T_F({\overline z})$. 
This, in particular, implies the {\em triplet} constraint on the 
supercurrent:
\begin{equation}\label{super}
 V^a_i+T^a_i=s_i~({\mbox{mod}}~1)~,~~~a=0,1,2,3~.
\end{equation}

{}The twists on $\psi^{a}$ can be written as shifts if we bosonize
the complex fermions:
\begin{eqnarray}
\psi^{a}&=& \exp(i \rho^{a}) ~, \nonumber \\
\psi^{a \dagger}&=& \exp(-i \rho^{a}) ~.
\end{eqnarray}
This latter form will be useful when we rewrite the {\em triplet}
constraint for the bosonic supercurrent. 

{}Because of the worldsheet $N=2$ superconformal field theory (SCFT) 
(which is
necessary for $N=1$ space-time supersymmetry), there is a conserved 
right moving $U(1)$ current
\begin{equation}
Y (\overline{z}) = i \sum_{a=1}^{3} \rho^{a}(\overline{z})~.
\end{equation} 
Therefore the supercurrent can be divided into two pieces 
$T_{F}(\overline{z})=T_{F}^{+}(\overline{z})+T_{F}^{-}(\overline{z})$ with
$U(1)$ charges $\pm 1$. The energy momentum tensor $T$, the supercurrent
$T_{F}^{\pm}$ together with 
$Y$ form a global $N=2$ superconformal algebra
\begin{eqnarray}\label{SCFT}
 T({\overline z}) T(0) &\sim& {{{3\over 4}\hat{c}}\over 
{{\overline z}^4}}+{{2T(0)}\over 
 {{\overline z}^2}}+
 {{\partial T(0)}\over {{\overline z}}}+ \cdots ~,\nonumber\\
 T({\overline z}) T^{\pm}_F (0) &\sim& {{{3\over 2}T^{\pm}_F (0)}\over 
{{\overline z}^2}} +
 {{\partial T^{\pm}_F (0)}\over {{\overline z}}} + \cdots ~,\nonumber\\
 T^+_F({\overline z}) T^-_F (0) &\sim& {{{1\over 8}\hat{c}}\over 
{{\overline z}^3}}+
 {{{1\over 4} Y(0)}\over  {{\overline z}^2}}+
 {{{1\over 4} T(0)}\over 
 {{\overline z}}}+ {{1\over 8} {\partial Y(0)}\over {{\overline z}}}+
 \cdots ~,\\
 Y({\overline z}) T^{\pm}_F (0) &\sim& {{\pm T^{\pm}_F (0)}\over 
{{\overline z}}}   
 + \cdots ~,\nonumber\\
Y({\overline z}) Y(0) &\sim& {{{1\over 2}\hat{c}}\over {{\overline z}^2}}+
 \cdots ~,\nonumber
\end{eqnarray}
where only the singular terms are shown. In our case, 
$\hat{c}={2 \over 3} c=8$. (The space-time part of the supercurrent
carries $\hat{c}=2$ and the internal part has $\hat{c}=6$.)

{}Let us consider the cases where 
the Kac-Moody algebra ${\cal G}^{\prime}_{R}$ corresponding to the 
right-moving part of the lattice is realized at level $1$ and has 
central charge 6. Then there exist six 
right-moving real bosons $\phi^I$ such that $i\partial \phi^I$ are  
the vertex operators for the Cartan generators of ${\cal G}^{\prime}_R$.
As discussed in the previous section, in order to rewrite the twists
in $\partial X^{a}$ as shifts in $\phi^{I}$, $\partial X^{a}$
must take the following form:
\begin{equation}\label{X^a}
 i\partial X^a =\sum_{{Q}^{2}=2} {\xi}^a ({Q}) 
J_{Q}~,~~~a=1,2,3~,
\end{equation}    
where 
\begin{equation}
J_{Q} ({\overline z})= \exp (i{Q} \cdot 
{\bf \phi}({\overline z}))~ c({Q}) ~.
\end{equation}
We have introduced six-dimensional real vectors 
${Q}=({Q}^1,...,{Q}^6)$ which are
root vectors of ${\cal G}_R$ with length 
squared 2. This ensures that $i\partial X^a$ has conformal dimension $1$. 
The $c({Q})$ are cocycle operators necessary in the 
Kac-Moody algebra. Note that $J_{Q}$ are Kac-Moody currents for root 
generators. The supercurrent is therefore a linear combination of terms with
different $H$ and $Q$ charges.

{}Here ${\xi}^a ({Q})$ are numerical coefficients constrained by the OPEs:
\begin{eqnarray}\label{XOPE}
\partial X^{a} (\overline{z}) \partial X^{b} (0) &\sim& {\mbox{regular}} 
\nonumber ~,\\
\partial X^{a} (\overline{z}) \partial X^{b \dagger} (0) &\sim& 
- \overline{z}^{-2} \delta^{ab} + {\mbox{regular}} ~.
\end{eqnarray} 

{}The monodromies $(V^a_i,T^a_i)$ for $\psi^{a}$ and $\partial X^{a}$ 
can now be 
translated into monodromies $(V^a_i, U^I_i)$ of $\rho^{a}$ and $\phi^{I}$:
\begin{eqnarray}
 \rho_a ({\overline z}e^{-2\pi i}) &=& \rho_a ({\overline z})-2\pi  V^a_i ~,
\nonumber\\
 \phi^I ({\overline z}e^{-2\pi i}) &=& \phi^I ({\overline z})-2\pi U^I_i~.
\end{eqnarray}

{}In this basis, the {\em triplet} constraint on the supercurrent becomes
\begin{equation}\label{triplet}
 \xi^a (Q)=0~{\mbox{unless}}~V^a_i+ U_i \cdot Q= s_i~({\mbox{mod}}~1)~.
\end{equation}

{}In terms of the chiral bosons $\rho_a$ and $\phi^I$, the energy momentum
tensor $T$ is given by:
\begin{equation}
T({\overline z}) = -{1\over 2} \sum_{a} (\partial \rho_a)^2 
                   -{1\over 2} \sum_{I} (\partial \phi^I)^2
                   -{1\over 2} \sum_{\mu} ( \partial X^{\mu}_R)^2 ~.
\end{equation}
The above form of $T$ together with the bosonic supercurrent $T_F^{\pm}$ 
and $Y$ satisfy the $N=2$ SCFT (\ref{SCFT}).

{}Having discussed the right-moving degrees of freedom, let us turn to the 
left-movers. Since ${\cal G}^{\prime}_L$ is realized at level $1$ and has 
central charge $22$, we have the following monodromy 
conditions on the fields $\varphi^A$:
\begin{equation}\label{varphi}
 \varphi^A (ze^{2\pi i})=\varphi^A (z)+2\pi U^A_i ~.
\end{equation}

{}The monodromies $V^a_i, U^I_i, U^A_I$ can be conveniently combined into 
$(10,22)$ dimensional Lorentzian vectors (with metric $((-)^{10},(+)^{22})$):
\begin{equation}\label{MONO}
 V_i=(V^a_i\vert U^I_i \vert\vert U^A_I)~.
\end{equation}
The monodromies $V_i$ can be viewed as fields $\Phi$ (where $\Phi$ is a 
collective notation for $\rho_a$, $\phi^I$ and $\varphi^A$) being 
periodic $\Phi (ze^{2\pi i} ,{\overline z}e^{-2\pi i})=\Phi(z,{\overline z})$
up to the identification $\Phi \sim g(V_i) \Phi g^{-1}(V_i)$, where 
$g(V_i)$ is an element of the orbifold group $G$. 
For $G$ to be a finite discrete group, the element 
$g(V_i)$ must have a finite order $m_i \in {\bf N}$, {\em i.e.}, $g^{m_i} 
(V_i)=1$. This implies that the vector $V_i$ must be a rational multiple 
of a vector in $\Delta^4 \otimes \Gamma^{6,22}$, that 
is, $m_i V_i \in \Delta^4 \otimes \Gamma^{6,22} $. Here 
$\Delta^4$ is the odd self-dual Euclidean lattice spanned by the 
four-dimensional vectors $p$ that correspond to the vector ${\bf v}$ 
and spinor ${\bf s}$ irreps of $SO(8)_1$. Thus, this lattice can be 
described as being spanned by vectors of the following form: 
$p=(p^0,p^1,p^2,p^3)$, 
where either $p^a \in {\bf Z}$ and $\sum_a p^a \in 2{\bf Z}+1$ 
(the momenta corresponding to the ${\bf v}$ irrep), or  
$p^a \in {\bf Z}+{1\over 2}$ and $\sum_a p^a \in 2{\bf Z}$ 
(the momenta corresponding to the ${\bf s}$ irrep). Here we note 
that we could have chosen $\Delta^4$ to be spanned by the the 
four-dimensional vectors $p$ that correspond to the vector ${\bf v}$ and 
conjugate ${\bf c}$ irreps of $SO(8)_1$. This freedom in choosing the 
lattice $\Delta^4$ is immaterial, and is related to the choice of the 
structure constant $k_{00}$ to be introduced below. 

{}To describe all the elements of the group $G$, it is convenient to 
introduce the set of generating vectors $\{V_i\}$ such that 
$\alpha V={\bf 0}$ if and only if $\alpha_i \equiv 0$. Here 
${\bf 0}$ is the null vector, {\em i.e.}, the vector of the form 
(\ref{MONO}) with all its entries being null:
\begin {equation}
 {\bf 0}=(0^4\vert 0^6\vert\vert 0^{22})~.
\end{equation}
Also, $\alpha V \equiv \sum_i \alpha_i V_i$ (the summation is defined as, 
say,  $(V_i+V_j)^a=V^a_i+V^a_j$), $\alpha_i$ being integers that take values 
from $0$ to $m_i -1$. The elements of the group $G$ are then in one-to-one 
correspondence with the vectors $\alpha V$ and will be denoted by 
$g(\alpha V)$. It is precisely the Abelian nature of $G$ that allows 
this correspondence (by simply taking all the possible linear 
combinations of the generating vectors $V_i$).  

{}Now we can identify the sectors of the model. They are labeled by 
the vectors $\alpha V$, and in a given sector $\alpha V$ the monodromies 
of the string degrees of freedom are given by 
$\Phi (ze^{2\pi i}, {\overline z}e^{-2\pi i})=g(\alpha V) \Phi (z, 
{\overline z}) g^{-1} (\alpha V)$. It is clear 
that the sectors with 
$\alpha s \in {\bf Z}$ give rise to the space-time bosons, whereas 
those with $\alpha s\in{\bf Z}+1/2$ give rise to the space-time fermions.

{}Note that a sector described by the vector 
\begin{equation} 
 V_0=((-{1\over 2})^4 \vert  0^6 \vert\vert 0^{22})
\end{equation}
is always present in any orbifold model. This is related to $N=4$ 
SUSY of the original Narain model \cite{narain}. 
In other words, the $V_0$ sector 
is the Ramond sector of the Narain model, 
whereas ${\bf 0}$ sector is the 
Neveu-Schwarz sector. It is convenient to include this sector $V_0$ into 
the set of generating vectors $\{V_i\}$ (then $i=0,1,2,...$). Since we 
have included $V_0$ into the set $\{V_i\}$, without loss of generality 
we can set $s_i=0$ for $i \not= 0$. Then the space-time bosons come from 
the sectors $\alpha V$ with $\alpha_0=0$, and the space-time fermions 
come from the sectors $\alpha V$ with $\alpha_0=1$. (Note that $m_0=2$.) 
With this definition, we can relax the constraint on $m_i V_i$, and 
require that $m_i V_i \in {\bf Z}^4 \otimes \Gamma^{6,22}$. The odd 
self-dual lattice  $\Delta^4$ for the Narain model will then emerge 
as a consequence of the spectrum generating formula (see below).

{}We use the light-cone gauge
in deriving the rules for computing the spectrum of a string model,
since it is manifestly ghost-free and so all the states constructed 
are physical. However, it is more
convenient to discuss scattering in the covariant gauge. Some of the
issues such as picture changing will be more clear in the covariant
approach. The translation from the light-cone gauge to the covariant 
gauge is straightforward. Given a vertex operator in the light-cone gauge,
we can construct a vertex operator in the covariant gauge
by simply covariantizing the light-cone coordinates and putting in the
ghosts.

{}In the light-cone gauge of $4$-dimensional heterotic string with
$N=1$ space-time supersymmetry (SUSY),
massless physical states are created by vertex operators of the form
$V(z,{\overline z})=V(z){\overline V}({\overline z})$, where $V(z)$ is
a left-moving vertex operator of conformal dimension $1$ (the vacuum energy
in the left-moving sector is $-1$ in the untwisted sector), 
and ${\overline V}({\overline z})$ is a
right-moving vertex operator of conformal dimension $1/2$ (the vacuum
energy in the right-moving sector, which has $N=1$ local world-sheet
supersymmetry and $N=2$ global SUSY, is $-1/2$ for bosons in the untwisted
sector).

{}In the covariant gauge, we have in addition to the light-cone degrees
of freedom: two longitudinal space-time coordinates,
two longitudinal components of the right-moving fermions,
reparametrization ghosts $b$ and $c$, and superconformal ghosts
$\beta$ and $\gamma$ \cite{FMS}. It is most convenient to
bosonize the $\beta,\gamma$
ghosts:
\begin{equation}
\beta = \partial \xi e^{-\phi}, ~~~ \gamma = \eta e^{ \phi}~,
\end{equation}
where $\xi$ and $\eta$ are auxiliary fermions and $\phi$ is a bosonic
ghost field obeying the OPE
$\phi(\overline{z}) \phi(\overline{w}) \sim
{\mbox{log}} ( \overline{z} - \overline{w})$. The conformal dimension of
$e^{q \phi}$
is $-{1\over 2} q (q+2)$.
{}In covariant gauge, vertex operators
are of the form $V(z,{\overline z})=V(z){\overline V}({\overline z})$
where $V(z)$ and $V(\overline{z})$ are both dimension $1$ operators
constructed from the conformal fields. These include the longitudinal
components as well as the ghosts. The vertex operators for space-time
bosons carry integral ghost charges ($q \in {\bf Z}$) whereas
for space-time fermions the ghost charges are half-integral
($q \in {\bf Z} + {1\over 2}$). Here, $q$ specifies the picture.
The canonical choice is
$q=-1$ for space-time bosons and $q=-{1\over 2}$ for space-time
fermions. We will denote the corresponding vertex operators by
$V_{-1} (z, \overline{z})$ and $V_{-{1\over 2}}(z, \overline{z})$
respectively. Vertex operators in the $q=0$ picture (with zero ghost charge)
is given by {\em picture-changing}~:
\begin{equation}
V_{0}(z,{\overline z})= \lim_{{\overline w}\rightarrow
{\overline z}}{ e^{\phi} T_F ({\overline z})
 V_{-1}(z,{\overline w})}~.
\end{equation}
Because the supercurrents in the $3$-family grand unified models have 
many terms, $V_{0}$ can be somewhat involved.

{}Having constructed the vertex operators for the massless states, one
can in principle compute the scattering amplitudes, or the corresponding
couplings in the superpotential. The coupling of $M$ chiral superfields in
the superpotential is given by the scattering amplitude of the component
fields in the limit when all the external momenta are zero.
Due to holomorphicity, one needs to consider only the scatterings of 
left-handed space-time fermions, with vertices $V_{-1/2}(z,{\overline z})$,
and their space-time superpartners.
Since the total $\phi$ ghost charge in any tree-level correlation 
function is $-2$, it is convenient to choose two of the vertex operators in 
the $-1/2$-picture,
one in the $-1$-picture, and the rest in the $0$-picture.
Using the $SL(2,{\bf C})$ invariance, the scattering amplitude is therefore
\begin{equation}\label{scattem}
 {\cal A}_{M} = g^{M-2}_{\mbox{st}}\int dz_{4} d \overline{z}_{4}
  \cdots dz_{M} d \overline{z}_{M}
        \langle V_{-{1\over 2}}(0,0)V_{-{1\over 2}}(1,1)
          V_{-1}(\infty,\infty) V_{0}(z_4,\overline{z}_{4})
          \cdots V_{0}(z_M,\overline{z}_{M}) \rangle ~,
\end{equation}
where we have normalized the $c$ ghost part of the correlation function
$\langle c(0,0) c(1,1) c(\infty,\infty) \rangle$ to $1$.

\subsection{Orbifold Rules for Level-1 Models}

{}We are now ready to give the rules for constructing consistent orbifold 
models with multiple twists using the bosonic supercurrent formalism. 
We will not give the derivation of these 
rules as they can be deduced from Ref \cite{kt} which contains more generic 
(and, therefore, more complicated) rules. Instead, we just list all the 
requirements that a consistent model must satisfy. In this subsection we 
will concentrate on models that have no non-trivial left-moving coset 
${\cal C}_L$. Models with left-moving twists (such as outer automorphisms) 
will be considered in the subsequent sections.

{}For a string model to be consistent, it must satisfy the following 
constraints which we impose:\\
(1) {\em Modular invariance}. One-loop partition function must be invariant 
under $S$ and $T$ modular transformations.\\
(2) {\em Physically sensible projection}. The physical states should 
appear in the partition function with proper weights and 
space-time statistics. The space-time bosons contribute $+1$
to the one loop vacuum amplitude while space-time fermions contribute
$-1$.\\
(3) {\em Worldsheet supersymmetry}. This is essential for space-time
Lorentz invariance. In order for the total supercurrent to have well
defined boundary condition, the space-time 
supercurrent and the internal supercurrent defined above should 
have the same monodromies. This is guaranteed by the triplet 
constraint (\ref{triplet}). Furthermore, the OPEs in (\ref{XOPE})
impose extra conditions on the coefficients ${\xi}^{a}({Q})$ in $T_{F}$.
This ensures the $N=2$ superconformal algebra is satisfied.

{}We start from a Narain model with the momenta of the internal bosons 
spanning an even self-dual Lorentzian lattice $\Gamma^{6,22}$. We introduce 
a set of generating vectors $\{V_i\}$ that includes the vector $V_0$. Next, 
we find the structure constants $k_{ij}$ that satisfy the following 
constraints:
\begin{eqnarray}
k_{ij}+k_{ji} &=& V_i \cdot V_j ~({\mbox{mod}}~1)~,\\
k_{ii}+k_{i0}+s_i -{1\over 2} V_i \cdot V_i &=& 0~({\mbox{mod}}~1)~,\\
k_{ij} m_j &=& 0~({\mbox{mod}}~1)~.
\end{eqnarray}
Note that there is no summation over the repeated indices here. The dot 
product of two vectors is defined with Lorentzian signature 
$((-)^{10},(+)^{22})$, {\em i.e.},
\begin{eqnarray} 
 V_i \cdot V_j &=& -\sum_a V^a_i V^a_j +{\vec U}_i \cdot {\vec U}_j \nonumber\\
   &=& -\sum_a V^a_i V^a_j -\sum_I U_i^I \cdot U_j^I
                         +\sum_A U_i^A \cdot U_j^A~.
\end{eqnarray}
Here we have combined the components $U^I_i$ and $U^A_i$ into a single 
$(6,22)$ dimensional vector ${\vec U_i}$. Note that this vector is a rational 
multiple of a vector in $\Gamma^{6,22}$, and, in particular, 
$m_i {\vec U}_i \in \Gamma^{6,22}$. The dot product of two vectors 
${\vec U}_i$ and ${\vec U}_j$ is then defined in the same way as for 
two vectors in $\Gamma^{6,22}$.

{}The above rules are not sufficient for the model to be consistent.
There must exist a supercurrent which satisfies the 
{\em triplet} constraint (\ref{triplet}) for all generating 
vectors $V_i$. Furthermore, the OPEs in (\ref{XOPE}) impose extra
conditions on the coefficients $\xi^{a}({Q})$ in $T_{F}$.
Thus, the rules that constrain the set $\{V_i ,k_{ij}\}$ 
together with the existence of the required supercurrent 
give the necessary and sufficient conditions 
for building a consistent string model.

{}The sectors of the model are $\alpha V$. In a given sector $\alpha V$ the 
states are nothing but the momentum states of ${\cal X}^\mu$, $\rho_a$, 
$\phi^I$ and $\varphi^A$. This means that the vertex operator for a given 
state has the form (in the covariant formalism where ${\cal X}^\mu$, 
$\mu=0,1,2,3$, are four space-time coordinates)
\begin{equation}\label{vertex} 
 V(z,{\overline z}) ={\tilde V}(z,{\overline z})
  \exp(i\sum_a H_a \rho_a ({\overline z}) +i\sum_I Q^I \phi^I ({\overline z}) +
 i\sum_A Q^A \varphi^A (z)) \exp(ik_\mu {\cal X}^\mu (z,{\overline z}))~,
\end{equation}
where ${\tilde V}(z,{\overline z})$ is a combination of 
ghost fields, derivatives of ${\cal X}^\mu$, $\rho_a$, $\phi^I$ and 
$\varphi^A$ ({\em i.e.},
this is the corresponding oscillator excitation contribution), and an
appropriate cocycle operator. The normal ordering is implicit here. 
Let us combine the 
$H$-charges $H_a$, $Q$-charges $Q^I$, and the gauge charges $Q^A$ into a 
$(10,22)$ dimensional momentum vector $P_{\alpha V}$. Then the physical 
states are those that satisfy the following spectrum generating formula:
\begin{equation} 
 V_i \cdot P_{\alpha V}=s_i +\sum_j k_{ij} \alpha_j ~({\mbox{mod}}~1)~,
\end{equation}
and the momenta $P_{\alpha V}\in {\bf Z}^4 \otimes \Gamma^{6,22}+\alpha V$. 
Thus, for example, $H_a \in {\bf Z}+ (\alpha V)^a$, and 
${\vec {\cal Q}} \in \Gamma^{6,22} +\alpha {\vec U}$, where we have combined 
the $Q$-charges $Q^I$ and the gauge charges $Q^A$ into a single $(6,22)$ 
dimensional vector ${\vec {\cal Q}}$. 

{}The above spectrum generating formula gives both on- and off-shell states. 
The on-shell states must satisfy the additional constraint that  the left- 
and right-moving energies be equal. In the $\alpha V$ sector they are given by
\begin{eqnarray}
 E^L_{\alpha V}&=&-1 +\sum_{q=1}^{\infty} q(\sum_\mu m^{\mu}_q +\sum_A n^A_q) +
 {1\over 2}(P^L_{\alpha V})^2~,\\
 E^R_{\alpha V}&=&-{1\over 2} +\sum_{q=1}^{\infty} 
 q(\sum_\mu {\tilde m}^{\mu}_q
 +\sum_I n^I_q+\sum_a 
 k^a_q) +{1\over 2}(P^R_{\alpha V})^2~,  
\end{eqnarray}
where $m^{\mu}_q$, ${\tilde m}^{\mu}_q$, $n^A_q$, $n^I_q$ and $k^a_q$ are the 
oscillator occupation numbers for the real bosons ${\cal X}^\mu_L$, 
${\cal X}^\mu_R$, $\varphi^A$, $\phi^I$ and $\rho_a$, respectively. 
Also, we note that $(P^L_{\alpha V})^2 =({\vec {\cal Q}}^L)^2$, and 
$(P^R_{\alpha V})^2 =({\vec {\cal Q}}^R)^2+\sum_a H^2_a$, where 
${\vec {\cal Q}}^L$ and ${\vec {\cal Q}}^R$ are the left- and right-moving 
parts of the vector 
${\vec {\cal Q}}$. (Here we note that for massless states that do not belong 
to the $N=1$ supergravity multiplet all the occupation numbers are zero and 
${\tilde V}(z,{\overline z})=1$ up to a factor that involves ghosts and a 
cocycle.)

{}As a simple illustration of these rules let us consider the model generated 
by a single vector $V_0$. This is nothing but a Narain model.  There is only 
one structure constant, namely, $k_{00}$. For $k_{00}=0$ we find that 
$\Delta^4$ is spanned by ${\bf v}$ and ${\bf c}$ momenta of $SO(8)_1$, 
whereas for the other choice $k_{00}=1/2$ we find that $\Delta^4$ is 
spanned by ${\bf v}$ and ${\bf s}$ momenta of $SO(8)_1$.

{}The above generating formula along with the constraints on the set 
$\{V_i, k_{ij} \}$ is all we need to construct the spectrum of any given 
model. For supersymmetric models, obtaining the spectrum is simplified 
further by space-time supersymmetry. For definiteness, let us concentrate on 
$N=1$ SUSY models. (Our discussion easily generalizes to models with larger 
SUSY.) In this case we have two SUSY generators $Q_L ({\overline z})$ and 
$Q_R ({\overline z})$, where the subscript indicates the space-time 
helicity. The vertex operators for these generators have the the same form 
as in (\ref{vertex}), but with ${\tilde V}(z,{\overline z})=1$ (up to ghosts
and a 
cocycle) and $Q^I = Q^A =k_\mu =0$, {\em i.e.}, only the $H$-charges are 
non-zero. Let us combine all of these charges into $(10,22)$ dimensional 
momenta ${\cal P}$. Then these momenta for the SUSY generators are 
determined by solving the following constraint:
\begin{equation}\label{SUSYgen} 
 V_i \cdot {\cal P}=-\sum_a V^a_i H_a ({\cal P}) =k_{i0}~({\mbox{mod}}~1)~,
\end{equation}
where $H_a ({\cal P})=\pm {1\over 2}$ are the corresponding $H$-charges for 
the SUSY generators. 
We will use the convention that the massless fermion states with $H_0=-1/2$ 
are left-handed, whereas those with $H_0=+1/2$ are right-handed. Then the 
solution of Eq. (\ref{SUSYgen}) with $H_0 ({\cal P})=-1/2$, which we will 
refer to as ${\cal P}_L$, corresponds to the left-handed SUSY generator 
$Q_L ({\overline z})$, whereas the solution with $H_0 ({\cal P})=+1/2$, 
which we will refer to as ${\cal P}_R$, corresponds to the right-handed 
SUSY generator $Q_R ({\overline z})$. (Note that ${\cal P}_L=-{\cal P}_R$.) 
These generators are useful in this case in the following way. Instead of 
working out the entire spectrum, we can confine our attention to 
left-moving fermion states in the $-1/2$-picture. These have $H_0=-1/2$ 
according to the above convention, and let the corresponding momenta be 
$P_{\alpha V}$. (Their CPT-conjugate states are right-handed with 
$H_0=+1/2$).  Their superpartners, which are boson states in the 
$-1$-picture, can be obtained by noting that their corresponding momenta 
would simply be $P_{\alpha^{\prime} V}={\cal P}_R +P_{\alpha V}$. Similarly, 
$P_{\alpha V}={\cal P}_L +P_{\alpha^{\prime} V}$. Note that 
$\alpha^{\prime }_i=\alpha_i$ for $i\not=0$, and $\alpha_0=1$, whereas 
$\alpha^{\prime}_0=0$. Thus, all we need to work out in this case is the 
spectrum for the left-moving fermion states. So by a vertex operator in the 
$-1/2$-picture we will always mean vertex operators for such states, whereas 
by those in the $-1$-picture we will mean vertex operators of the 
corresponding bosonic superpartners. Here we comment that the number of 
supersymmetries in a general case is given by a half of the number of 
solutions of Eq. (\ref{SUSYgen}).

\section{Orbifold Rules for Higher Level Models}

{}In this section we generalize the rules for the orbifold construction 
presented in section III. This generalization will allow us to construct 
models with reduced rank, in particular, models that contain gauge groups 
realized via higher level Kac-Moody algebras. The necessity of such a 
generalization can be seen from the fact that gauge symmetry in $N=1$ 
heterotic string models arises from the left-moving sector of the theory. 
On the other hand, the monodromies (\ref{varphi}) for the left-moving bosons 
$\varphi^A$ are such that they cannot project out Cartan generators of the 
original Kac-Moody algebra ${\cal G}_L$ of the Narain model that we orbifold.
Thus, the final gauge group ${\cal G}^{\prime}_L$ is always realized via a 
level-$1$ Kac-Moody algebra and has rank $22$. Thus, to obtain models with 
reduced rank, we must project out some of the original Cartan generators, 
that is, we have to twist some of the $22$ real left-moving bosons 
$\varphi^A$. Here we note that such twisting does not guarantee rank 
reduction. Sometimes it can happen that the model with such twists still 
has ${\cal G}^{\prime}_L$ with rank $22$. In this case it is always possible 
to rewrite the model so that all the left-moving bosons are shifted but not 
twisted. 

\subsection{Framework}

{}In this subsection we will set up the framework for the remainder of 
this section. We will borrow the notation from section III. 
The right-moving degrees of freedom are the same. The left-movers come in 
two varieties. There are $22-2d$ real bosons $\varphi^A$, $A=1,...,22-2d$, 
and also $d$ complex bosons $\Phi^r$, $r=1,...,d$. (These can be viewed 
as complexifications of the original $2d$ real bosons 
$\varphi^{23-2d},...,\varphi^{22}$ via 
$\Phi^r =(\varphi^{21-2d+2r} + i \varphi^{22-2d+2r})/\sqrt{2}$.) The string 
sectors are labeled by the monodromies of the string degrees of freedom:  
\begin{eqnarray}
 \rho_a ({\overline z}e^{-2\pi i})&=&\rho_a ({\overline z})-2\pi  V^a_i ~,
\nonumber\\
 \phi^I ({\overline z}e^{-2\pi i})&=& \phi^I ({\overline z})-2\pi  U^I_i~,
\nonumber\\
 \varphi^A (ze^{2\pi i})&=&\varphi^A (z) +2\pi  U^A_i~,\\
 \partial \Phi^r  (ze^{2\pi i})&=&\exp (-2\pi i T^r_i) \partial \Phi^r (z)~.
\nonumber
\end{eqnarray}
These monodromies can be combined into a single vector
\begin{equation}
 V_i=(V^a_i \vert U^I_i \vert\vert U^A_i \vert T^r_i )~.
\end{equation}
Without loss of generality we can restrict the values of $T^r_i$ as follows: 
$0\leq T^r_i <1$. This restriction is actually necessary for correctly 
identifying the sectors of the orbifold model in what follows.

{}The monodromies $V_i$ can be viewed as fields $\Phi$ (where $\Phi$ is a 
collective notation for $\rho_a$, $\phi^I$, $\varphi^A$ and $\Phi^r$) being 
periodic $\Phi (ze^{2\pi i} ,{\overline z}e^{-2\pi i})=\Phi(z,{\overline z})$
up to the identification $\Phi \sim g(V_i) \Phi g^{-1}(V_i)$, where 
$g(V_i)$ is an element of the orbifold group $G$. (Since we are considering 
Abelian orbifolds, we have excluded shifts from the monodromies of the 
bosons $\Phi^r$.) For $G$ to be a finite discrete group, the element 
$g(V_i)$ must have a finite order $m_i \in {\bf N}$, {\em i.e.}, $g^{m_i} 
(V_i)=1$. This implies that the vector $V_i$ must be a rational multiple 
of a vector in ${\bf Z}^4 \otimes \Gamma^{6,22} \otimes {\bf N}^d$, that 
is, $m_i V_i \in {\bf Z}^4 \otimes \Gamma^{6,22} \otimes {\bf N}^d$. In 
the component form we have: $m_i V^a_i \in {\bf Z}$, $m_i {\vec U}_i \in 
\Gamma^{6,22}$, and $m_i T^r_i \in {\bf N}$. Here we have combined the 
components $U^I_i$, $I=1,...,6$, and $U^A_i$, $A=1,...,22-2d$, along 
with $2d$ null entries into a single $(6,22)$ dimensional vector 
${\vec U}_i =(U^I_i \vert\vert U^A_i \vert 0^{2d})$.
Later we will use 
the dot product of such vectors defined in the same way as for two 
vectors in $\Gamma^{6,22}$. 

{}To describe all the elements of the orbifold group $G$, it is convenient 
to introduce the set of generating vectors$\{V_i\}$ such that 
${\overline {\alpha V}}={\bf 0}$ if and only if $\alpha_i \equiv 0$. 
Here ${\bf 0}$ is the null vector
\begin{equation}
 {\bf 0}=(0^4 \vert 0^6 \vert\vert 0^{22-2d} \vert 0^d )~.
\end{equation}
Also, ${\alpha V}\equiv\sum_i \alpha_i V_i$ (the summation is defined 
as, say, $(V_i+V_j)^r=T^r_i+T^r_j$), $\alpha_i$ being integers that take 
values from $0$ to $m_i-1$. The overbar notation is defined as follows:
$({\overline {\alpha V}})^{r}=(\alpha V)^{r}~({\mbox{mod}}~1)$ and
$0\leq ({\overline {\alpha V}})^{r} <1$. 

{}Now we can identify the sectors of the model. They are labeled by the 
vectors ${\overline{\alpha V}}$, and in a given sector 
${\overline {\alpha V}}$ the monodromies of the string degrees of freedom 
are given by 
$\Phi (ze^{2\pi i}, {\overline z}e^{-2\pi i})=g({\overline {\alpha V}}) 
\Phi (z, {\overline z}) g^{-1} ({\overline {\alpha V}})$. It is clear from 
the supercurrent constraint (\ref{super}), (\ref{triplet}) that the sectors 
with $\alpha s \in {\bf Z}$ give rise to the space-time bosons, whereas 
those with $\alpha s\in{\bf Z}+1/2$ give rise to the space-time fermions.

{}Note that a sector described by the vector 
\begin{equation} 
 V_0=((-{1\over 2})^4 \vert  0^6 \vert\vert 0^{22-2d}\vert 0^d)
\end{equation}
is always present in any orbifold model. As before, the $V_0$ sector is 
the Ramond sector of the Narain model, whereas ${\bf 0}$ sector is the 
Neveu-Schwarz sector. Without loss of generality we can set $s_i=0$ for 
$i\not=0$ (Recall that $s_i \equiv V^0_i$.) Then the space-time bosons 
come from the sectors ${\overline{\alpha V}}$ with $\alpha_0=0$, and 
the space-time fermions come from the sectors ${\overline {\alpha V}}$ 
with $\alpha_0=1$. 

{}We finish this subsection by noting that we have not modified the action 
of the orbifold on the right-moving degrees of freedom, so that all the 
rules concerning the supercurrent construction, in particular, the 
constraints (\ref{XOPE}) and (\ref{triplet}) remain unchanged and must be 
satisfied by a consistent orbifold model with left-moving twists within 
this framework just as in the case of models with no left-moving twists. 
The spectrum generating formula and the left-moving energy are different 
in the case of left-moving twists, and we turn to these issues next. 

\subsection{Orbifold Rules}

{}We will not attempt to give the most general rules in this subsection 
as the rules for constructing a rather large class of orbifolds can be 
found in Ref \cite{kt}. Rather, we will confine our attention to the case 
where we have only one generating vector, which we choose to be $V_1$, with 
$T^r_1\not=0$. For all the other vectors $V_i$, $i\not=1$, we require that 
$T^r_i\equiv 0$. (Thus, without loss of generality, we can assume that 
$T^r_1\not=0$ for all values of $r$.) Moreover, we only consider the 
cases where $m_1$ is a prime number.

{}Next, we start from a Narain model with the momenta of the internal bosons 
spanning an even self-dual Lorentzian lattice ${\Gamma}^{6,22}$. We introduce 
a set of generating vectors $\{V_i\}$ with the above properties. This set 
includes the $V_0$ vector, and also the $V_1$ vector with a left-moving 
twist. Here we assume that the lattice ${\Gamma}^{6,22}$ possesses 
${\bf Z}_{m_1}$ symmetry generated by the twist part of the orbifold 
group element $g(V_1$). Next, we find the structure constants $k_{ij}$ 
that satisfy the following constraints 
\begin{eqnarray}
 k_{ij}+k_{ji} &=& V_i \cdot V_j ~({\mbox{mod}}~1)~,~~~i \not=j~, \\
 k_{ii}+k_{i0}+s_i -t_i -{1\over 2} V_i \cdot V_i &=& 0~({\mbox{mod}}~1)~,\\
 k_{ij} m_j &=& 0~({\mbox{mod}}~1)~.
\end{eqnarray}
Note that there is no summation over the repeated indices here. The dot 
product of two vectors is defined as
\begin{eqnarray} 
 V_i \cdot V_j &=& -\sum_a V_i^a V_j^a + \vec{U}_i \cdot \vec{U}_{j} 
 \nonumber \\
              &=& -\sum_a V^a_i V^a_j 
                   -\sum_I U_i^I \cdot U_j^I
               +\sum_A U_i^A \cdot U_j^A ~.
\end{eqnarray}
Also, we have introduced the following notation:
\begin{equation}
 t_i \equiv {1\over 2} \sum_r T^r_i (1-T^r_i)~.
\end{equation}
Note that $t_i=0$ for $i\not=1$.

{}Let $I(V_1)$ be the sublattice of $\Gamma^{6,22}$ invariant under the 
twist part of the orbifold group element $g(V_1)$, and let ${\tilde I}(V_1)$ 
be the lattice dual to $I(V_1)$. Then for the model to be consistent 
(in particular, for level-matching dictated by modular 
invariance) we must have
\begin{equation}
 m_1 {\vec P}^2 \in {\bf Z}~{\mbox{for all}}~{\vec P}\in {\tilde I}(V_1)~. 
\end{equation}
Furthermore, for the sake of simplicity we will confine our attention to 
models with 
only one fixed point, so that we will require the following constraint to be 
satisfied:
\begin{equation}\label{fixed}
 \prod_r [2\sin (\pi T^r_1)] =\sqrt{{\mbox{Vol}}(I(V_1))}~. 
\end{equation}
Here ${\mbox{Vol}}(I(V_1))$ is the volume (or, equivalently, the determinant 
of the metric) of the lattice $I(V_1)$. (Note that Eq. (\ref{fixed}) can be 
relaxed so that the r.h.s. also contains a factor which is some integer 
power of $m_1$. Then this factor is nothing but the number of fixed points 
for the twist given by $(T^r_1)$. Here we only consider the models with one 
fixed point as we already mentioned.) 

{}Now we turn to describing the sectors of the theory. They are labeled by 
${\overline {\alpha V}}$. Let us start with the sectors with $\alpha_1=0$. 
To describe the vertex operators of states in these sectors, we will need 
to distinguish two different cases. They arise as follows. Note that there 
are two types of momenta $P\in {\Gamma}^{6,22}$: those that belong to the 
invariant sublattice $I(V_1)$, and those that do not. Let us consider the 
latter type first. Thus consider a set of momentum vectors 
$N(V_1) \subset \Gamma^{6,22}$ such that if $P\in N(V_1)$, then 
$P\not\in I(V_1)$. Let $\vert P\rangle$ be the corresponding momentum 
states. It is clear that we can always decompose $P$ into 
$P=P^\perp +P^\parallel$, where $P^\parallel \in I(V_1)$, and 
$P^\perp \cdot P^\prime =0$ for all $P^\prime \in I(V_1)$. Then 
$\vert P\rangle =\vert P^\parallel\rangle \otimes \vert P^\perp \rangle$. 
Let $N^*(V_1)\subset N(V_1)$ be a set of momenta spanned by all $P^\perp$, 
and let ${\cal H}$ be the corresponding Hilbert space,
 {\em i.e.}, ${\cal H}$ is spanned by states $\vert P^\perp \rangle$, 
$P^\perp \in N^*(V_1)$. This space can be represented as 
${\cal H}=\otimes_{\ell=0}^{m_1-1} {\cal H}_\ell$, where 
${\cal H}_\ell$ is spanned by the states of the form 
\begin{equation}
 \vert P^\perp; \ell\rangle =
 {1\over \sqrt{m_1}}\sum_{k=0}^{m_1-1} \exp (-2\pi ik\ell /m_1) g^k 
\vert P^\perp \rangle~,
\end{equation}    
where $g=\exp(-2\pi i\sum_r T^r_1 J^r)$, and $J^r$ is the angular momentum 
operator (that acts on the momenta) for the complex boson $\Phi^r$ (or, 
equivalently, for the real bosons $\phi^{21-2d+2r}$ and $\phi^{22-2d+2r}$, 
and in this language $J^r$ is the generator of $SO(2)$ rotations in the 
plane of these bosons). Note that $g \vert P^\perp; \ell\rangle=\exp(2\pi 
i \ell/ m_1) \vert P^\perp; \ell\rangle$. Later, we will identify $\ell$ as
part of a discrete charge ($D$-charge).

{}In the sectors with  $\alpha_1=0$ there are two kinds of vertex 
operators. The first kind are momentum states (in the covariant formalism):
\begin{equation}\label{vertex1} 
 V (z,{\overline z}) ={\tilde V}(z,{\overline z})
  \exp(i\sum_a H_a \rho_a ({\overline z}) +i\sum_I Q^I \phi^I 
({\overline z}) +
 i\sum_A Q^A \varphi^A (z)) \exp(ik_\mu {\cal X}^\mu (z,{\overline z}))~,
\end{equation}
with ${\vec {\cal Q}}=(Q^I\vert\vert Q^A)\in I(V_1)$. (Note that 
${\vec {\cal Q}}$ is a $(6,22-2d)$ dimensional vector.) Here 
${\tilde V}(z,{\overline z})$ is a combination of derivatives of 
${\cal X}^\mu$, $\rho_a$, $\phi^I$, $\varphi^A$ and $\Phi^r$ ({\em i.e.}, 
this is the corresponding oscillator excitation contribution), ghosts, certain 
cocycles, and the normal ordering is implicit here. Let us combine the 
$H$-charges $H_a$, $Q$-charges $Q^I$, and the gauge charges $Q^A$ into a 
$(10,22-2d)$ dimensional momentum vector $P_{\overline {\alpha V}}$. Then 
the physical states are those that satisfy the following spectrum 
generating formula:
\begin{equation}\label{SGFU1}
 V_i \cdot P_{\overline {\alpha V}} + {\delta_{i1} \over m_1} D \equiv
 V_i \cdot P_{\overline {\alpha V}} +\sum_r T^r_i N^r=s_i +\sum_j k_{ij} 
\alpha_j~({\mbox{mod}}~1)~,
\end{equation}
and the momenta $P_{\overline {\alpha V}}\in {\bf Z}^4\otimes I(V_1) +
{{\alpha V}}$. Thus, for example, $H_a \in {\bf Z}+
({\overline {\alpha V}})^a$, and ${\vec {\cal Q} }\in I(V_1) +
{\alpha U}$. Here we have introduced the boson number operators $N^r$ 
for the bosons $\Phi^r$. It can be expressed in terms of the occupation 
number operators $s^r$ and ${\tilde s}^r$ for these bosons: 
$N^r_{\overline {\alpha V}}=\sum_q^{\infty} (s^r_q-{\tilde s}^r_q)$.
The discrete $D$-charge is defined by the above equation. The origin
of this $D$-charge (defined modulus $m_1$) is the 
${\bf Z}_{m_1}$ twist acting on the 
$d$ complex bosons $\partial \Phi^r$. 
This, however, is not the most general form of $D$ as it has 
contributions only from the oscillators. We will give the general form of
$D$ in a moment. 

{}The vertex operators (in the covariant formalism) for the 
second kind of states have the same form as (\ref{vertex1}), but now,
in addition to the derivatives of 
${\cal X}^\mu$, $\rho_a$, $\phi^I$, $\varphi^A$ and $\Phi^r$, ghosts and 
cocycles, ${\tilde V} (z,{\overline z})$ contains also 
the vertex operator for a state $\vert P^\perp_{\overline 
{\alpha V}}; \ell\rangle$. The latter can be written as
\begin{equation}
 {1\over \sqrt{m_1}}\sum_{k=0}^{m_1-1} \exp (-2\pi ik\ell /m_1) g^k 
\exp(i \sum_r [q^r  
 (\Phi^r)^\dagger (z)+(q^r)^* \Phi^r (z)])~,
\end{equation}
where $q^r$ is the (complex) momentum of the $\Phi^r$ boson (and the $2d$ 
real components of all $d$ momenta $q^r$ give the momentum vector 
$P^\perp_{\overline {\alpha V}}$). The physical states 
are those that satisfy the following spectrum generating formula: 
\begin{equation}\label{SGFU2}
 V_i \cdot P^\parallel_{\overline {\alpha V}} +{\delta_{i1} \over m_1}D \equiv
 V_i \cdot P^\parallel_{\overline {\alpha V}} +\sum_r T^r_i 
N^r_{\overline {\alpha V}} +
 {\delta_{i1} \over m_1} \ell =s_i +\sum_j k_{ij} \alpha_j~({\mbox{mod}}~1)~.
\end{equation}
Here we have combined the charges $H_a$, $Q^I$ and $Q^A$ into a single 
vector $P^\parallel_{\overline {\alpha V}}$. The lattice momenta
in this sector are given by 
$P_{\overline {\alpha V}}=P^\parallel_{\overline {\alpha V}}+
P^\perp_{\overline {\alpha V}}$.

{}The general form of the $D$-charge is given by:
\begin{equation}
D \equiv \ell + m_1 \sum_r T^r_1 N^r_{\overline{\alpha V}} \pmod{m_1}
\end{equation}
which has contributions from both the lattice momentum and the oscillators.
We recover the previous case if we set $\ell = 0$ 
({\em i.e.} $P_{\overline{\alpha V}} \in I(V_1)$).

{}Finally, the vertex operators for states in the
sectors ${\overline {\alpha V}}$ with $\alpha_1\not=0$ read:
\begin{eqnarray}\label{vertex3} 
 V (z,{\overline z}) =&&
 {\tilde V}(z,{\overline z}) \sigma_{\overline {\alpha V}} (z) 
\times\nonumber\\
 &&\exp(i\sum_a H_a \rho_a ({\overline z}) +i\sum_I Q^I \phi^I 
({\overline z}) +
 i\sum_A Q^A \varphi^A (z)) \exp(ik_\mu {\cal X}^\mu (z,{\overline z}))~,
\end{eqnarray}
where ${\tilde V}(z,{\overline z})$ is as defined in 
(\ref{vertex1})
and $\sigma_{\overline {\alpha V}} (z)$ is a vertex operator for 
the twisted ground state with conformal dimension $\sum_r {1\over 2} 
({\overline {\alpha V}})^r (1-({\overline {\alpha V}})^r)$. The physical 
states are those that satisfy the following spectrum generating formula:
\begin{equation}\label{SGFT}
 V_i \cdot P_{\overline {\alpha V}} =s_i +\sum_j k_{ij} 
\alpha_j~({\mbox{mod}}~1)~,~~~i\not=1~.
\end{equation}
Here $P_{\overline {\alpha V}}\in {\bf Z}^4 \otimes {\tilde I}(V_1)+
{\overline {\alpha V}}$. (Here we just take the momentum part of 
${\overline {\alpha V}}$.)
Note that here we are not imposing the constraint with respect to the 
vector $V_1$ as the latter is automatically satisfied in these sectors.

{}The above spectrum generating formulas give both on- and off-shell states. 
The on-shell states must satisfy the additional constraint that the left- 
and right-moving energies be equal. In the ${\overline {\alpha V}}$ sector 
they are given by
\begin{eqnarray}
 E^L_{\overline {\alpha V}}=&&
 -1 +\sum_{q=1}^{\infty} (q\sum_\mu m^{\mu}_q +q\sum_A n^A_q \nonumber\\
 &&+ \sum_r [(q+({\overline {\alpha V}})^r-1)s_q +
(q-({\overline {\alpha V}})^r){\tilde s}_q])+
 {1\over 2}(P^L_{\overline {\alpha V}})^2 ~,\\
 E^R_{\overline {\alpha V}}=&&
 -{1\over 2} +\sum_{q=1}^{\infty} q(\sum_\mu {\tilde m}^{\mu}_q +\sum_I  
 n^I_q+\sum_a k^a_q) +{1\over 2}(P^R_{\overline {\alpha V}})^2~,  
\end{eqnarray}
where $m^{\mu}_q$, ${\tilde m}^{\mu}_q$, $n^A_q$, $n^I_q$ and $k^a_q$ are 
the oscillator occupation numbers for the real bosons ${\cal X}^\mu_L$, 
${\cal X}^\mu_R$, $\varphi^A$, $\phi^I$ and $\rho_a$, respectively. Also, 
we note that $(P^L_{\overline {\alpha V}})^2 =({\vec {\cal Q}}^L)^2$, and 
$(P^R_{\overline {\alpha V}})^2 =({\vec {\cal Q}}^R)^2+\sum_a H^2_a$, 
where ${\vec {\cal Q}}^L$ and ${\vec {\cal Q}}^R$ are the left- and 
right-moving parts of the vector ${\vec {\cal Q}}$.

\subsection{Scattering Amplitudes}

{}From the scattering amplitudes ${\cal A}_{M}$ (\ref{scattem}), where the 
external space-time momenta are set to zero, we can read off the terms in 
the superpotential. For a non-zero coupling, the corresponding scattering 
amplitude must be present. Having taken care of the ghost factors,
this means that all the gauge and discrete symmetries must be satisfied.
In particular, a necessary condition is that the sum of all the lattice 
momenta must be zero in ${\cal A}_{M}$ (\ref{scattem}). These selection 
rules impose very tight constraints on the possible terms that can 
appear in the superpotential of any model. 

\noindent
(1) {\em Gauge Invariance}. This local symmetry must be conserved. 
In the $4$-dimensional $N=1$ heterotic string models, the gauge 
symmetries come from left-movers only. We will refer to these as 
$G$-charges. Note that picture-changing does not touch the
$G$-charges so each state carries well-defined gauge quantum numbers.

\noindent
(2) {\em $H$- and $Q$-Charge Conservation}. They must be conserved
in the scattering amplitude. Note that the supercurrent carries terms 
with different $H$- and $Q$-charges. Because of picture
changing, $H$- and $Q$- charges are not global charges even though
they must be conserved exactly in ${\cal A}_{M}$. This is consistent
with the fact that string theory has no global continuous symmetries
\cite{global}. Point group and space 
group selection rules follow from these conservation laws.

\noindent
(3) {\em Invariance under Discrete Symmetries}. In higher level models,
there is a discrete gauge charge (or quantum number) associated with the 
twist field. We shall call this a $D$-charge. As we shall see, in the 
models we are interested in, the selection rule coming from this discrete 
symmetry is subsumed in the other selection rules.

In principle, we can calculate the couplings in the superpotential 
explicitly, since we know all the vertex operators. However, in this paper, 
we shall consider only the selection rules coming from the conservation of
the above $G$-, $Q$-, $H$- and $D$-charges.
Note that space-time superpartners have identical $G$-, $Q$- and $D$-charges,
but different $H$-charges.

\section{Simple Level-1 Examples}

{}In this section we use some simple examples to illustrate 
our rules for constructing level-1 models and calculating scattering
amplitudes. After a 
detailed discussion of the familiar asymmetric ${\bf Z}_3$ orbifold, we  
will briefly discuss the original symmetric $Z$-orbifold. 

{}Consider the Narain model with 
$\Gamma^{6,22} =\Gamma^{2,2} \otimes \Gamma^{2,2} \otimes \Gamma^{2,2} 
\otimes \Gamma^8 \otimes \Gamma^8$, where $\Gamma^8$ is the $E_8$ root 
lattice, whereas $\Gamma^{2,2}=\{(p_R \vert\vert p_L)\}$ with 
$p_L,p_R \in{\tilde {\Gamma}}^2$ ($SU(3)$ weight lattice) and 
$p_L-p_R \in{\Gamma}^2$ ($SU(3)$ root lattice). This model has $N=4$ SUSY 
and $SU(3) \otimes SU(3) \otimes SU(3)\otimes E_8 \otimes E_8$ gauge 
symmetry (counting only the gauge bosons coming from the left-moving 
sector of the string; the right-moving sector contributes $6$ $U(1)$ 
vector bosons that are part of the $N=4$ supergravity multiplet). 
Let us write the corresponding $V_0$ vector as
\begin{equation}
 V_0 =((-{1\over 2})^4 \vert 0^3\vert\vert 0^3\vert 0^8 \vert 0^8)~.
\end{equation} 
Here the first four entries stand for the right-moving complex world-sheet 
fermions $\psi^a$, $a=0,1,2,3$, next three stand for $3$ right-moving 
complex bosons $X^a$, $a=1,2,3$ (each corresponding to a factor 
$\Gamma^{2,2}$). Double vertical line separates the right-movers from 
the left-movers. The first three left-moving entries correspond to the 
left-moving counterparts of the $X^a$ bosons. The next $8+8$ entries 
correspond to the $E_8 \otimes E_8$ lattice which we will describe using 
$16$  real bosons, and we will use the ${\mbox{Spin}}(16)/{\bf Z}_2$ basis 
for each of the $E_8$ factors ({\em i.e.}, the $E_8$ roots are described as 
those of $SO(16)$ plus $128$ additional roots in the corresponding irrep of 
$SO(16)$; thus, for example, $(+1, 0, -1, 0,0,0,0,0)$ is a root of $SO(16)$, 
and the roots in the $128$ irrep are those with all eight entries equal 
$+1/2$ or $-1/2$ and total number of positive signs being even).

\medskip

\noindent {\em (1) An asymmetric $Z$-orbifold model.} 

{}Consider the following asymmetric ${\bf Z}_3$ orbifold of the 
$SU(3)^3 \otimes (E_8)^2$ Narain model described above:
\begin{equation}
 V_1 =(0(-{1\over 3})^3 \vert (e_1/3)^3\vert\vert 0^3\vert 
 {1\over 3}{1\over 3}{2\over 3} 
 0^5 \vert 0^8)~.
\end{equation}
This is the original asymmetric $Z$-orbifold model of Ref \cite{NSV}. 
It has $N=1$ SUSY and $SU(3)\otimes E_6 \otimes E_8 \otimes SU(3)^3$ 
gauge symmetry. 
Here $e_1$ is a simple root of $SU(3)$. We denote the other simple 
root by $e_2$, and define $e_3 \equiv -e_1 -e_2$. Note that 
$e_\alpha \cdot e_\beta =2$ for $\alpha=\beta$ and $-1$ for 
$\alpha\not=\beta$ ($\alpha,\beta=1,2,3$). 
Let $\phi^a$, $a=1,2,3$, be the two-component real bosonic fields 
corresponding to each of the three $SU(3)$ lattices in the model with 
$e_1/3$ shifts. The $i \partial X ^a$ in the supercurrent $T_{F}$ is given by
\begin{equation}\label{SU3}
 i \partial X ^a = {1 \over {\sqrt 3}} \sum_\alpha 
e^{-ie_\alpha \phi ^a} c(-e_\alpha)  ~.
\end{equation}
It is easy to see that the {\em triplet} constraint is satisfied:
\begin{equation}
V^a_1 + U_1 \cdot Q =
-{1 \over 3} + {e_{1} \over 3} \cdot (-e_{\alpha}) 
= 0 \pmod{1}
\end{equation}

{}We can now compute the massless spectrum of the model. 
First, let us choose $k_{00}=0$ (note that there is a freedom in choosing 
$k_{00}$ to be $0$ or $1/2$ which only reflects in flipping the chirality 
of the states). The other structure constants are fixed: $k_{10}=1/2$, 
$k_{01}=0$ and $k_{11}=1/3$. It is convenient to define  
\begin{equation}
P_{\alpha V}=(H_{0},\cdots,H_{3},{\bf Q}^{R}_{1},{\bf Q}^{R}_{2},
{\bf Q}^{R}_{3} \vert {\bf Q}^{L}_{1}, {\bf Q}^{L}_{2}, {\bf Q}^{L}_{3},
Q_{4}, \cdots, Q_{19})
\end{equation}
where ${\bf Q}^{L,R}$ are charges under $U(1)^{2}$ of $SU(3)$ and all other 
$Q$'s are $U(1)$ charges. 

{}First, consider the untwisted sector, the spectrum generating formulae 
read:
\begin{eqnarray}
{1\over 2} (H_0 + H_1 + H_2 + H_3) &=& {1\over 2} \pmod{1} \\
{1\over 3} (Q_4 + Q_5 + 2 Q_6) + {1\over 3} (H_1+ H_2 + H_3)
-{e_{1} \over 3} \cdot ({\bf Q}^{R}_{1} + {\bf Q}^{R}_{2} +{\bf Q}^{R}_{3}) 
&=& 0 \pmod{1}
\end{eqnarray}
The first spectrum generating formula requires that one of the $H$-charges 
must be equal to $1$. It then follows from the massless condition that
${\bf Q}^R=0$ which implies that ${\bf Q}^{L} \in \Gamma^{2}$ 
($SU(3)$ root lattice). The choice $a=0$ gives rise to gauge bosons
of $SU(3) \otimes E_{6} \otimes E_{8} \otimes SU(3)^{3}$. For 
$a=1,2, {\mbox{or}}~3$,
the second spectrum generating formula gives $Q_4=-1$ or $Q_5=-1$ or 
$Q_6=1$. Therefore, ${\bf Q}^{L}=0$ or else $({\bf Q}^{L})^2 \geq 2$ and the 
state is massive. Note that even though the orbifold group does not act
on ${\bf Q}^{L}$, constraints such as ${\bf Q}^{L}-{\bf Q}^{R} \in \Gamma^{2}$
and massless condition restrict the possible values of ${\bf Q}^{L}$. 
The fields that survive the projection are $\chi_{a}$ 
which transform in the $({\bf 3}, {\bf 27}, {\bf 1})$ 
irrep of $SU(3)\otimes E_6 \otimes E_8$, and are neutral under the other 
$SU(3)^3$. 
The index $a$ labels the choice of $H$-charges: $H=(1,0,0)$ for $a=1$,
$H=(0,1,0)$ for $a=2$ and $H=(0,0,1)$ for $a=3$. 

{}We now turn to the twisted sectors. Note that in the $-1$ picture,
$H_{a} \in {\bf Z} - {1\over 3}$ in $V_1$ sector while
$H_{a} \in {\bf Z} + {1\over 3}$ in $2V_1$ sector. The possible
choices of $H_{a}$ which do not give rise to massive states are
$H_{a}=(-{1\over 3},-{1\over 3},-{1\over 3})$ in $V_1$ sector and 
$H_{a}=({1\over 3},{1\over 3},{1\over 3})$ in $2V_1$ sector. 
The left-handed chiral supermultiplets all come from $2V_{1}$ and
$V_{0}+ 2 V_{1}$ sectors. 
In $2V_1$ sector, the spectrum generating formulae give:
\begin{eqnarray}
{1\over 2} (H_0 + H_1 + H_2 + H_3) &=& {1\over 2} \pmod{1} \\
{1\over 3} (Q_4 + Q_5 + 2 Q_6) + {1\over 3} (H_1+ H_2 + H_3)
-{e_{1} \over 3} \cdot ({\bf Q}^{R}_{1} + {\bf Q}^{R}_{2} +{\bf Q}^{R}_{3}) 
&=& {2\over 3} \pmod{1}
\end{eqnarray}
where $Q_4,Q_5 \in {\bf Z}-{1\over 3}$, $Q_6 \in {\bf Z}+ {1\over 3}$
and ${\bf Q}^{R} \in \tilde{\Gamma}^{2} - e_{1}/3$. For massless states,
$\sum_{j} ({\bf Q}_{j}^{R})^{2}=2/3$ and therefore, there are $27$ choices of
${\bf Q}^{R}$:
\begin{equation}
{\bf Q}^{R}= \left( -{e_{\alpha} \over 3},-{e_{\beta} \over 3},
       -{e_{\gamma} \over 3} \right) ~~~{\mbox{for}}~\alpha,\beta,\gamma=1,2,3
\end{equation} 
The $Q$ charges ${\bf Q}^{L}$ and ${\bf Q}^{R}$ are correlated by
$p_{L}-p_{R} \in \Gamma^{2}$. For simplicity, let us consider for the
moment only the first $SU(3)$. We have
\begin{equation}
\begin{array}{llllc}
& & & & \underline{{\mbox{irrep of }}~SU(3)} \\
\alpha=1  & \phantom{6} &{\bf Q}^{L}_{1}=0 & \phantom{6} &{\bf 1} \\
\alpha=2  & &{\bf Q}^{L}_{1}=\tilde{e}_{2},-\tilde{e}_{1},
\tilde{e}_{1}- \tilde{e}_{2}  & & {\bf 3}  \\
\alpha=3 & &{\bf Q}^{L}_{1}=-\tilde{e}_{2},\tilde{e}_{1},
\tilde{e}_{2} - \tilde{e}_{1} & & {\bf {\overline{3}}} \\
\end{array}
\end{equation}
where $\tilde{e}_{1}={2 \over 3} e_{1} + {1 \over 3} e_{2}$ and
$\tilde{e}_{2}={1 \over 3} e_{1} + {2 \over 3} e_{2}$ are $SU(3)$ weights 
such that
$\tilde{e}_{i} \cdot e_{j} = \delta_{ij}$. The conformal dimension
of ${\bf 3}$ (and ${\bf {\overline{3}}}$) of $SU(3)$ is  
${1 \over 2} ({\bf Q}^{L}_{1})^2= {1\over 3}$
as expected.
 
{}Massless states are created by conformal fields with total left-moving 
conformal dimension $1$. 
Therefore, they must transform non-trivially under $SU(3)$ as well as
other gauge group such as $E_6$. For instance, a field that transforms as 
${\bf 3}$ of $SU(3)$ and $\overline{\bf 27}$ of $E_6$ (which has conformal 
dimension $2/3$) has the right conformal dimension. It remains to
check if the spectrum generating formulae are satisfied.  
To summarize, we have the following left-handed chiral supermultiplets
from the twisted sectors:\\
$\bullet$ The twisted sector field ${\overline \chi}$.\\
This field transforms in the $({\bf 3}, {\overline {\bf 27}}, {\bf 1}, 
{\bf 1},{\bf 1},{\bf 1})$ irrep of $SU(3) \otimes E_6 \otimes E_8 \otimes 
SU(3)^3$. The $Q$-charges for this field read $(-e_1/3,-e_1/3,-e_1/3)$.\\
$\bullet$ The twisted sector fields ${\chi}_{A\pm}$.\\
The field ${\chi}_{A+}$ transforms in the $({\bf 1}, {{\bf 27}},{\bf 1}, 
{\bf x}, {\bf y},{\bf z})$ irrep of 
$SU(3) \otimes E_6 \otimes E_8 \otimes SU(3)^3$ with ${\bf x}={\bf 3}$, 
${\bf y}={\bf z}={\bf 1}$ for $A=1$, ${\bf y}={\bf 3}$, 
${\bf x}={\bf z}={\bf 1}$ for $A=2$, and ${\bf z}={\bf 3}$, 
${\bf x}={\bf y}={\bf 1}$ for $A=3$. Similarly, the field 
${\chi}_{A-}$ transforms in the $({\bf 1}, {{\bf 27}},{\bf 1}, {\bf x}, 
{\bf y},{\bf z})$ irrep of  $SU(3) \otimes E_6 \otimes E_8 \otimes SU(3)^3$ 
with ${\bf x}={\overline {\bf 3}}$, ${\bf y}={\bf z}={\bf 1}$ for $A=1$, 
${\bf y}={\overline {\bf 3}}$, ${\bf x}={\bf z}={\bf 1}$ for $A=2$, and 
${\bf z}={\overline {\bf 3}}$, ${\bf x}={\bf y}={\bf 1}$ for $A=3$. The 
$Q$-charges for the fields ${\chi}_{A+}$ are given by 
$(-e_2/3,-e_1/3, -e_1/3)$ for $A=1$,  $(-e_1/3,-e_2/3, -e_1/3)$ for $A=2$, 
and  $(-e_1/3,-e_1/3, -e_2/3)$ for $A=3$. Similarly, the $Q$-charges for the 
fields ${\chi}_{A-}$ are given by $(-e_3/3,-e_1/3, -e_1/3)$ for $A=1$,  
$(-e_1/3,-e_3/3, -e_1/3)$ for $A=2$, and  $(-e_1/3,-e_1/3, -e_3/3)$ for 
$A=3$.\\
$\bullet$ The twisted sector fields ${T}_{{\bf x}{\bf y}{\bf z}}$.\\
The field ${T}_{{\bf x}{\bf y}{\bf z}}$ transforms in the 
$({\overline {\bf 3}}, {{\bf 1}},{\bf 1}, {\bf x}, {\bf y},{\bf z})$ 
irrep of 
$SU(3) \otimes E_6 \otimes E_8 \otimes SU(3)^3$ 
where ${\bf x},{\bf y},{\bf z}$ are irreps of $SU(3)$ such that only one
of them is a singlet and the others can be either ${\bf 3}$ or
${\bf {\overline{3}}}$. The $Q$-charges are correlated with the
$SU(3)$ irrep as described above. For example, if ${\bf x}={\bf 1},
{\bf y}={\bf z}={3}$, the $Q$-charges are given by
$(-e_{1}/3,-e_{2}/3,-e_{2}/3)$.

{}The gauge quantum numbers as well as the $Q$- and $H$-charges
of the massless spectrum are summarized in Table I.

{}Having described the bosonic supercurrent and the massless spectrum 
of the model, we are ready to calculate scattering amplitudes. 
Let us start with the three-point Yukawa interactions of the chiral 
families in ${\bf 27}$ of $E_6$. 
The three-point 
Yukawa coupling $\chi_a \chi_b \chi_c$ is non-zero only for 
$a\not=b\not=c\not=a$ as only 
in this case are the $H$-charges conserved. For illustrative purposes 
let us consider this case in more detail. If $a=1$, $b=2$ and $c=3$, then 
in the $-1$-picture the $H$-charges are $(+1,0,0)$, $(0,+1,0)$ and 
$(0,0,+1)$, respectively. Two of the fields must be in the $-1/2$-picture, 
and the third one must be in the $-1$-picture, however. Let the $\chi_a$ and 
$\chi_b$ fields in the $-1/2$-picture. Their $H$-charges then are given by 
$(+1/2,-1/2,-1/2)$ and $(-1/2,+1/2,-1/2)$. Then we see that the total 
$H$-charge is $(+1/2,-1/2,-1/2)+(-1/2,+1/2,-1/2)+(0,0,+1)=(0,0,0)$, and the 
process is allowed by the $H$-charge conservation. All the other quantum 
numbers are also conserved in this case. It is easy to see that in all the 
other cases $\chi_a \chi_b \chi_c$ Yukawa coupling vanishes.
 
{}Now consider the three-point couplings of the twisted sector fields. 
Naively, just from the conservation of gauge charges, one might expect that, 
say, the coupling $\chi_{A+} \chi_{B+} \chi_{C+}$ is non-zero for $A=B=C$. 
This is, however, not the case. Indeed, to have a singlet of the 
corresponding $SU(3)$ (note that each of these three fields transform in the 
irrep ${\bf 3}$ of the corresponding $SU(3)$), we must completely 
antisymmetrize these fields. This means that we must take the completely 
antisymmetric combination of the three ${{\bf 27}}$s (note that each of 
these three fields transform in the irrep ${ {\bf 27}}$ of $E_6$). The 
latter antisymmetric product does not contain a singlet of $E_6$, so that 
the trace of this product vanishes. The same conclusion can be drawn from 
the $Q$-charge non-conservation. Note
that the $Q$-charges in this case are not 
conserved. (Recall that for the three-point couplings there is no picture 
changing insertions of the supercurrent.) In fact, it is easy to see that 
for the same reason all three-point couplings involving at least one 
twisted sector field vanish. Some higher point couplings, however, are 
non-zero. For example, consider the following six-point couplings: 
${\overline \chi}{\overline \chi}{\overline \chi}\chi_a \chi_b \chi_c$. It 
can be easily checked that $H$-, $Q$- and $G$-charge conservations in this 
case imply that this coupling is non-zero if and only if 
$a\not=b\not=c\not=a$.  This is a typical situation for asymmetric orbifolds: 
lower point couplings typically vanish because of the $Q$-charge 
non-conservation, whereas there are (usually) somewhat higher point couplings 
that are non-zero. This observation will play an important role for the 
superpotentials of the three-family grand unified string theories which we 
will discuss in the subsequent sections.

{}For illustrative purpose, we give the lowest order non-vanishing terms 
in the superpotential of this asymmetric ${\bf Z}_3$ orbifold:
\begin{equation}
W= (\lambda_1  + \lambda_2 \overline{\chi}^3)
 \sum_{a\not=b\not=c\not=a} \chi_a \chi_b \chi_c + \dots ~.
\end{equation}

{}In passing, we remark that the above model was originally constructed
in the twist basis:
\begin{equation}
 V_1 =(0(-{1\over 3})^3 \vert \theta^3\vert\vert 0^3\vert {1\over 3}{1\over 3}
{2\over 3} 
 0^5 \vert 0^8)~.
\end{equation}
Here $\theta$ denotes a $2 \pi/3$ rotation in the corresponding $SU(3)$ 
lattice. The above generating vector defines the same asymmetric $Z_{3}$ 
orbifold model.
The relation between the complex bosons $i \partial X^a$ in the twist 
formalism and the real bosons $\phi^a$ in the shift formalism is given 
in Eq.(\ref{SU3}).

\medskip

\noindent {\em (2) A symmetric ${Z}$-orbifold model}. 

{}Consider the following ${\bf Z}_3$ orbifold of the above Narain model:
\begin{equation}
 V_1 =(0(-{1\over 3})^3 \vert (e_1/3)^3\vert\vert (e_1/3)^3\vert 
 {1\over 3}{1\over 3}{2\over 3} 0^5 \vert 0^8)~.
\end{equation} 
This model has $N=1$ SUSY, $U(1)^6 \otimes SU(3) \otimes E_6 \otimes E_8$ 
gauge group, and $36$ chiral families of fermions in the ${\bf 27}$ of $E_6$. 
Nine of these come from the untwisted sector, and the other $27$ from the 
twisted sector. This orbifold model is nothing but the original $Z$-orbifold 
model of Ref \cite{DHVW} at the special values of the moduli of the 
compactification torus. This is precisely the reason for the enhanced 
$U(1)^6$ gauge symmetry.
Since the group actions are identical on the right-movers for
these symmetric and asymmetric orbifolds, we can use the same
supercurrent $T_{F}$.

{}Let us fix the structure constants of the model. They
are the same as in the asymmetric case except for $k_{11}=-1/3$.
We have the following left-handed chiral supermultiplets:\\
$\bullet$ The untwisted sector fields $S_{aA\alpha}$.\\
The fields $S_{aA\alpha}$ are $SU(3)\otimes E_6 \otimes E_8$ singlets that 
are charged under $U(1)^6$. Thus, for $A=1$ the $U(1)$ charges are given by 
$(e_\alpha,0,0)$, for  $A=2$ the $U(1)$ charges are given by 
$(0, e_\alpha,0)$, and for  $A=3$ the $U(1)$ charges are given by 
$(0,0,e_\alpha)$. There is no correlation between the gauge quantum 
numbers and the index $a$. The latter is related to the $H$-charges 
(here we only give $H_{1,2,3}$ charges in the $-1$ picture; 
the corresponding $H$-charges for $Q_L$ left-handed SUSY generator are 
given by $(-1/2,-1/2,-1/2)$). For $a=1$ they are $(+1,0,0)$, for $a=2$ 
they are $(0,+1,0)$, and for $a=3$ they are $(0,0,+1)$.\\
$\bullet$ The untwisted sector fields $\chi_{a}$.\\
These are the same as in the asymmetric case. \\
$\bullet$ The twisted sector fields $\chi_{\alpha \beta \gamma}$.\\
The $27$ fields $\chi_{\alpha \beta \gamma}$ transform in the irrep 
$({\bf 1}, {\bf 27}, {\bf 1})$ of $SU(3)\otimes E_6 \otimes E_8$. Their 
$U(1)^6$ charges are given by $(-e_\alpha/3, -e_\beta/3, -e_\gamma/3)$. 
Their $Q$-charges are the same as their $U(1)^6$ charges in the spectrum 
picture. For illustrative purposes we give the part of the vertex operator 
for the field $\chi_{\alpha \beta \gamma}$ that corresponds to the 
right-moving internal bosons:
\begin{equation}
 \exp (-i[e_\alpha \cdot \phi^1 ({\overline z})
 +e_\beta \cdot \phi^2 ({\overline z}) +e_\gamma \cdot \phi^3 
({\overline z})]/3)~c((e_\alpha , e_\beta , e_\gamma ))~.
\end{equation}
$\bullet$ The twisted sector fields $T_{A\alpha \beta \gamma}$.\\
The $81$ fields $T_{A\alpha \beta \gamma}$ transform in the $({\overline 
{\bf 3}}, {\bf 1}, {\bf 1})$
irrep of $SU(3)\otimes E_6 \otimes E_8$. Their $Q$-charges are given by 
$(-e_\alpha/3, -e_\beta/3, -e_\gamma/3)$. Their $U(1)^6$ charges are given 
by $(+2e_\alpha/3, -e_\beta/3, -e_\gamma/3)$ for $A=1$, 
$(-e_\alpha/3, +2e_\beta/3, -e_\gamma/3)$ for $A=2$, and 
$(-e_\alpha/3, -e_\beta/3, +2e_\gamma/3)$ for $A=3$. Note that in the 
twisted sector all the left-handed fermions have $H$-charges 
$(-1/6,-1/6,-1/6)$, whereas their bosonic superpartners have $H$-charges 
$(+1/3,+1/3,+1/3)$.

{}We are now ready to calculate scattering amplitudes in this model.
Let us start 
with the three-point Yukawa interactions of the chiral families in ${\bf 27}$ 
of $E_6$. Note that terms like 
$\chi_a \chi_{\alpha \beta \gamma} \chi_{\alpha^{\prime} \beta^{\prime} 
\gamma^{\prime}}$ and $\chi_a \chi_b \chi_{\alpha \beta \gamma}$ are not 
allowed by $H$-charge and also $G$-charge conservation. The three-point 
Yukawa coupling $\chi_a \chi_b \chi_c$ is the same as in the asymmetric
case and is non-zero only for 
$a\not=b\not=c\not=a$.

{}Next, we turn to the scattering 
$\chi_{\alpha \beta \gamma} \chi_{\alpha^{\prime} \beta^{\prime} 
\gamma^{\prime}} \chi_{\alpha^{\prime\prime} \beta^{\prime\prime} 
\gamma^{\prime\prime}}$. 
The $G$-charge and $Q$-charge conservation (which in this case give the 
same selection rules as the $Q$-charges for these fields are the same as 
the $U(1)^6$ $G$-charges) give the following selection rules for the 
scattering: $\alpha\not=\alpha^{\prime}\not=\alpha^{\prime\prime}\not=\alpha$,
and similarly for the $\beta$ and $\gamma$ indices.

{}We will not give all the couplings for this model. 
We will finish our discussion here by considering the couplings of the type 
$S_{aA\delta} \chi_{\alpha \beta \gamma} \chi_{\alpha^{\prime} 
\beta^{\prime} \gamma^{\prime}} \chi_{\alpha^{\prime\prime} 
\beta^{\prime\prime} \gamma^{\prime\prime}}$. 
Since all the other couplings are similar, for definiteness let us consider 
the case $a=1$ and $A=1$. Let us take $S_{aA\delta}$ to be in the $0$-picture,
two other fields in the $-1/2$-picture, and the last one in the $-1$-picture.
Note that for the $H$-charges to be conserved, we must have $H_a=0$ for the 
field $S_{aA\delta}$ in the $0$-picture. This means that the term in the OPE 
$T_F S_{aA\delta}$ that can possibly contribute into this scattering must be 
of the form $\sim \exp(-i\rho_1)\sum_\alpha \exp (ie_\alpha\cdot \phi^1)$. 
Then the $Q$-charge and $G$-charge conservation together tell us that the 
above four-point coupling is non-zero if and only if 
$\alpha=\alpha^{\prime}=\alpha^{\prime\prime}=\delta$, 
$\beta\not=\beta^{\prime}\not=\beta^{\prime\prime}\not=\beta$, and
$\gamma\not=\gamma^{\prime}\not=\gamma^{\prime\prime}\not=\gamma$.

{}Let us compare our results with those of Ref \cite{DFMS}. To do so let us 
first note that this $Z$-orbifold is a factorized orbifold, and it is 
convenient to carry out the discussion on the example of a ${\bf Z}_3$ 
orbifold of one complex boson. Then $27$ families of chiral fermions in 
the twisted sector come from $3 \times 3\times 3$ fixed points. Thus, 
within the first set of $3$,
fixed points in our notation are labeled by $\alpha$. So we will discuss 
the couplings only for one index $\alpha$, and just point out that the other 
two indices $\beta$ and $\gamma$ obey similar selection rules. 
We see that the Yukawa couplings in our case are non-zero only for 
$\alpha\not=\alpha^{\prime}\not=\alpha^{\prime\prime}\not=\alpha$. 
All the other Yukawa couplings are zero. In Ref \cite{DFMS} there are also 
additional non-zero Yukawa couplings with 
$\alpha=\alpha^{\prime}=\alpha^{\prime\prime}$ (although all the others, 
with say, $\alpha=\alpha^{\prime}\not=\alpha^{\prime\prime}$, are zero). 
The reason why we do not have the latter couplings is because we are 
considering a special point in the moduli space with enhanced gauge symmetry, 
and the $G$-charge conservation forbids these couplings. Note that if we move 
away from this point of enhanced gauge symmetry, and consider generic points 
as in Ref \cite{DFMS}, we will account for additional non-vanishing 
couplings. This can be seen from the four-point couplings 
$S_{aA\delta} \chi_{\alpha \beta \gamma} \chi_{\alpha^{\prime} 
\beta^{\prime} \gamma^{\prime}} \chi_{\alpha^{\prime\prime} 
\beta^{\prime\prime} \gamma^{\prime\prime}}$. As we move away from the 
special point in the moduli space we break the $U(1)^2$ (for one complex 
boson) gauge symmetry. In terms of effective field theory this corresponds 
to giving vevs to the fields $S_{aA\delta}$. Then effectively we generate 
three-point Yukawa couplings for 
$\alpha=\alpha^{\prime}=\alpha^{\prime\prime}$, but the couplings with, 
say, $\alpha=\alpha^{\prime}\not=\alpha^{\prime\prime}$, remain zero, 
because there were no corresponding higher point couplings in the 
superpotential to begin with. The latter was due to the $Q$-charge 
non-conservation. On the other hand, note that the absence of the couplings 
of the $\alpha=\alpha^{\prime}\not=\alpha^{\prime\prime}$ type in the 
orbifold language is due to the orbifold space-group selection rules. 
Thus, the $Q$-charge conservation has this space-group discrete symmetry 
encoded in it, just as $H$-charge conservation guarantees that the orbifold 
point-group selection rules are satisfied. 

{}Again, the above model can be written in the twist basis. The following 
generating vector:
\begin{equation}
 V_1 =(0(-{1\over 3})^3 \vert \theta^3\vert\vert \theta^3\vert 
{1\over 3}{1\over 3}{2\over 3} 
 0^5 \vert 0^8)~.
\end{equation} 
produces the same symmetric ${\bf Z}_3$ orbifold 
as above.

\section{Three-Family $E_6$ Model}

{}In this and the following sections, we describe the construction of 
some of the 
three-family grand unified string theories previously presented in Refs 
\cite{kt}. In particular, we will rewrite these 
models in the bases where the supercurrent is in the bosonized form. 
As explained previously, scattering amplitudes are 
most easily calculated in such a representation.
This section describes the $E_6$ model. It is organized as follows. 
In subsection A we set up the notation and 
describe the construction of the Narain model to be orbifolded. In 
subsection B we construct the unique three-family $E_6$ model by 
${\bf Z}_6$ orbifolding this Narain model. This $E_6$ model can be 
realized as two different ${\bf Z}_6$ orbifolds which we refer to as
$E1$ and $E2$ models. In subsection C we present 
the shift construction for $E1$ model and compute the 
superpotential
using the bosonic supercurrent. The shift construction and
the superpotential for $E2$ model will be given in subsection D. 
In section VII, we will discuss a three-family $SO(10)$ model.
We use this example to illustrate how one can calculate 
the superpotential 
for models that do not admit a bosonic supercurrent. 
In section VIII, we will discuss two three-family $SU(6)$ models which are 
obtained by adding a 
${\bf Z}_3$ Wilson line to the $E_6$ model.
Finally, we will briefly comment on other three-family grand unified string
models. 

\subsection{Narain Model} 

{}Consider the Narain model, which we will refer to as $N(1,1)$, with 
$\Gamma^{6,22}=\Gamma^{6,6} \otimes \Gamma^{16}$, where $\Gamma^{16}$ is 
the ${\mbox{Spin}}(32)/{\bf Z}_2$ lattice, and ${\Gamma}^{6,6}$ is the 
lattice spanned by the vectors $(p_R \vert\vert p_L)$, where 
$p_L,p_R \in \Gamma^6$ ($E_6$ weight lattice), and $p_L-p_R\in \Gamma^6$ 
($E_6$ root lattice). Recall that, under $E_6\supset SU(3)^3$,
\begin{eqnarray}
 {\bf 27} &=& ({\bf 3},{\bf 3},{\bf 1})+ ({\overline {\bf 3}},{\bf 1},
  {\bf 3}) + ({\bf 1}, {\overline {\bf 3}}, {\overline {\bf 3}}) ~,\\
 {\bf 78} &=& ({\bf 8},{\bf 1},{\bf 1}) +({\bf 1},{\bf 8},{\bf 1}) +
 ({\bf 1},{\bf 1},{\bf 8})+ ({\bf 3},{\overline {\bf 3}},{\bf 3})+
 ({\overline {\bf 3}},{\bf 3}, {\overline {\bf 3}}) ~, 
\end{eqnarray}
so we can write ${\bf p} \in \Gamma^6$ as
\begin{equation}
 \label{psu3}
 {\bf p}=({\bf q}_1+ \lambda {\bf w}_1,{\bf q}_2+ \lambda {\overline 
{\bf w}}_2,
  {\bf q}_3+ \lambda {\bf w}_3) ~,
\end{equation}
where ${\bf q}_i \in \Gamma^2$ ($SU(3)$ root lattice), ${\bf w}_i$
(${\overline {\bf w}}_i$) is in the ${\bf 3}$ (${\overline {\bf 3}}$)
weight of $SU(3)$, and $ \lambda=0,~1,~2$.

{}Next, consider the model, which we will refer to as $N1(1,1)$, generated 
from the $N(1,1)$ model by adding the following Wilson lines:
\begin{eqnarray}
 &&V_1 =(0^4\vert 0^3 \vert\vert e_1/2~0~0 \vert {\bf s}~{\bf 0}~{\bf 0} 
\vert {\overline S}) ~,
 \nonumber\\
 \label{Wilson}
 &&V_2 =(0^4\vert 0^3 \vert\vert e_2/2~0~0 \vert {\bf 0}~{\bf s}~{\bf 0} 
\vert {\overline S}) ~.
\end{eqnarray}
Here the first four entries correspond to the right-moving world-sheet 
fermions, the next three right-moving entries stand for the three 
right-moving complex bosons $X^a$, $a=1,2,3$ (each corresponding to one of 
the three $SU(3)$s). The double vertical line separates the right-movers 
from the left-movers. The first three left-moving entries correspond to the 
left-moving counterparts of the $X^a$ bosons. The remaining 16 left-moving 
world-sheet bosons generate the ${\mbox{Spin}}(32)/{\bf Z}_2$ lattice. 
The $SO(32)$ shifts are given in the $SO(10)^3
\otimes SO(2)$ basis. In this basis, ${\bf 0}$ stands for the null vector, 
${\bf v}$($V$) is the vector weight, whereas
${\bf s}$($S$) and ${\overline {\bf s}}$(${\overline S}$) are the
spinor and anti-spinor weights of $SO(10)$($SO(2)$). (For $SO(2)$, $V=1$,
$S=1/2$ and ${\overline S}=-1/2$). The 
untwisted sector provides gauge bosons of 
$SU(3)^{2} \otimes U(1)^{2} \otimes SO(10)^{3} \otimes SO(2)$. There
are additional gauge bosons from the new sectors. Recall that under
$E_6 \supset SO(10) \otimes U(1)$,
\begin{equation}
{\bf 78}={\bf 1}(0)+{\bf 45}(0)+{\bf 16}(3)+\overline{\bf 16}(-3)~.
\end{equation}
It is easy to see that the $V_1$, $V_2$ and $V_1+V_2$ sectors provide
the necessary ${\bf 16}(3)$ and $\overline{\bf 16}(-3)$ gauge
bosons to the three $SO(10)$'s respectively.
The resulting Narain $N1(1,1)$ model has $N=4$ SUSY and gauge group 
$SU(3)^2 \otimes (E_6)^3$ provided that we set $k_{10}=k_{20}=0$.  
The permutation symmetry of the three $E_6$ factors should be clear from the 
above construction. Since there is only one $N=4$ SUSY 
$SU(3)^2 \otimes (E_6)^3$ model in $4$-dimensional heterotic string theory,
this permutation symmetry may also be explicitly checked by looking at its 
one-loop modular invariant partition function.

\subsection{$E_6$ Model: Twist Construction}

{}Before we describe the ${\bf Z}_6$ asymmetric orbifold that leads to the 
three-family $E_6$ model, we will introduce some notation. By $\theta$ we 
will denote a 
$2\pi/3$ rotation of the corresponding complex (or, equivalently, two real) 
chiral world-sheet boson(s). Thus, $\theta$ is a ${\bf Z}_3$ twist. 
Similarly, by $\sigma$ we will denote a $\pi$ rotation of the 
corresponding complex chiral world-sheet boson. 
Thus, $\sigma$ is a ${\bf Z}_2$ twist. By ${\cal P}$ we will denote
the outer-automorphism of the three $SO(10)$s that arise in the 
breaking $SO(32)\supset SO(10)^3 \otimes SO(2)$. 
Note that ${\cal P}$ is a ${\bf Z}_3$ twist. Finally, by $(p_1,p_2)$ 
we will denote the outer-automorphism of the corresponding two complex 
chiral world-sheet bosons. Note that $(p_1,p_2)$ is a ${\bf Z}_2$ twist. 

{}The $E_6$ model can be constructed by performing the following
asymmetric ${\bf Z}_6$ orbifolds on the $N=4$ SUSY 
$SU(3)^2 \otimes (E_6)^3$ model ({\em i.e.}, $N1(1,1)$ model): \\ 
$\bullet$ The $E1$ model. 
Start from the $N1(1,1)$ model and perform the following twists:
\begin{eqnarray}\label{E1twist}
 &&T_3 =(0 (-1/3)^3\vert \theta,\theta,\theta\vert\vert \theta,e_1/3,0 \vert
 {\cal P} \vert  2/3)~,\nonumber\\
 &&T_2=(0~(-1/2)^2~0\vert\sigma, p_1,p_2\vert\vert  0,e_1/2,e_1/2 \vert
 0^{15} \vert 0)~.
\end{eqnarray}  
This model has $SU(2)_1 \otimes (E_6)_3 \otimes U(1)^3$ gauge symmetry. 
The massless spectrum of the $E1$ model is given in Table \ref{E6spectra}. 
They are grouped according to where they come from, namely, the untwisted 
sector U,
the ${\bf Z}_3$ twisted ({\em i.e.}, $T_3$ and $2T_3$) sector T3, 
the ${\bf Z}_6$ twisted ({\em i.e.}, $T_3+T_2$ and $2T_3+T_2$) sector T6,
and ${\bf Z}_2$ twisted ({\em i.e.}, $T_2$) sector T2.
Note that all particles have integer $U(1)$ charges. The 
normalization, or compactification radius $r$, of each left-moving
world-sheet boson is given at the bottom of the table. The $U(1)$ charge
of a particle with charge $n$ contributes $n^2 r^2/2$ to its conformal
highest weight. That is, the corresponding part of the vertex operator 
has momentum $nr$.\\
$\bullet$ The $E2$ model. Start from the $N1(1,1)$ model and perform the 
following twists:
\begin{eqnarray}\label{E2twist}
 &&T_3 =(0~0~(-1/3)^2\vert 0,\theta,\theta\vert\vert \theta,e_1/3,0 \vert
 {\cal P} \vert  2/3)~,\nonumber\\
 &&T_2=(0~(-1/2)^2~0\vert \sigma, p_1,p_2\vert\vert  0,e_1/2,e_1/2\vert
 0^{15} \vert 0)~.
\end{eqnarray}  
This model has $SU(2)_1 \otimes (E_6)_3 \otimes U(1)^3$ gauge symmetry. 
The massless spectrum of the $E2$ model is given in Table \ref{E6spectra}.

{}Note that the spin structures of the world-sheet fermions in the 
right-moving sector are fixed by the world-sheet supersymmetry consistency. 
The string consistency
conditions impose tight constraints on the allowed twists. Using the
approach given in Ref \cite{kt} one can check that both sets of twists 
presented above
are consistent provided that the appropriate choices of the structure 
constants $k_{ij}$ are made. It is then straightforward, but somewhat 
tedious, to work out the massless spectrum of the model. (Again, more 
details can be found in Ref \cite{kt}). We will give an alternative way 
of constructing this model in the following subsections, and working out the 
massless spectrum there is somewhat easier.

{}The models $E1$ and $E2$ have the same tree-level massless 
spectra. We will show that interactions on the
two orbifolds are the same even though naively the fields 
in the two models seem to have different origins (in particular, the same 
states come from different twisted sectors of the two orbifolds). They 
are possibly a $T$-dual pair.  

{}In the following subsections, we will give the shift representation for 
the twists given in models 
$E1$ and $E2$. To be able to do so it is necessary and sufficient, as we 
already discussed in the previous sections, that the corresponding 
Kac-Moody algebras ${\cal G}^\prime_R$ are realized at level $1$ and have 
central charge $6$. This is indeed the case for the $E1$ and $E2$ models. 
Let us first consider the $E1$ model. The original Kac-Moody algebra 
${\cal G}_R$ before orbifolding was $(E_6)_1$.  The ${\bf Z}_3$ twist 
reduces it to $(SU(3)_1)^3$, whereas the ${\bf Z}_2$ twist breaks it to 
$SU(6)_1\otimes SU(2)_1$. The combined action of the ${\bf Z}_3$ and 
${\bf Z}_2$ twists, {\em i.e.}, the corresponding ${\bf Z}_6$ twist breaks 
$(E_6)_1$ to ${\cal G}^\prime_R= [SU(2)_1 \otimes U(1)]^3$, which is realized 
at level one. Similarly, in the $E2$ model we have 
${\cal G}^\prime_R= (SU(2)_1)^4 \otimes U(1)^2$. Even though the
right moving Kac-Moody algebra is not $SU(3)^3$ for both cases, we
can always write the Kac-Moody charges in this basis. The translation
of the charges in different bases are given in Table \ref{convert}.

\subsection{$E1$ Model: Shift Construction}

{}Let us present the generating vectors in the shift 
formalism which produce the $E1$ model. Here $V_1$ and 
$V_2$ correspond to the $T_3$ and $T_2$ twists acting on the 
$N1(1,1)$ model:
\begin{eqnarray}
 &&V_1 =(0 (-1/3)^3\vert e_1/3,e_1/3,e_1/3\vert\vert \theta,e_1/3,0 \vert
 {\cal P} \vert  2/3)~,\nonumber\\
 \label{ME1}
 &&V_2=(0~(-1/2)^2~0\vert e_1/2,0,0\vert\vert  0,e_1/2,e_1/2 \vert
 0^{15} \vert 0)~.
\end{eqnarray}  

{}We may rewrite the vector $V_1$ in the following way.
Consider the branching of $SO(32)$ to $SO(10)^3 \otimes SO(2)$.
The three $SO(10)$s are permuted by the action of the
${\bf Z}_3$ outer-automorphism twist ${\cal P}$:
$\phi^I_1 \rightarrow
\phi^I_2 \rightarrow \phi^I_3 \rightarrow \phi^I_1$, where the real bosons
$\phi^I_p$, $I=1,...,5$, correspond to the $p^{\mbox{th}}$ $SO(10)$
subgroup, $p=1,2,3$. We can define new bosons 
$\varphi^I \equiv {1\over \sqrt{3}}(\phi^I_1 +\phi^I_2 +\phi^I_3)$; 
the other ten real bosons are complexified via the linear combinations 
$\Phi^I \equiv {1\over\sqrt{3}}(\phi^I_1 +\omega\phi^I_2 +\omega^2 \phi^I_3)$ 
and
$(\Phi^I)^\dagger \equiv {1\over \sqrt{3}}(\phi^I_1 +\omega^2\phi^I_2
+\omega \phi^I_3)$, where $\omega =\exp(2\pi i /3)$.
Under ${\cal P}$, $\varphi^I$ is invariant,
while $\Phi^I$ ($(\Phi^I)^\dagger$) are eigenstates with
eigenvalue $\omega^2$ ($\omega$), i.e., ${\cal P}$ is equivalent to
a ${\bf Z}_3$ twist $\theta$ on each $\Phi^I$. 
Finally, string consistency requires the
inclusion of the $2/3$ shift in the $SO(2)$ lattice of the boson $\eta$. 
This simply changes the
radius of this left-moving world-sheet boson. So, in the $\Phi^I$, $\varphi^I$
and $\eta$ basis, $V_1$ becomes 
\begin{eqnarray}
V_1 =(0 (-1/3)^3\vert e_1/3,e_1/3,e_1/3\vert\vert \theta, e_1/3,0 \vert
 \theta ^5 0_r^5 \vert  (2/3)_r)~
\end{eqnarray}
where the subscript $r$ indicates real bosons.

{}Including the $V_0$ vector, the generating vectors for $E1$ model 
can then be written as:
\begin{eqnarray}
V_0 &=& ( (-1/2)^4 \vert 0^6 \vert \vert 0^{22} ) 
\nonumber \\
V_1 &=& (0 (-1/3)^3\vert e_1/3,e_1/3,e_1/3\vert\vert (1/3)_t, e_1/3,0 \vert
 (1/3)^5_t 0_r^5 \vert  (2/3)_r)~, \nonumber \\
V_2 &=& (0~(-1/2)^2~0\vert e_1/2,0,0\vert\vert  0,e_1/2,e_1/2 \vert
 0^{15} \vert 0)~. 
\end{eqnarray}
where the subscript $t$ indicates that the corresponding complex boson is 
twisted.

{}The dot products $V_i \cdot V_j$ and the choice of the structure 
constants $k_{ij}$
are given by:
\begin{equation}
 V_i \cdot V_j = \left( \begin{array}{ccc}
                        -1   & -1/2  & -1/2 \\
                        -1/2 & -1/3  & -1/3 \\
                        -1/2 & -1/3  & 0
                        \end{array}
                 \right) ~,
\quad \quad
         k_{ij}= \left( \begin{array}{ccc}
                         0   & 0   & 0 \\
                         1/2 & 0   & 0 \\
                         1/2 & 2/3 & 1/2
                        \end{array}
                 \right) ~.
\end{equation}
This defines a consistent string model, provided that a supercurrent
satisfying (\ref{XOPE}) and (\ref{triplet}) can be found.
 
\subsubsection{Bosonic Supercurrent}

{}Before we compute the spectrum and calculate the superpotential, 
it is important to construct
the supercurrent. First of all, the supercurrent satisfying the
constraints (\ref{XOPE}), (\ref{triplet}) must exist in order
for the model to be consistent. Furthermore, in calculating higher point
couplings, picture-changing requires insertions of the supercurrent.
In the following, we will construct the supercurrent for the $E1$ model 
in detail.

{}The currents $\partial X^{a}$ can be written as linear combinations of 
vertex operators for the root generators of the original 
${\cal G}_{R}= (E_{6})_1$
Kac-Moody algebra. It is convenient to express the roots of $E_6$ in 
$SU(3)^3$ basis (Eq.(\ref{psu3})). 
As before, we define $\pm e_1,\pm e_2$ and $\pm e_3=\mp (e_1+e_2)$ as 
the roots of $SU(3)$. The weight vectors $\tilde{e}^1,\tilde{e}^2$ are
defined by $\tilde{e}^i e_j = \delta^i_j$ and 
$\tilde{e}^0=\tilde{e}^2-\tilde{e}^1$. Therefore, the weight ${\bf 3}$ of
$SU(3)$ is represented by $\tilde{e}^2,-\tilde{e}^1,-\tilde{e}^0$.
The triplet constraints 
(Eq.(\ref{triplet})) from the $V_1$ and $V_2$ vectors restrict the
possible terms in $\partial X^{a}$. The currents $\partial X^{a}$
and $\partial X^{b \dagger}$ must also obey the OPEs (Eq.(\ref{XOPE})).
A solution satisfying all the constraints is given by:
\begin{eqnarray}\label{E1T_F}
i \partial X^{1} &=& {1 \over \sqrt{12}} \left( J_1 + \sqrt{2} J_2 + 
\sqrt{3} J_3 + \sqrt{2} J_4 + J_5 + \sqrt{2} J_6 + J_7 \right) ~,
\nonumber \\
i \partial X^{2} &=& {1 \over \sqrt{12}} \left( L_1 - \sqrt{2} L_2 + 
\sqrt{3} L_3 - \sqrt{2} L_4 - L_5 - \sqrt{2} L_6 + L_7 \right) ~, \\
i \partial X^{3} &=& {1 \over \sqrt{6}} \left( K_1 - K_2 - K_3 
+K_1^{'} - K_{2}^{'} + K_{3}^{'} \right) ~, \nonumber
\end{eqnarray}
where $J_{i}= \exp(-i a_i \phi) c(-a_i)$, $L_{i}= \exp(-i l_i \phi) c(-l_i)$,
$K_i = \exp(-i k_i \phi) c(-k_i)$ and 
$K_{i}^{'}= \exp(-i k_{i}^{'} \phi) c(-k_{i}^{'})$.
Here $a_i$, $l_i$, $k_i$ and $k_i^{'}$ are roots of $E_6$ 
defined in Table \ref{E1super}. We will define the cocycle
operators $c(K)$ in a moment.

{}Notice that out of the $72$ roots of $E_6$, only $7$ of them contribute
to $\partial X^{1}$.
It is easy to see that 
$a_1,\cdots,a_6$ form a set of simple roots of $E_6$
and $-a_7=a_1+2a_2+3a_3+2a_4+a_5+2a_6$ is the highest root. 
(Note that the coefficients are simply the co-marks of $E_6$ simple roots).
Similarly,
$l_1,\cdots,l_6$ which contribute to $\partial X^{2}$ form another set of 
simple roots of $E_6$ with $-l_7$
being the highest root. 
Since the ${\bf Z}_2$ action ($V_2$ vector)
does not act on $\partial X^{3}$ and $\psi^{3}$, $\partial X^{3}$
can be expressed in terms of the root generators of $SU(3)^{2}$.
We see that $k_1$ and $k_2$ form a set of simple roots of $SU(3)$ and $-k_3$
is the highest root. Similarly for the $k_{i}^{'}$.  The co-marks of 
the simple roots of $E_6$ and $SU(3)$ are also listed in Table \ref{E1super}.
Other choices of the supercurrent are equivalent to this one 
(involving only a change of basis). In this sense, the supercurrent 
consistent with the twists-shifts of this model is unique.

{}Any weight vector $K$ of $E_6$ can then be expressed in terms of the basis
vectors $a_1,\cdots,a_6$, {\em i.e.}, $K=\sum_{j} n_j a_j$. We can define an 
ordered product of weight vectors $K$ and $K^{'}$,
\begin{equation}
K \ast K^{'} = \sum_{i>j} n_{i} n_{j}^{'} a_{i} \cdot a_{j}
\end{equation}
The cocycle operator and the cocycle structure constant can then be given by
\cite{cocycle}:
\begin{eqnarray}
c(K)&=&(-1)^{{\bf p} \ast K} \\
\epsilon(K,K^{'})&=&(-1)^{K \ast K^{'}} 
\end{eqnarray}
where ${\bf p}$ is the momentum operator, i.e., 
${\bf p} \vert P \rangle = P \vert P \rangle$. The cocycles defined 
above depend crucially on our choice of the basis vectors.

{}The coefficient of each term in $\partial X^{a}$ ({\em i.e.\/} 
$\sqrt{m_i/ h}$
where $m_i$ is the co-mark and $h=1+\sum_{i} m_{i}^{2}$ is the Coxeter
number) are determined up to
a phase by the OPE of $\partial X^{1}$ with $\partial X^{1 \dagger}$.
The phases we have chosen in (\ref{E1T_F}) ensures that the 
OPEs of $\partial X^{a}$ and $\partial X^{b}$ for $a \not= b$ are non-singular.

{}The supercurrent $T_F$ (the internal part) is therefore a linear 
combination of $40$ terms,
each with different right-moving quantum numbers 
(given in Table \ref{E1super}):
\begin{eqnarray}
T_{F} &=& {i \over 2} \sum_{a=1}^{3} \psi^a \partial X^a +{\mbox{h.c.}}
\nonumber \\
      &=& {1 \over 2} \left( {1\over \sqrt{12}} e^{i H_1 \rho_1} 
         \sum_{i=1}^{7} \sqrt{m_i} e^{-ia_i \phi} c(-a_i) 
       + {1\over \sqrt{12}} e^{i H_2 \rho_2} 
         \sum_{i=1}^{7} \sqrt{m_i} e^{-il_i \phi} c(-l_i) 
\right. \nonumber \\
      &&+\left. {1\over \sqrt{6}} e^{i H_3 \rho_3} 
         \sum_{i=1}^{3} ( e^{-ik_i \phi} c(-k_i) 
                   +e^{-ik^{\prime}_i \phi} c(-k^{\prime}_i)) \right) 
+{\mbox{h.c.}}~.
\end{eqnarray}
where $m_i$ are the co-marks of the $E_6$ simple roots and $h=12$ for
$E_6$.                      

\subsubsection{Spectrum}

{}We are now ready to compute the massless spectrum. 
Let us recall what the left- and the right-moving quantum numbers are.
For the left-movers, we have the $H$-charges $H_0,\cdots,H_3$, and the 
$Q^R$ charges $({\bf Q}^R_1,{\bf Q}^R_2,{\bf Q}^R_3)$ in the $SU(3)^3$ basis.
The left-moving charges are: $SU(3)$ charges ${\bf Q}^L_2,{\bf Q}^L_3$, 
a $SO(10)$ charge ${\bf q}$, a $U(1)$ charge $Q$ and a discrete charge $D$
coming from the ${\bf Z}_3$ twist. Here, the charge $Q$ is 
the $U(1)$ charge of the real boson which is shifted by $2/3$.

{}Let us first consider the untwisted sector. In the ${\bf 0}$ sector, the 
spectrum generating formulae (\ref{SGFU1}),(\ref{SGFU2}) read:
\begin{eqnarray}
V_{0} \cdot N &=& {1\over 2} (H_0 + H_1 +H_2 + H_3) = {1\over 2} \pmod{1} \\ 
V_{1} \cdot N &=& {1\over 3} (H_1 + H_2 + H_3) 
- {e_{1} \over 3} \cdot ({\bf Q}^R_1 + {\bf Q}^R_2 + {\bf Q}^R_3) 
 \nonumber \\
&&+ {e_{1} \over 3} \cdot {\bf Q}^L_2  
+ {1 \over 3} D + {2\over 3} Q = 0 \pmod{1} \\
V_{2} \cdot N &=& {1\over 2} (H_1+H_2) - {e_1 \over 2} \cdot {\bf Q}^R_1 
+ {e_1 \over 2} \cdot ( {\bf Q}^L_2 + {\bf Q}^L_3 ) = 0 \pmod{1}
\end{eqnarray}
Here, the quantum number $D$ is the eigenvalue of the twist operator
on the $6$ twisted complex bosons. Since the twist is ${\bf Z}_3$,
$D=0,1,2 \pmod{3}$.
For gauge bosons, $H_0=1$. The right-moving energy given by
\begin{equation}
E_R=-{1\over 2} + {1\over 2} \sum_{j=0}^3 H_j^2 
+ {1 \over 2} \sum_{j=1}^{3} ({\bf Q}^R_j)^2 + {\mbox{oscillators}}
\end{equation}
implies that ${\bf Q}^R=(0,0,0)$. The original Narain model has
gauge symmetry $SU(3)^2 \otimes (E_6)^3$. It is easy to see from the 
last spectrum generating formula that 
all the $SU(3)^2$ roots are projected out
except ${\bf Q}^L_3=\pm e_1$. As a result, $SU(3)^2$ is broken to
$SU(2) \otimes U(1)^3$. On the other hand, $(E_6)^3$ is broken to
the diagonal $(E_6)_3$ by the ${\bf Z}_3$ outer-automorphism. 
The resulting gauge group is therefore 
$SU(2)_1 \otimes (E_6)_3 \otimes U(1)^3$.

{}There are other massless states in the ${\bf 0}$ sector with their fermionic
partners in $V_0$ sector. To
determine the chiralities of the fields, let us consider the $V_0$ sector.
We choose the convention that $H_0=-1/2$ for left handed fermions.
The spectrum generating formulae (\ref{SGFU1}),(\ref{SGFU2}) read:
\begin{eqnarray}
V_{0} \cdot N &=& {1\over 2} (H_0 + H_1 +H_2 + H_3) = {1\over 2} \pmod{1} \\ 
V_{1} \cdot N &=& {1\over 3} (H_1 + H_2 + H_3) 
- {e_{1} \over 3} \cdot ({\bf Q}^R_1 + {\bf Q}^R_2 + {\bf Q}^R_3) 
 \nonumber \\
&&+ {e_{1} \over 3} \cdot {\bf Q}^L_2  
+ {1 \over 3} D + {2\over 3} Q = {1\over 2} \pmod{1} \\
V_{2} \cdot N &=& {1\over 2} (H_1+H_2) - {e_1 \over 2} \cdot {\bf Q}^R_1 
+ {e_1 \over 2} \cdot ( {\bf Q}^L_2 + {\bf Q}^L_3 ) = {1\over 2} \pmod{1}
\end{eqnarray}
Again, massless states have ${\bf Q}^R=(0,0,0)$. First, consider 
states that are charged only under ${\bf Q}^L_2$ and ${\bf Q}^L_3$
and hence do not carry $D$ and $Q$ quantum number. The left-moving
energy is zero only if $({\bf Q}^L_2)^2+({\bf Q}^L_3)^2=2$. The 
spectrum generating formulae further restrict the choices to
$({\bf Q}^L_2,{\bf Q}^L_3)=(e_1,0)$ for the field $U_0$, $(e_2,0)$ for
$U_{\pm +}$, and $(e_3,0)$ for $U_{\pm -}$. The definition of the
fields as well as their $H$-charges are given in Table \ref{E1charges}.

{}Now we turn to states that has non-trivial $D$ quantum numbers.
The adjoint scalar $\Phi$
has ${\bf Q}^L_2={\bf Q}^L_3=0$. The spectrum generating formulae
give $(H_1,H_2,H_3)=(+1/2,+1/2,+1/2)$ and 
\begin{equation}\label{D}
D= 2 (1-Q) \pmod{3}
\end{equation}

{}Upon decomposition of ${\bf 78}$ of $E_6$ into $SO(10) \otimes U(1)$
representations:
\begin{equation}
{\bf 78} = {\bf 1}(0) + {\bf 45}(0) + {\bf 16}(+3) + {\overline{\bf 16}}(-3)~,
\end{equation}
we notice that the component fields of ${\bf 78}$ carry different 
$U(1)$ charges.
($Q=0,0,1/2,-1/2$ respectively). It then follows from (\ref{D})
that the component fields of
a $E_6$ multiplet can carry different discrete ${\bf Z}_3$ charges.
We will return to this point when we discuss the
superpotential.

{}Now, let us turn to the twisted sectors. In $\overline{V_0+V_1}$
sector,
\begin{equation}
\overline{V_0+V_1}=(-1/2~(1/6)^3 \vert (e_1/3)^3 
\vert \vert 
(1/3)_t (e_1/3) 0 \vert (1/3)^5 0_r^5 \vert (2/3)_r) ~,
\end{equation}
the spectrum generating formulae (\ref{SGFT}) give:
\begin{eqnarray}
V_{0} \cdot N &=& {1\over 2} (H_0 + H_1 +H_2 + H_3) = {1\over 2} \pmod{1} \\ 
V_{1} \cdot N &=& {1\over 3} (H_1 + H_2 + H_3) 
- {e_{1} \over 3} \cdot ({\bf Q}^R_1 + {\bf Q}^R_2 + {\bf Q}^R_3) 
 \nonumber \\
&&+ {e_{1} \over 3} \cdot {\bf Q}^L_2  
+ {1 \over 3} D + {2\over 3} Q = {1\over 2} \pmod{1} \\
V_{2} \cdot N &=& {1\over 2} (H_1+H_2) - {e_1 \over 2} \cdot {\bf Q}^R_1 
+ {e_1 \over 2} \cdot ( {\bf Q}^L_2 + {\bf Q}^L_3 ) = +{1\over 6} \pmod{1}
\end{eqnarray}
Since
$H_a \in {\bf Z}+ {1\over 6}$ for $a=1,2,3$, for massless states,
$(H_1,H_2,H_3)=(1/6,1/6,1/6)$ and hence $H_0=1/2$. Therefore, this
sector provides antiparticle states whereas 
$\overline{V_0+2V_1}$ provides the
corresponding particles. The $H$-charges (and also $Q$-charges) of particle 
and 
antiparticle states have opposite sign and so it suffices to consider only
$\overline{V_0+V_1}$ sector. 
In this sector, ${\bf Q}^R_j$ for $j=1,2,3$ and ${\bf Q}^L_{2}$ are 
shifted by $e_1/3$.
The simplest choice ${\bf Q}^R=(e_1/3,e_1/3,e_1/3)$ and 
$({\bf Q}^L_2,{\bf Q}_3)=(e_1/3,0)$ 
has $E_R=0$ and
\begin{eqnarray}
E_L &=& -1+ {1\over 2} \sum_{r=1}^{6} T^r(1-T^r) 
+ {1\over 2} \left( ({\bf Q}^L_2)^2 +({\bf Q}^L_3)^2 \right)
+ {1\over 2} ({{\bf q} \over \sqrt{3}})^2 
+ {1 \over 2}({2\over 3}+Q)^2 +\dots \nonumber\\
&=& -{2\over 9} +  {1\over 6} {\bf q}^2 + {1 \over 2}({2\over 3}+Q)^2
+ \dots 
\end{eqnarray}
The only possible massless
states come from $({\bf q},Q)=({\bf 0},0)$, $({\bf v},-1)$ and 
$({\bf c},-1/2)$. Here ${\bf 0}$, ${\bf v}$ and ${\bf c}$ are the singlet,
vector and spinor representation of $SO(10)$ respectively, and together they
form ${\overline{\bf 27}}$ of $E_6$. It is easy to see that the 
spectrum generating formulae are satisfied.

{}This, however, is not the only choice of $({\bf Q}^R,{\bf Q}^L)$
which gives rise to massless states. The appearance of the other massless 
states requires a careful explanation. Recall that the 
outer-automorphism 
$\theta -{\cal P}$ in (\ref{ME1}) does not commute 
with the Wilson lines (\ref{Wilson}) and the combined action can be 
viewed as a non-Abelian orbifold corresponding to modding out 
the original Spin(32)/${\bf Z}_2$ Narain model (which we refer to as
$N(1,1)$ model) by the 
tetrahedral group ${\cal T}$ \cite{KST}. 
This group 
is generated by three elements $\Theta, R_1, R_2$, where $\Theta^3=1$, 
$(R_1)^2=(R_2)^2=1$, and $R_2=\Theta R_1$. In our case, 
the Wilson lines $V_1$ and $V_2$ in (\ref{Wilson}) correspond to the elements 
$R_1$ and $R_2$, respectively, whereas the ${\bf Z}_3$ twist in (\ref{ME1}) 
corresponds to the element $\Theta$. The ${\bf Z}_2$ twist in (\ref{ME1}) 
commutes with all the elements $\Theta, R_1, R_2$ and will not be important 
for this discussion. The resulting $E1$ model is therefore a 
${\cal T} \times {\bf Z}_2$ orbifold. 
The spectrum generating formulae are written in the basis in which
the Wilson lines (\ref{Wilson}) are diagonal. Therefore,
the ${\bf Z}_3$ action $\theta$ in (\ref{ME1}) is represented as a twist 
but not a shift. However, in determining the ${\bf Q}^L$ charges, 
we can always go to the basis where the $\theta$ twist is replaced 
by a shift
${\bf Q}^L_1 \rightarrow {\bf Q}^L_1 + e_1/3$. (The Wilson lines in this
basis are not diagonal). Notice that the conformal dimension of the
momentum state $h={1\over 2}(e_1/3)^2=1/9$ is the same as that of a
${\bf Z}_3$ twist field. We therefore have 
${\bf Q}^R=p_R + (e_1/3,e_1/3,e_1/3)$ and ${\bf Q}^L=p_L + (e_1/3,e_1/3,0)$ 
where $p_R,p_L \in \tilde{\Gamma}^6$ ($E_6$ weight lattice), and
$p_L-p_R \in \Gamma^6$ ($E_6$ root lattice).
We can add weight
vectors $(p_R,p_L)$ to the above $({\bf Q}^R,{\bf Q}^L)$ provided
that the lengths 
$({\bf Q}^R)^2$ and $({\bf Q}^L_2)^2+({\bf Q}^L_3)^2$ are preserved
(hence $E_R=E_L=0$) and the spectrum generating formulae are satisified.
It turns out there are 4 choices:
\begin{equation}
\begin{array}{ccccc}
\chi_{++}: \quad &{\bf Q}^R= & (e_3/3,e_3/3,e_1/3) \quad
&{\bf Q}^L= &(e_3/3,e_3/3,0) \\
\chi_{-+}: \quad & &(e_2/3,e_1/3,e_3/3) \quad & &(e_3/3,e_3/3,0) \\
\chi_{+-}: \quad & &(e_3/3,e_1/3,e_2/3) \quad & &(e_2/3,e_2/3,0) \\
\chi_{--}: \quad & &(e_2/3,e_2/3,e_1/3) \quad & &(e_2/3,e_2/3,0) 
\end{array}
\end{equation} 
For example, the $({\bf Q}^R,{\bf Q}^L)$ charges of $\chi_{++}$ are
obtained by taking $p_R=p_L=(-\tilde{e}^1,-\tilde{e}^1,0)$.

{}The other twisted sector fields coming from $T6$ and $T2$ sectors can 
be worked out in a similar way. The $Q$- and $H$- charges for the 
$E1$ model are summarized in Table \ref{E1charges}. The discrete 
$D$-charges are given in Table \ref{discrete}.

\subsubsection{Superpotential}

{}Now we are ready to calculate scattering amplitudes and deduce the 
non-vanishing terms in the superpotential for $E1$ model.
Here we are only interested in whether a given term is 
vanishing or not, not in the actual numerical values of these couplings. 
That is, we are only concerned with the selection rules for the scattering. 
Since the coefficients $\xi^a ({Q})$ in the supercurrent are completely
determined, calculating the actual numerical values of the 
coupling is rather straightforward, although certain couplings 
might be tedious to work out. We will return to such a calculation in future 
publications if there is a necessity for it to be done for phenomenological 
purposes.

{}Let us recall our selection rules. First, 
all the terms in the superpotential must be gauge singlets. This is dictated 
by gauge invariance ($G$-charge conservation). Next, the $H$-charges must be 
conserved. Here one should be careful as the $H$-charges are altered by 
picture changing. Similarly, the $Q$-charges also must be conserved. These 
are also altered by picture changing, and Table \ref{E1super} provides the 
quantum numbers of terms that are relevant when doing the picture changing 
by inserting the supercurrent $T_F$. Finally, the discrete charges 
($D$-charges) coming from the left-moving coset ${\cal C}_L$ (recall 
that this is a level-$3$ model with reduced rank) must also be conserved. 
As we will see, $D$-charge conservation is guaranteed by $G$-, $H$-, and
$Q$-charge conservation. We will see this explicitly in an $SO(10)$ 
model in the next section.

{}Let us first focus on gauge symmetries which are common for the 
$E1$ and $E2$ models. Gauge invariance implies that some of the fields 
cannot enter into the superpotential by themselves. For instance,
$D_{+}$ must couple with $D_{-}$ due to the $SU(2)$ symmetry.
Similarly, the only fields that are charged under the first $U(1)$
are $\tilde{U}_{\pm}$ and $\tilde{\chi}_{\pm}$. Therefore, they must enter
into the superpotential in the invariant combinations 
$\tilde{U}_{+}\tilde{U}_{-}$, 
$\tilde{\chi}_{+}\tilde{\chi}_{-}$ and 
$\tilde{U}_{\pm} \tilde{\chi}_{\mp}^3$. 
In the same fashion, $U_{i\pm}$ and $\chi_{i\pm}$, $i=+,-$, are
bound to form the invariant combinations 
$U_{i+}U_{j-}$, $\chi_{i+}\chi_{j-}$ and
$U_{i\pm}\chi_{j_1 \mp} \chi_{j_2 \mp} \chi_{j_3 \mp}$
as a consequence of invariance under the first $U(1)$.

{}Naively, the three-point couplings 
$\chi_{0} \chi_{i\pm}\chi_{j\mp}$ are allowed by $G$-charge conservation. 
However, the $Q$-charges are not conserved. Therefore, there is no
three-point coupling in the superpotential. Nevertheless, conservation of
$Q$- and $H$-charges allows a limited set of 
four-point couplings: $\chi_{0} \chi_{++}\chi_{--} \Phi$ and 
$\chi_{0} \chi_{+-} \chi_{-+} \Phi$. To see this, let us consider
the $H$- and $Q$-charges of $\chi_{0}$, $\chi_{++}$ and $\chi_{--}$
in the $-1$, $-1/2$ and $-1/2$ picture respectively. The total $H$-charge
is then 
$(1/3,1/3,1/3)+(-1/6,-1/6,-1/6)
+(-1/6,-1/6,-1/6)=(0,0,0)$, whereas the total
$Q$-charge is $(-e_1/3,-e_1/3,-e_1/3)+(-e_3/3,-e_3/3,-e_1/3)
+(-e_2/3,-e_2/3,-e_1/3)=(0,0,-e_1)$. The adjoint $\Phi$ has 
$H=(0,0,1)$ and $Q=(0,0,0)$ in the $-1$ picture. Using the supercurrent
we have constructed in the previous section, it is easy to see that $\Phi$
in the $0$-picture contains a term with $H=(0,0,0)$ and $Q=(0,0,e_1)$.
Hence, the four-point coupling $\chi_{0} \chi_{++}\chi_{--} \Phi$
is allowed. Similarly, one can show that 
$\chi_{0} \chi_{+-} \chi_{-+} \Phi$ is allowed
but other four-point couplings such as 
$\chi_{0} \chi_{++} \chi_{+-} \Phi$ and $\chi_{0} \chi_{-+} \chi_{--}$
are forbidden. This analysis has been performed in the $SU(3)^3$ basis, but 
we could have equally successfully used, say, the $SU(2)^3 \otimes U(1)^3$ 
basis for the $E1$ model, and obtained the same result.
Notice that $\Phi$ carries $H$-charges in the $-1$ picture
and so $\Phi^{n}$ by itself for any integer $n$ is not allowed in the
superpotential due to $H$-charge conservation. 
Hence, the adjoint $\Phi$ is a moduli.
However, in the $0$-picture,
$\Phi$ contains terms with $H=(0,0,0)$ and $Q=(0,e_{\alpha},0)$ or
$(0,0,e_{\alpha})$. Therefore, $\Phi^{3}$ contains terms with no
$H$- and $Q$-charges when all the $\Phi$ are in the $0$-picture.
An immediate consequence is that if $w$ is an allowable $N$-point
coupling ($N \geq 3$) in the superpotential, then $w \Phi^{3n}$ is
also allowable.

{}Let us now turn to the $D$-charge. As mentioned before, in the
branching $E_6 \supset SO(10) \otimes U(1)$, the component fields
({\em i.e.}, different representations of $SO(10)$) of a 
$E_6$ multiplet carry different $D$-charges. They are summarized in
Table \ref{discrete}. This $D$-charge is a ${\bf Z}_3$ symmetry charge
and so it must be conserved modulus $3$.
The origin of this $D$-charge is the $S_3$ permutation symmetry of
the three $(E_6)^3 \supset SO(10)^3 \otimes U(1)^3$.
A priori, $D$-charge
conservation imposes an extra constraint on the superpotential. However,
it turns out that this $D$-charge conservation is subsumed in
the other selection rules. To see this, consider the example of
$\chi_{0} \chi_{++} \chi_{--} \Phi$ coupling. 
In terms of $SO(10) \otimes U(1)$
representations,
\begin{eqnarray}
\chi_{0} \chi_{++} \chi_{--} \Phi = & &
[ Q_{0}Q_{++}H_{--}+Q_{0}Q_{--}H_{++} + Q_{++}Q_{--}H_{0} 
\nonumber \\
&+&H_{0}H_{++}S_{--}+H_{0}H_{--}S_{++} + H_{++}H_{--}S_{0} ] 
(\Phi + \phi) \nonumber \\
&+& [Q_{0}H_{++}S_{--}+Q_{0}H_{--}S_{++}+ Q_{++}H_{0}S_{--}
    +Q_{++}H_{--}S_{0} \nonumber \\
&+& Q_{--}H_{0}S_{++}+ Q_{--}H_{++}S_{0} + Q_{0}Q_{++}Q_{--} ] Q 
\nonumber \\
&+& [ Q_{0}H_{++}H_{--} + Q_{++}H_{0}H_{--} + Q_{--}H_{0}H_{++} ] 
\overline{Q}
\end{eqnarray}
 From Table \ref{discrete}, one can easily check that the 
$D$-charge is conserved for every term on 
the right-handed side. The same is true for other terms in the 
superpotential.
 
{}It is, therefore, useful to organize the fields and the invariant
combinations according to their $U(1)^3$ $G$-charges, $Q$-charges and 
$H$-charges when deducing the selection rules. Instead of trying to write 
down the most general form of the superpotential, for illustrative purposes 
we will give the lowest order non-vanishing couplings here:
\begin{eqnarray}
 W=&\phantom{+}&\lambda_1(\Phi^3,D_{+}D_{-}) 
      \chi_{0} (\chi_{++}\chi_{--} + \chi_{+-}\chi_{-+}) 
      \Phi \nonumber\\
 &+& \lambda_2 (\Phi^3,D_{+}D_{-})\tilde{\chi}_{+}\tilde{\chi}_{-} 
               (\chi_{++}\chi_{--} + \chi_{+-}\chi_{-+}) \nonumber\\
 &+& \lambda_{3} (\Phi^3,D_{+}D_{-}) U_0
               ( U_{++}U_{--} + U_{+-}U_{-+}) \nonumber\\
 &+& \lambda_{4} (\Phi^3,D_{+}D_{-}) 
       [ \chi_{++}^3 U_{--} + \chi_{+-}^3 U_{-+} + 
         \chi_{-+}^3 U_{+-} + \chi_{--}^3 U_{++} ] \Phi^2 \nonumber\\
 &+& \lambda_{5} (\Phi^3,D_{+}D_{-}) (\chi_{0})^3 U_{0}  D_{+}D_{-} \Phi^2
       \nonumber\\
 &+& \lambda_{6} (\Phi^3,D_{+}D_{-})  
   [ (\tilde{\chi}_{+})^3 \tilde{U}_{-} + \tilde{\chi}_{-}^3 \tilde{U}_{+} ]
       D_{+}D_{-} \Phi +...~,
\end{eqnarray}
where traces over the irreps of the 
gauge group are implicit. Here, $\lambda_{k}$ are 
certain polynomials of their respective 
arguments such that $\lambda_{k} (0)\not=0$, {\em i.e.},
\begin{equation}
\lambda_{k}(\Phi^{3},D_{+}D_{-})= \sum_{m,n} \lambda_{kmn} \Phi^{3m} 
(D_{+}D_{-})^{n}
\end{equation}
It is clear that all discrete (local) symmetries impose stringent
constraints on the couplings. This is clearly an important property
of string theory.

\subsection{$E2$ Model: Shift Construction}

{}After a detail discussion of the $E1$ model, our discussion of 
the $E2$ model will be brief. The generating vectors in the shift 
formalism which produce the $E2$ model are: 
\begin{eqnarray}
 &&V_1 =(0~0~(-1/3)^2\vert 0,e_1/3,e_1/3\vert\vert \theta,e_1/3,0 \vert
 {\cal P} \vert  2/3)~,\nonumber\\
 \label{ME2}
 &&V_2=(0~(-1/2)^2~0\vert e_1/2,0,0\vert\vert  0,e_1/2,e_1/2\vert
 0^{15} \vert 0)~.
\end{eqnarray}  
As in $E1$ model, we can rewrite the generating vectors as follows:
\begin{eqnarray}
V_1 &=& (0~0~(-1/3)^2\vert 0,e_1/3,e_1/3\vert\vert (1/3)_t, e_1/3,0 \vert
 (1/3)^5_t 0_r^5 \vert  (2/3)_r)~, \nonumber \\
V_2 &=& (0~(-1/2)^2~0\vert e_1/2,0,0\vert\vert  0,e_1/2,e_1/2 \vert
 0^{15} \vert 0)~. 
\end{eqnarray}
The structure constants can then be determined: $k_{00}=k_{10}=k_{02}=0$, 
$k_{12}=k_{20}=k_{22}=1/2$, and $k_{01}=k_{11}=k_{21}=2/3$. 

{}The currents $i \partial X^a$ that satisfy the triplet constraint 
(\ref{triplet}) and the OPEs (\ref{XOPE}) are given by
\begin{eqnarray}
i \partial X^1 &=& {1\over 2} \left( J_1 + J_2 + J_3 - J_4 \right) ~,
\nonumber \\
i \partial X^2 &=& {1\over \sqrt{6}} \left( L_1+L_2+L_3+L_4+L_5+L_6 \right)~,\\
i \partial X^3 &=& {1\over \sqrt{6}} 
\left( K_1+K_2+K_3-K^{\prime}_1+K^{\prime}_2 + K^{\prime}_3 \right) ~,
\nonumber
\end{eqnarray}
where $J_i=\exp(-i \alpha_i \phi) c(-\alpha_i)$, 
$L_i=\exp(-i \beta_i \phi) c(-\beta_i)$, 
$K_i=\exp(-ik_i \phi) c(-k_i)$ and 
$K^{\prime}_i=\exp(-ik^{\prime}_i \phi) c(-k^{\prime}_i)$.
Here, $\alpha_i$, $\beta_i$, $k_i$ and $k^{\prime}_i$ are the roots
of $E_6$ defined in Table \ref{E2super}. It is easy to see that
$\alpha_1,\dots,\alpha_4$ are simple roots of $SU(2)^4$ whereas
$\beta_1,\dots,\beta_5$ and $-\beta_6$ form the simple roots and
the highest weight of $SU(6)$. The roots of $E_6$ that enter into
$i \partial X^3$ are the same as that in $E1$ model because again
the ${\bf Z}_2$ action does not act on $\partial X^3$ and $\psi^3$
and so $i \partial X^3$ can be expressed in terms of the root
generators of $SU(3)^2$.

{}The massless spectrum of the $E2$ model can be obtained in a similar
way. The $G$-, $Q$- and 
$H$-charges for the $E2$ model are given in Table \ref{E2charges}.
Notice that $E1$ and $E2$ models have the same massless spectrum
with some of the twisted and untwisted sector fields interchanged.
By examining some of the lowest order non-vaishing couplings, one sees that
the $E1$ and $E2$ models have the same tree level superpotential.

\section{$SO(10)$ Model}

{}We have seen in the previous section that for models which admit
a bosonic supercurrent, 
scattering amplitudes become straightforward to compute.
However, there are other three-family grand unified models classified in 
\cite{kt} which do not admit a basis where the supercurrent can be bosonized.
It is, nonetheless, 
possible to deduce the superpotential in an indirect way. We will use 
the $T1(1,1)$ and $T2(1,1)$ models of Ref \cite{kt} to illustrate how
this can be done.

{}Let us begin with the construction of the models.
Start from $N(1,1)$, {\em i.e.}, the $N=4$ SUSY Spin(32)/${\bf Z}_2$
model, and add the following 
Wilson lines:
\begin{eqnarray}
 &&V_1 =(0^4\vert 0~e_1/2~e_1/2
       \vert\vert e_1/2~0~0 \vert {\bf s}~{\bf 0}~{\bf 0} \vert 
{\overline S}) ~,\\
 &&V_2 =(0^4\vert 0~e_2/2~e_2/2
       \vert\vert e_2/2~0~0 \vert {\bf 0}~{\bf s}~{\bf 0} \vert 
{\overline S}) ~.
\end{eqnarray}
The resulting model, which we will refer to as $N2(1,1)$, is a $N=4$ SUSY
model 
with gauge symmetry $SU(3)^2\otimes SO(10)^3\otimes U(1)^3$ provided that we 
set $k_{10}=k_{20}=0$.\\
$\bullet$ The $T1(1,1)$ model. Start from the $N2(1,1)$ model and 
perform the same twists $T_3$ and $T_2$ as in the $E1$ model 
(\ref{E1twist}). This model has 
$SU(2)_1 \otimes SO(10)_3 \otimes U(1)^4$ gauge symmetry. The massless 
spectrum of the $T1(1,1)$ model is given in Table \ref{SO(10)spectra}.\\
$\bullet$ The $T2(1,1)$ model. Start from the $N2(1,1)$ Narain model and 
perform the same twists $T_3$ and $T_2$ as in the $E2$ model 
(\ref{E2twist}). This model has $SU(2)_1 \otimes SO(10)_3 \otimes U(1)^4$ 
gauge symmetry. The massless spectrum of the $T2(1,1)$ model is given in 
Table \ref{SO(10)spectra}.

{}The models $T1(1,1)$ and $T2(1,1)$ are possibly a $T$-dual pair
just as $E1$ and $E2$ are. We will compute the superpotentials 
for these models, and show that they are the same.

{}Let us briefly discuss the underlying conformal field theory, or the 
Kac-Moody algebra ${\cal G}^\prime_R$, for the $T1(1,1)$ and $T2(1,1)$ 
models. Without going into details, we simply state the results: 
${\cal G}^\prime_R=SU(2)_3 \otimes U(1)$ for the $T1(1,1)$ model, 
and ${\cal G}^\prime_R=SU(2)_3 \otimes SU(2)_1$ for the $T2(1,1)$ model. 
Thus, what happens is that the first three $SU(2)_1$s in both models 
(meaning starting from $E1$ and $E2$ and arriving at $T1(1,1)$ and 
$T2(1,1)$, respectively) are broken down to their diagonal subgroup 
$SU(2)_3$. The last two $U(1)s$ are completely broken. The first 
$U(1)$ in the $E1$ model, and the last $SU(2)_1$ in the $E2$ model 
are untouched. In the process of this breaking one encounters additional 
cosets ${\cal C}_R$, which give rise to certain discrete symmetries. 

{}Therefore, the $SO(10)$ models $T1(1,1)$ and $T2(1,1)$
do not 
admit bosonization of the supercurrent because the corresponding 
right-moving Kac-Moody algebras ${\cal G}^\prime_R$ have reduced rank 
which is less than $6$. 
There is, however, a way to deduce their superpotentials which is due 
to the fact that they are connected  
by (classically) flat directions to the $E_6$ model.

{}Comparing the massless spectrum of the $SO(10)$ model to that of the
$E_6$ model, given in Table \ref{E6spectra}, we see that the two spectra 
are very similar. In particular, the spectrum of the $SO(10)$ model is the 
same as that of the $E_6$ model with a non-zero vev of the adjoint $\Phi$ 
of $(E_6)_3$ such that the $(E_6)_3$ is broken down to 
$SO(10)_3 \otimes U(1)$ (the last $U(1)$ in the first and second columns 
of Table \ref{SO(10)spectra}). Therefore, the ${\bf 27}$ of 
$(E_6)_3$ branches into 
${\bf 16}(-1)+{\bf 10}(+2)+ {\bf 1}(-4)$ of $SO(10)_3 \otimes U(1)$. 
The adjoint ${\bf 78}$ branches into 
${\bf 45}(0)+{\bf 1}(0)+{\bf 16}(+3)+{\overline {\bf 16}}(-3)$. 
Note that the ${\bf 45}(0)$ and ${\bf 1}(0)$ are present in the $SO(10)$ 
model, the ${\bf 16}(+3)$ and ${\overline {\bf 16}}(-3)$ are missing, 
however. This is due to the fact that the latter have been eaten by the 
corresponding gauge bosons of  $(E_6)_3$ in the super-Higgs mechanism. 
Thus, in the effective field theory language, the $SO(10)$ model is the 
same as the $E_6$ model with the adjoint vev turned on. There is a subtlety 
here. The two models are equivalent only if we also turn on the vev of the 
singlet $\phi$ in the $SO(10)$ model correspondingly. This can be seen
from the fact that once the adjoint of $(E_6)_3$ acquires a vev, 
there is an effective three-point Yukawa coupling in the $SO(10)$ model 
which has the form $H_0 Q_+ Q_-$ (here we define $Q_{\pm}$ and 
$Q^\prime_{\pm}$ in the same way as $\chi_{\pm}$ and $\chi^\prime_{\pm}$). 
In the $SO(10)$ model without the vev of $\phi$ turned on there are only 
four-point couplings  $H_0 Q_+ Q_- \phi$ and $H_0 Q_+ Q_- \Phi$. Thus, 
to get the three-point coupling, we have to turn on the vev of $\phi$ 
(whereas turning on the vev of $\Phi$ in the $SO(10)$ model would break 
the $SO(10)$ symmetry further, say, to $SU(5)\otimes U(1)$). From 
this discussion it is 
clear how to get the superpotential for the $SO(10)$ model from that of 
the $E_6$ model. One simply starts from the latter and turns on the vev of 
the adjoint of $(E_6)_3$. In practice, just to get the selection rules for 
the $SO(10)$ model, simply replace $\chi$s by $(Q+H+S)$s (and similarly for 
${\tilde \chi}s$), and $\Phi$ by $\Phi +\phi$. Let us write down the
first few lowest order terms in the superpotential:
\begin{eqnarray}
 W=& & \lambda_1((\Phi + \phi^{\prime} )^3,D_{+}D_{-}) 
   \left[ Q_{0} (Q_{++}H_{--} + Q_{+-}H_{-+} + Q_{-+}H_{+-} + Q_{--}H_{++})
\nonumber \right. \\
      &+& H_{0} (Q_{++}Q_{--}+Q_{+-}Q_{-+}) 
      + H_{0} (H_{++}S_{--}+H_{+-}S_{-+}+H_{-+}S_{+-}+H_{--}S_{++})
\nonumber \\
&+&\left.  S_{0} (H_{++}H_{--}+H_{+-}H_{-+}) \right]  
      (\Phi + \phi^{\prime}) \nonumber\\
&+&\lambda_2 ((\Phi + \phi^{\prime} )^3,D_{+}D_{-})
   \left[ \tilde{Q}_{+}\tilde{Q}_{-} (Q_{++}Q_{--}+Q_{+-}Q_{-+}) \right.
\nonumber \\
&+& \tilde{Q}_{+}\tilde{Q}_{-} ( H_{++}S_{--}+H_{+-}S_{-+}
                               + H_{-+}S_{+-}+H_{--}S_{++} )
\nonumber \\
&+&(\tilde{Q}_{+}\tilde{S}_{-}+\tilde{Q}_{-}\tilde{S}_{+})
          (Q_{++}S_{--}+Q_{+-}S_{-+} + Q_{-+}S_{+-} + Q_{--}S_{++}) 
               \nonumber\\
&+&(\tilde{H}_{+} \tilde{S}_{-} +\tilde{H}_{-} \tilde{S}_{+})
   (H_{++}S_{--}+H_{+-}S_{-+}+H_{-+}S_{+-}+H_{--}S_{++}) \nonumber \\
&+&(\tilde{H}_{+} \tilde{S}_{-} +\tilde{H}_{-} \tilde{S}_{+})
   (Q_{++}Q_{--}+Q_{+-}Q_{-+}) \nonumber \\
&+&(\tilde{Q}_{+}\tilde{H}_{-}+\tilde{Q}_{-}\tilde{H}_{+})
   (Q_{++}H_{--} + Q_{+-}H_{-+} + Q_{-+}H_{+-} + Q_{--}H_{++})
\nonumber \\
&+& \left. \tilde{H}_+ \tilde{H}_- (H_{++}H_{--}+H_{+-}H_{-+})
   + \tilde{S}_+ \tilde{S}_- (S_{++}S_{--}+S_{+-}S_{-+}) \right] 
\nonumber\\
 &+& \lambda_{3} ((\Phi + \phi^{\prime} )^3,D_{+}D_{-}) U_0
               ( U_{++}U_{--} + U_{+-}U_{-+}) \nonumber\\
 &+& \lambda_{4} ((\Phi + \phi^{\prime} )^3,D_{+}D_{-}) 
     \left[ (Q_{++}^2 H_{++} + H_{++}^2 S_{++}) U_{--} + 
         (Q_{+-}^2 H_{+-} + H_{+-}^2 S_{+-}) U_{-+} \right. \nonumber \\ 
 &+& \left.  (Q_{-+}^2 H_{-+} + H_{-+}^2 S_{-+}) U_{+-} 
        +(Q_{--}^2 H_{--} + H_{--}^2 S_{--}) U_{++} \right] 
(\Phi + \phi^{\prime} )^2 +...~,
\end{eqnarray}
where traces over the irreps of the 
gauge group are implicit and the Clebsch-Gordan coefficients are
of order $1$. 
The field $\phi^{\prime}$ is defined to be
\begin{equation}
\phi^{\prime}= \phi + \langle \phi \rangle
\end{equation}
Here, $\lambda_{k}$ are 
certain polynomials of their respective 
arguments such that $\lambda_{k} (0)\not=0$, {\em i.e.},
\begin{equation}
\lambda_{k}((\Phi + \phi^{\prime} )^3,D_{+}D_{-})
= \sum_{m,n} \lambda_{kmn} (\Phi + \phi^{\prime})^{3m} (D_{+}D_{-})^{n}
\end{equation}

\section{$SU(6)$ Models}

{}In this section we give the construction of two three-family $SU(6)$ 
models. Both of them can be obtained from the $E_6$ models discussed in 
section VI by adding the following Wilson line:
\begin{equation}
 V_3 =(0^4\vert 0^3\vert\vert 0,{\tilde e}^2,{\tilde e}^2\vert 
 ({1\over 3}{1\over 3} {1\over 3} {1\over 3} {2\over 3})^3  \vert  0)~.
\end{equation}
Since the Wilson line $V_3$ does not act on the right-movers, the
supercurrent is the same as that of the original $E_6$ model.\\
$\bullet$ The $S1$ model. Add the $V_3$ Wilson line to the set (\ref{ME1}) 
and set $k_{13}=0$ or $k_{13}=1/3$ (both choices give the same model). This 
model has 
$SU(3)_1 \otimes SU(6)_3 \otimes U(1)^3$ gauge symmetry. 
The $Q$- and $H$-charges of the massless spectrum are 
given in Table \ref{S1charges}.\\
$\bullet$ The $S2$ model. Add the $V_3$ Wilson line to the set (\ref{ME1}) 
and set $k_{13}=2/3$. This model has 
$SU(2)_1 \otimes SU(2)_1 \otimes SU(6)_3 \otimes U(1)^3$ gauge symmetry.
The $Q$- and $H$-charges of the massless spectrum are given in 
Table \ref{S2charges}.\\
$\bullet$ The $S3$ model. Add the $V_3$ Wilson line to the set (\ref{ME2}) 
and set $k_{13}=0$ or $k_{13}=1/3$ (both choices give the same model). This 
model has 
$SU(3)_1 \otimes SU(6)_3 \otimes U(1)^3$ gauge symmetry.
The $Q$- and $H$-charges of the massless spectrum are 
given in Table \ref{S3charges}.\\
$\bullet$ The $S4$ model. Add the $V_3$ Wilson line to the set (\ref{ME2}) 
and set $k_{13}=2/3$. This model has 
$SU(2)_1 \otimes SU(2)_1 \otimes SU(6)_3 \otimes U(1)^3$ gauge symmetry. 
The $Q$- and $H$-charges of the massless spectrum are given in 
Table \ref{S4charges}.

{}The models $S1$ and $S3$ have the same tree-level massless spectra. 
We will show that interactions are also the same. These two models are
possibly 
$T$-dual to each other. Similarly, $S2$ and $S4$ are a possible 
$T$-dual pair. 

\subsection{Superpotential For $SU(6)$ Model $S1=S3$}.

{}Next we give the superpotential for the $S1=S3$ model. We will be brief 
here as the techniques involved in deducing the selection rules should be 
clear by now after we have given the example of the $E_6$ model. 

{}The superpotential for the $SU(3)_1 \otimes SU(6)_3 \otimes U(1)^3$ 
model ($S1=S3$) reads:
 \begin{eqnarray}
 W = &\phantom{+}& \lambda_{1} (\Phi,\phi) U_0
               ( U_{++}U_{--} +U_{+-}U_{-+} )
      +  \lambda_{2} (\Phi,\phi) T_0
               ( {\tilde T}_{+} U_{-+} +{\tilde T}_{-} U_{++}) \nonumber\\
     &+& \lambda_{3} (\Phi,\phi) {\tilde T}_0 T_- U_0 +
           \lambda_{4} (\Phi,\phi) {\tilde T}_0 T_0 {\tilde U}_- \nonumber\\
     &+& \lambda_{5} (\Phi,\phi) \sum_{A,B,C} y_{ABC}
 [ U_{++} ( F^A_{+}F^B_{+}F^C_{-} + {1 \over 3} F^A_{-}F^B_{-}F^C_{-} )
 + U_{-+} ( F^A_{+}F^B_{-}F^C_{-} + {1 \over 3} F^A_{+}F^B_{+}F^C_{+} )] 
\nonumber\\
     &+& [\lambda_{6} (\Phi,\phi)  \Phi^2 
         +\lambda_{7} (\Phi,\phi) \Phi \phi
         +\lambda_{8} (\Phi,\phi)  \phi^2] \nonumber \\
     && ~~~~~~\times \sum_A \{ U_{++} [(F^A_{-})^3 + F^A_{-}(F^A_{+})^2] 
                        +U_{-+} [(F^A_{+})^3 + F^A_{+}(F^A_{-})^2] \}
\nonumber\\
     &+& \lambda_{9} (\Phi,\phi) \sum_A 
            [F^A_{+} {\tilde S}^A_{-} + F^A_{-} {\tilde S}^A_{+}] 
      {\tilde S}^A_0 \nonumber\\
     &+& [\lambda_{10} (\Phi,\phi) \Phi + \lambda_{11} (\Phi,\phi) \phi ]
      \sum_{A,B,C} y_{ABC} (F^A_{+} {\tilde S}^B_{-}+F^A_{-} {\tilde S}^B_{+})
      {\tilde S}^C_0  \nonumber\\
      &+& \lambda_{12} (\Phi,\phi)  {\tilde U}_+ T_+ {\tilde T}_0 
       \sum_{A,B,C} y_{ABC} {\tilde F}^A {\tilde F}^B 
        {\tilde F}^C  \nonumber\\
      &+& [\lambda_{13} (\Phi,\phi)\Phi + \lambda_{14} (\Phi,\phi)\phi ] 
{\tilde U}_+ T_+ {\tilde T}_0
      \sum_{A}  ({\tilde F}^A)^3   \nonumber\\
      &+& \lambda_{15}(\Phi,\phi) \sum_{A,B} z_{AB} {\tilde F}^A S^B
         ( F^A_+ {\tilde S}^B_{-} + F^A_- {\tilde S}^B_{+}  
          +F^B_+ {\tilde S}^A_{-} + F^B_- {\tilde S}^A_{+} )        
\nonumber\\
      &+& [\lambda_{16}(\Phi,\phi) \Phi + \lambda_{17}(\Phi,\phi)\phi]  
      \sum_{A,B,C} y_{ABC} {\tilde F}^A S^A 
        (F^B_{+} {\tilde S}^C_{-}+ F^B_{-} {\tilde S}^C_{+})  \nonumber\\
      &+& [\lambda_{18}(\Phi,\phi) \Phi^2 + 
     \lambda_{19}(\Phi,\phi) \Phi  \phi+
      \lambda_{20}(\Phi,\phi) \phi^2]
     \sum_{A,B,C} y_{ABC} [F^A_{+} {\tilde S}^A_{-} +F^A_{-} {\tilde S}^A_{+}]
       {\tilde F}^B S^C  
      \nonumber\\
     &+& [\lambda_{21}(\Phi,\phi) \Phi^3 + 
     \lambda_{22}(\Phi,\phi) \Phi^2 \phi+
      \lambda_{23}(\Phi,\phi) \Phi \phi^2+\lambda_{24}(\Phi,\phi) \phi^3]
      \nonumber\\
     &&~~~~~~\times \sum_{A}[F^A_{+} {\tilde S}^A_{-}+F^A_{-}{\tilde S}^A_{+}]
        {\tilde F}^A S^A 
      \nonumber\\
     &+& \lambda_{25}(\Phi,\phi) \sum_{A,B,C} y_{ABC} 
                      [ (F^{A}_{+})^2+(F^{A}_{-})^2 ] S^B S^C 
                      ( U_{++}U_{+-} + U_{-+}U_{--} ) 
      \nonumber\\
     &+& \lambda_{26}(\Phi,\phi) \sum_{A,B,C} y_{ABC} 
                F^A_{+}F^A_{-}S^B S^C ( U_{++}U_{--}+U_{+-}U_{-+} )
      \nonumber\\
     &+& [ \lambda_{27}(\Phi,\phi) \Phi + \lambda_{28}(\Phi,\phi) \phi ]
                 \sum_{A,B} z_{AB}
                F^A_{+}F^B_{+}S^A S^B ( U_{++}U_{+-} + U_{-+}U_{--} ) 
      \nonumber\\
     &+& [ \lambda_{29}(\Phi,\phi) \Phi^2 
          +\lambda_{30}(\Phi,\phi) \Phi \phi 
          +\lambda_{31}(\Phi,\phi) \phi^2 ]
      \nonumber \\         
     &&~~~~~~\times \sum_{A,B,C} y_{ABC} 
                F^A_{+}F^B_{+} (S^C)^2 ( U_{++}U_{+-} + U_{-+}U_{--} ) 
      \nonumber\\
     &+& [ \lambda_{32}(\Phi,\phi) \Phi + \lambda_{33}(\Phi,\phi) \phi ]
           \sum_{A,B} z_{AB} F^A_{+}F^B_{-}S^A S^B 
            ( U_{++}U_{--}+U_{+-}U_{-+} )
      \nonumber\\
     &+& [ \lambda_{34}(\Phi,\phi) \Phi^2 + \lambda_{35}(\Phi,\phi) \Phi \phi
          +\lambda_{36}(\Phi,\phi) \phi^2 ]
      \nonumber\\
     &&~~~~~~\times
           \sum_{A,B,C} y_{ABC} F^A_{+}F^B_{-}(S^C)^2
            ( U_{++}U_{--}+U_{+-}U_{-+} )   
      \nonumber\\
     &+&\lambda_{37}(\Phi,\phi) \sum_{A,B,C} y_{ABC} 
               \tilde{S}^A_{+} \tilde{S}^B_{-} (\tilde{F}^C)^2 U_{0}
      \nonumber\\
     &+& [ \lambda_{38}(\Phi,\phi) \Phi + \lambda_{39}(\Phi,\phi) \phi ] 
       \sum_{A,B,C} y_{ABC} 
          \tilde{S}^A_{+} \tilde{S}^A_{-} \tilde{F}^B \tilde{F}^C U_{0}
      \nonumber\\
     &+& [ \lambda_{40}(\Phi,\phi) \Phi^2 + \lambda_{41}(\Phi,\phi) \Phi \phi 
          + \lambda_{42}(\Phi,\phi) \phi^2 ] \sum_{A}
            \tilde{S}^A_{+} \tilde{S}^A_{-} (\tilde{F}^A)^2 U_{0}     
     \nonumber\\
     &+& [ \lambda_{43}(\Phi,\phi) \Phi^2 + \lambda_{44}(\Phi,\phi) \Phi \phi
          + \lambda_{45}(\Phi,\phi) \phi^2 ] \sum_{A,B} z_{AB}
            \tilde{S}^A_{+} \tilde{S}^B_{-} \tilde{F}^A \tilde{F}^B U_{0}
      +...~,
\end{eqnarray}
where $\lambda_{k}$, $k=1,...,45$, are certain polynomials of their 
respective arguments, which combine into the terms of the form 
$\Phi^{3n-m} \phi^m$, $n,m\in {\bf N}$,  such that 
$\lambda_k (0,\phi)\not\equiv 0$, and $\lambda_k (\Phi,0)\not\equiv 0$, 
and traces over the irreps of the gauge group are implicit here. The 
coefficients $y_{ABC}$ and $z_{AB}$ are defined as follows: 
$y_{ABC}=\epsilon_{ABC}$, and $z_{AB}=1-\delta_{AB}$.

\subsection{Superpotential For $SU(6)$ Model $S2=S4$}.

{}The superpotential for the $SU(2)_1^2 \otimes SU(6)_3 \otimes U(1)^3$
model ($S2=S4$) reads:
\begin{eqnarray}
 W= &\phantom{+}& \lambda_{1} (\Phi,\phi) U_0 ( d_{++}d_{--} +d_{+-}d_{-+})
      +\lambda_{2} (\Phi,\phi)
        (D_+ {\tilde d}_- \Delta_- + D_- {\tilde d}_+ \Delta_+)  \nonumber\\
    &+&\lambda_{3} (\Phi,\phi) U_0 \Delta_+ \Delta_- \sum_{A,B,C} y_{ABC}
       F^A F^B F^C \nonumber\\
    &+&[\lambda_{4} (\Phi,\phi) \Phi^2 +\lambda_{5} (\Phi,\phi)\Phi \phi +
      \lambda_{6} (\Phi,\phi) \phi^2 ]  U_0 \Delta_+ \Delta_- \sum_{A}
       (F^A)^3 \nonumber\\
    &+& \lambda_{7} (\Phi,\phi) \sum_A F^A 
       [ {\tilde S}^A_{++}{\tilde S}^A_{--} 
        +{\tilde S}^A_{+-}{\tilde S}^A_{-+} ]  \nonumber\\
    &+& [\lambda_{8} (\Phi,\phi)\Phi + \lambda_{9} (\Phi,\phi)\phi ]
      \sum_{A,B,C} y_{ABC} F^A ({\tilde S}^B_{++}{\tilde S}^C_{--} 
                              +{\tilde S}^B_{-+}{\tilde S}^C_{+-}) \nonumber\\ 
    &+& \lambda_{10} (\Phi,\phi) \sum_{A,B} z_{AB} S^A_{+}S^B_{-}
     ({\tilde S}^A_{++} {\tilde S}^B_{--}+{\tilde S}^A_{-+} {\tilde S}^B_{+-}
     +{\tilde S}^A_{+-} {\tilde S}^B_{-+}+{\tilde S}^A_{--} {\tilde S}^B_{++})
       \nonumber \\ 
    &+& [\lambda_{11}(\Phi,\phi) \Phi^3 + 
         \lambda_{12}(\Phi,\phi) \Phi^2 \phi+
         \lambda_{13}(\Phi,\phi) \Phi \phi^2+\lambda_{14}(\Phi,\phi) \phi^3]
   \nonumber\\
     &&~~~~~~\times \sum_{A} S^A_+ S^A_- 
    ({\tilde S}^A_{++} {\tilde S}^A_{--} + {\tilde S}^A_{+-}{\tilde S}^A_{-+})
   \nonumber\\
    &+& [\lambda_{15}(\Phi,\phi) \Phi + \lambda_{16}(\Phi,\phi)  \phi]
     \sum_{A,B,C} y_{ABC} S^A_+ S^A_- 
      ({\tilde S}^B_{++}{\tilde S}^C_{--}+{\tilde S}^B_{-+}{\tilde S}^C_{+-}) 
\nonumber\\
  &+&[\lambda_{17}(\Phi,\phi) \Phi^2 + \lambda_{18}(\Phi,\phi) \Phi \phi+
   \lambda_{19}(\Phi,\phi) \phi^2 ] \nonumber\\  
    &&~~~~~~\times \sum_{A,B,C} y_{ABC} S^B_{+}S^C_{-}
    ({\tilde S}^A_{++}{\tilde S}^A_{--}+{\tilde S}^A_{+-}{\tilde S}^A_{-+}) 
    +...~.
\end{eqnarray}
The couplings $\lambda_k$ and the coefficients $y_{ABC}$ and $z_{AB}$ are
defined in the same way as in the $S1=S3$ model. 

\section{Remarks}

{}Using the bosonic supercurrent formalism, we give a prescription
how to calculate the correlation functions in the $3$-family grand 
unification models. We use this approach to determine the quantum
numbers of the massless spectra of some of these models. This gives us
selection rules for the allowed couplings in the respective superpotentials. 
Many couplings that are allowed by gauge symmetries are forbidden by
stringy discrete symmetries.
The explicit values of the couplings are not hard to determine; however we 
will leave them 
for the future. Even without their explicit determination, there are still
plenty of phenomenological issues one can address. This will also be
discussed elsewhere. One question that might need clarification is whether 
there are additional discrete quantum numbers associated with the 
coset (${\cal C}_L$) coming from the rank reduction. 

{}In this paper, we discuss explicitly some $3$-family grand 
unified string models: the unique $E_6$ model, an $SO(10)$ model and 
two $SU(6)$ models. 
There are other $SO(10)$, $SU(5)$ and $SU(6)$ three-family grand unified 
string models classified in Refs \cite{kt}. 
Most of these models are connected to the unique $E_6$ model by classically 
flat moduli. Some $SU(5)$ models are connected to the two $SU(6)$ models. 
Finally, there are a few $SU(5)$ models and one unique $SO(10)$ model that 
are isolated from the $E_6$ and two $SU(6)$ models we have described here. 
The superpotentials of the models that are connected to the above models
via flat moduli can be easily obtained simply by giving the flat moduli
appropriate vevs. These vevs can be of the order of string scale.

{}As we have already mentioned, some of these models do not 
admit bosonization of the supercurrent because their corresponding 
right-moving Kac-Moody algebras ${\cal G}^\prime_R$ have reduced rank 
which is less than $6$. Thus, deducing the discrete symmetries is 
complicated by the presence of the right-moving cosets ${\cal C}_R$. 
There is, however, a way to deduce their superpotentials 
when 
they are connected
by (classically) flat directions to the above three models.
We have illustrated how this can be done with the $SO(10)$ model 
$T1(1,1)=T2(1,1)$.

{}We note that all the couplings (in string units) 
$\lambda_k$ are of order one for vanishing values of the $\Phi$ 
and $\phi$ vevs. The latter fields are (classically) flat moduli in these 
models. In the $E_6$ and the two $SU(6)$ models we have studied in this 
paper, there are no other completely flat moduli of a geometric origin, 
but such flat directions are present in other three-family grand unified 
string models classified in Refs \cite{kt}. Their stabilizations can only be 
achieved via non-perturbative dynamics.

\acknowledgments

\bigskip

{}We would like to thank Damiano Anselmi, Ignatios Antoniadis, 
Costas Bachas, Alexander Bais, Riccardo Barbieri, Zurab Berezhiani, 
Michael Bershadsky, Keith Dienes, Gia Dvali, Alon Faraggi, Matthias Gaberdiel,
Luis Ib{\'a}{\~n}ez, Andrei Johansen, Elias Kiritsis, Costas Kounnas, 
Pran Nath, Peter Nilles, Michal Spalinski, Zurab Tavartkiladze,
Tom Taylor, Angel Uranga, Cumrun Vafa, Erik Verlinde, Herman Verlinde, 
Yan Vtorov-Karevsky, 
Mikhail Vysotsky and Barton Zwiebach 
for discussions. The research of G.S. and S.-H.H.T. was partially 
supported by the National Science Foundation. G.S. would also like to thank
Joyce M. Kuok Foundation for financial support. The work of Z.K. was 
supported in part by the grant NSF PHY-96-02074, and the DOE 1994 OJI award. 
Z.K. would like to thank the High Energy Theory Group at Cornell University 
for their kind hospitality while parts of this work were completed.  
Z.K. would also like to thank Mr. Albert Yu and Mrs. Ribena Yu for 
financial support.


\begin{table}[t]
\begin{tabular}{|c|l|l|l|l|l|}
 & Field &  $SU(3) \otimes E_6 \otimes E_8 \otimes SU(3)^{3}$ 
 & $Q^R$-charges in $SU(3)^3$ & $(H_1,H_2,H_3)_{-1}$ 
 & $(H_1,H_2,H_3)_{-1/2}$
      \\ \hline
  & & & & & \\
  & $\chi_{1}$ &$({\bf 3},{\bf 27},{\bf 1},{\bf 1},{\bf 1},{\bf 1})$
  & $(0,0,0)$ &$(+1,0,0)$ & $(+{1\over 2},-{1\over 2},-{1\over 2})$ \\
$U$
  & $\chi_{2}$ &$({\bf 3},{\bf 27},{\bf 1},{\bf 1},{\bf 1},{\bf 1})$
  & $(0,0,0)$ &$(0,+1,0)$ & $(-{1\over 2},+{1\over 2},-{1\over 2})$ \\
  & $\chi_{3}$ &$({\bf 3},{\bf 27},{\bf 1},{\bf 1},{\bf 1},{\bf 1})$
  & $(0,0,0)$ &$(0,0,+1)$ & $(-{1\over 2},-{1\over 2},+{1\over 2})$ \\
 & & & & & \\
  \hline
 & & & & & \\
 & $\tilde{\chi}$ 
 & $({\bf 3},\overline{\bf 27},{\bf 1},{\bf 1},{\bf 1},{\bf 1})$
 & $(-e_{1}/3,-e_{1}/3,-e_{1}/3)$
 & $({1\over 3},{1\over 3},{1\over 3})$ 
 & $(-{1\over 6},-{1\over 6},-{1\over 6})$  \\
 & $\chi_{1+}$
 & $({\bf 1},{\bf 27},{\bf 1},{\bf 3},{\bf 1},{\bf 1})$
 & $(-e_{2}/3,-e_{1}/3,-e_{1}/3)$
 & $({1\over 3},{1\over 3},{1\over 3})$ 
 & $(-{1\over 6},-{1\over 6},-{1\over 6})$  \\
 & $\chi_{1-}$
 & $({\bf 1},{\bf 27},{\bf 1},\overline{\bf 3},{\bf 1},{\bf 1})$
 & $(-e_{3}/3,-e_{1}/3,-e_{1}/3)$
 & $({1\over 3},{1\over 3},{1\over 3})$ 
 & $(-{1\over 6},-{1\over 6},-{1\over 6})$  \\
$T3$
 & $\chi_{2+}$
 & $({\bf 1},{\bf 27},{\bf 1},{\bf 1},{\bf 3},{\bf 1})$
 & $(-e_{1}/3,-e_{2}/3,-e_{1}/3)$
 & $({1\over 3},{1\over 3},{1\over 3})$ 
 & $(-{1\over 6},-{1\over 6},-{1\over 6})$  \\
 & $\chi_{2-}$
 & $({\bf 1},{\bf 27},{\bf 1},{\bf 1},\overline{\bf 3},{\bf 1})$
 & $(-e_{1}/3,-e_{3}/3,-e_{1}/3)$
 & $({1\over 3},{1\over 3},{1\over 3})$ 
 & $(-{1\over 6},-{1\over 6},-{1\over 6})$  \\
 & $\chi_{3+}$
 & $({\bf 1},{\bf 27},{\bf 1},{\bf 1},{\bf 1},{\bf 3})$
 & $(-e_{1}/3,-e_{1}/3,-e_{2}/3)$
 & $({1\over 3},{1\over 3},{1\over 3})$ 
 & $(-{1\over 6},-{1\over 6},-{1\over 6})$  \\
 & $\chi_{3-}$
 & $({\bf 1},{\bf 27},{\bf 1},{\bf 1},{\bf 1},\overline{\bf 3})$
 & $(-e_{1}/3,-e_{1}/3,-e_{3}/3)$
 & $({1\over 3},{1\over 3},{1\over 3})$ 
 & $(-{1\over 6},-{1\over 6},-{1\over 6})$  \\
 & $T_{{\bf x}{\bf y}{\bf z}}$
 & $(\overline{\bf 3},{\bf 1},{\bf 1},{\bf x},{\bf y},{\bf z})$
 & $(Q({\bf x}),Q({\bf y}),Q({\bf z}))$
 & $({1\over 3},{1\over 3},{1\over 3})$ 
 & $(-{1\over 6},-{1\over 6},-{1\over 6})$  \\
 & & & & & \\
\end{tabular}
\caption{The $G$-, $Q^R$- and $H$-charges (in the $-1$ and $-1/2$ pictures) of 
the massless fields in the asymmetric ${\bf Z}_3$ orbifold model.
Here, ${\bf x}$, ${\bf y}$ and ${\bf z}$ are irreps of $SU(3)$ such that only
one of them is a singlet and the others can be either ${\bf 3}$ or
$\overline{\bf 3}$. The charges $Q$ as a function of irrep is given
by $Q({\bf 1})=-e_{1}/3$, $Q({\bf 3})=-e_{2}/3$ and 
$Q(\overline{\bf 3})=-e_{3}/3$.}  
\end{table}

\begin{table}[t]
\begin{tabular}{|c|l|l||l|l|}
 &$E1$ & & $E2$ &\\
 M & $SU(2) \otimes E_6 \otimes U(1)^3$ & Field 
   & $SU(2) \otimes E_6 \otimes U(1)^3$ & Field
      \\ \hline
 & & & &\\
   & $ ({\bf 1},{\bf 78})(0,0,0)_L$ & $\Phi$ 
   & $ ({\bf 1},{\bf 78})(0,0,0)_L$ & $\Phi$ \\
$U$   
   & $ ({\bf 1},{\bf 1})(0,+6,0)_L$ & $U_{0}$ 
   & $ ({\bf 1},{\bf 1})(0,+6,0)_L$ & $U_{0}$\\
   & $ 2 ({\bf 1},{\bf 1})(0,-{3},\pm 3)_L$ & $U_{+ \pm}, U_{- \pm}$ 
   & $({\bf 2}, {\bf 1})(0,0,\pm 3)_L$ & $D_{\pm}$\\
 & & & $({\bf 1},{\bf 1})(\pm 3,+{3},0)_L$ & $\tilde{U}_{\pm}$\\
  & & & &\\
  \hline
 & & & & \\
$T3$
   &  $ ({\bf 1},{\bf 27})(0,-{2},0)_L$ & $\chi_{0}$ 
   &  $ ({\bf 1},{\bf 27})(0,-{2},0)_L$ & $\chi_{0}$\\
   & $2({\bf  1}, {\bf 27})(0,+1,\pm1)_L$ & $\chi_{+ \pm}, \chi_{- \pm}$ 
   & $({\bf 1},{\overline {\bf 27}}) (\pm 1,-1,0)_L$ &$\tilde{\chi}_{\pm}$\\
 & & & & \\
 \hline
  & & & & \\
  T6  &  $ ({\bf 1},{\overline {\bf 27}}) (\pm 1,-1,0)_L$ 
      &  $ \tilde{\chi}_{\pm}$
      &  $2({\bf  1}, {\bf 27})(0,+1,\pm1)_L$  &$\chi_{+ \pm}, \chi_{- \pm}$\\
    & & & & \\
 \hline
 & & & & \\
   $T2$ & $({\bf 2},{\bf 1})(0,0,{\pm 3})_L$ & $D_{\pm}$
        & $2 ({\bf 1},{\bf 1})(0,-{3},\pm 3)_L$ & $U_{+ \pm},U_{- \pm}$\\
    &  $ ({\bf 1},{\bf 1})(\pm {3},+{3},0)_L$ & $\tilde{U}_{\pm}$ & &\\
  & & & & \\
   \hline
 & & & & \\
 $U(1)$ & $(1/ \sqrt{6}, ~1/{3\sqrt{2}}, ~1/\sqrt{6})$ & 
        & $(1/ \sqrt{6}, ~1/{3\sqrt{2}},  ~1/\sqrt{6})$ &\\
\end{tabular}
\caption{The massless spectra of the $T$-dual pair of $E_6$ models $E1$ and 
$E2$ 
both with gauge symmetry $SU(2)_1 \otimes (E_6)_3\otimes U(1)^3$.
The $U(1)$ normalization
radii are given at the bottom of the Table.
The gravity, dilaton and gauge supermultiplets are not shown.}
\label{E6spectra}
\end{table}
 
\begin{table}[t]
\begin{tabular}{|c|l|l|l|l|l|}
 $E1$ & Field &  $SU(2) \otimes E_6 \otimes U(1)^{3}$ &
 $Q^R$-charges in $SU(3)^3$  
 & $(H_1,H_2,H_3)_{-1}$ & $(H_1,H_2,H_3)_{-1/2}$
      \\ \hline
  & & & & &  \\
  & $\Phi$ & $({\bf 1},{\bf 78})(0,0,0)_L$
  &$(0,0,0)$ & $(0,0,+1)$ 
  &$(-{1\over 2}, -{1\over 2}, +{1\over 2})$ \\    
  $U$ & $U_0$ & $({\bf 1},{\bf 1})(0,+6,0)_L$
  &$(0,0,0)$ & $(0,0,+1)$ 
  &$(-{1\over 2}, -{1\over 2}, +{1\over 2})$ \\    
  & $U_{+ \pm}$ & $({\bf 1},{\bf 1})(0,-{3},\pm 3)_L$
  &$(0,0,0)$ & $(+1,0,0)$ 
  & $(+{1\over 2}, -{1\over 2}, -{1\over 2})$ \\    
  & $U_{- \pm}$ & $({\bf 1},{\bf 1})(0,-{3},\pm 3)_L$
  &$(0,0,0)$ & $(0,+1,0)$ 
  & $(-{1\over 2}, +{1\over 2}, -{1\over 2})$ \\
 & & & & & \\
  \hline
 & & & & & \\
 & $\chi_0$ & $({\bf 1},{\bf 27})(0,-{2},0)_L$
 & $-(e_1/3,e_1/3,e_1/3)$
 & $(+{1\over 3}, +{1\over 3},+{1\over 3})$ 
 &  $(-{1\over 6}, -{1\over 6},-{1\over 6})$\\ 
 & $\chi_{++}$ & $({\bf  1}, {\bf 27})(0,+1,+1)_L$
 & $-(e_3/3,e_3/3,e_1/3)$ 
 & $(+{1\over 3}, +{1\over 3},+{1\over 3})$ 
 & $(-{1\over 6}, -{1\over 6},-{1\over 6})$ \\
$T3$
 & $\chi_{-+}$ & $({\bf  1}, {\bf 27})(0,+1,+1)_L$
 & $-(e_2/3,e_1/3,e_3/3)$ 
 & $(+{1\over 3}, +{1\over 3},+{1\over 3})$ 
 & $(-{1\over 6}, -{1\over 6},-{1\over 6})$ \\
 & $\chi_{+-}$ & $({\bf  1}, {\bf 27})(0,+1,-1)_L$
 & $-(e_3/3,e_1/3,e_2/3)$ 
 & $(+{1\over 3}, +{1\over 3},+{1\over 3})$
 & $(-{1\over 6}, -{1\over 6},-{1\over 6})$\\
 & $\chi_{--}$ & $({\bf  1}, {\bf 27})(0,+1,-1)_L$
 & $-(e_2/3,e_2/3,e_1/3)$ 
 & $(+{1\over 3}, +{1\over 3},+{1\over 3})$
 & $(-{1\over 6}, -{1\over 6},-{1\over 6})$\\
 & & & & &  \\ 
 \hline
 & & & & &  \\
 $T6$ & $\tilde{\chi}_{+}$ & $({\bf 1},{\overline {\bf 27}}) (+1,-1,0)_L$
 & $(-e_1/6,e_3/3,e_3/3)$ 
 & $(+{1\over 6},+{1\over 6},+{2\over 3})$
 &$(-{1\over 3},-{1\over 3},+{1\over 6})$ \\ 
 & $\tilde{\chi}_{-}$ & $({\bf 1},{\overline {\bf 27}}) (-1,-1,0)_L$
 & $(-e_1/6,e_2/3,e_2/3)$ 
 &  $(+{1\over 6},+{1\over 6},+{2\over 3})$
 &  $(-{1\over 3},-{1\over 3},+{1\over 6})$\\
 & & & & &\\
 \hline
 & & & & &\\ 
  $T2$ & $D_{\pm}$ &$({\bf 2},{\bf 1})(0,0,{\pm 3})_L$ 
 & $(e_1/2,0,0)$
 & $(+{1\over 2},+{1\over 2},0)$ 
 & $(0,0,-{1\over 2})$ \\
 & $\tilde{U}_{\pm}$ &$ ({\bf 1},{\bf 1})(\pm {3},+{3},0)_L$
 & $-(e_1/2,0,0)$ 
 & $(+{1\over 2},+{1\over 2},0)$ 
 & $(0,0,-{1\over 2})$ \\
  & & & & &\\
  \hline
 & & & & &\\
  $U(1)$ & & $(1/ \sqrt{6}, ~1/{3\sqrt{2}}, ~1/\sqrt{6})$
  & &$(1,1,1)$ & $(1,1,1)$
\end{tabular}
\caption{The $G$-, $Q^R$- and $H$-charges (in the $-1$ and $-1/2$ pictures) of 
the massless fields in the $E1$ model. The $U(1)$ normalization radii for 
the $G$- and $H$-charges are given at the bottom of the Table. 
The $Q^R$-charges (in the $-1$ picture which are the same as in the $-1/2$ 
picture) are written in $SU(3)^3$ basis.}
\label{E1charges}
\end{table}

\begin{table}[t]
\begin{tabular}{|c|l|c|l|}
 $E1$ & $Q^R$-charges in $SU(3)^3$ basis &co-marks  
      &$(H_1,H_2,H_3)$  \\
  \hline
 & & & \\
 & $a_1=(-\tilde{e}^{1},\tilde{e}^{0},-\tilde{e}^{0})$ 
          & $1$ &\\
 & $a_2=(-\tilde{e}^{0},\tilde{e}^{1},\tilde{e}^{2})$ 
          & $2$ &\\
 & $a_3=(e_2,0,0)$ 
          & $3$ &\\
$i\partial X^{1}$ 
 & $a_4=(-\tilde{e}^{0},\tilde{e}^{0},-\tilde{e}^{1})$ 
          &$2$ &$(-1,0,0)$\\
 & $a_5=(-\tilde{e}^{1},-\tilde{e}^{2},\tilde{e}^{2})$ 
          & $1$ &\\
 & $a_6=(-\tilde{e}^{0},-\tilde{e}^{2},-\tilde{e}^{0})$ 
          & $2$ &\\
 & $a_7=(-\tilde{e}^{1},\tilde{e}^{1},-\tilde{e}^{1})$ 
          & &\\
 & $~~~=-a_1-2a_2-3a_3-2a_4-a_5-2a_6$ 
          & $1$ &\\
 & & & \\ 
 \hline
 & & & \\
 & $l_1=(\tilde{e}^{0},-\tilde{e}^{0},\tilde{e}^{0})=-a_1-a_2-a_3-a_4-a_6$
 &$1$ &\\
 & $l_2=(\tilde{e}^{1},\tilde{e}^{2},\tilde{e}^{1})=a_1+2a_2+2a_3+a_4+a_6$
 & $2$ & \\
 & $l_3=(e_3,0,0)=-a_2-2a_3-a_4-a_6$ 
 & $3$ &\\ 
$i\partial X^{2}$ 
 & $l_4=(\tilde{e}^{1},-\tilde{e}^{1},\tilde{e}^{0})=a_2+2a_3+2a_4+a_5+a_6$
 & $2$ &$(0,-1,0)$ \\
 & $l_5=(\tilde{e}^{0},\tilde{e}^{2},-\tilde{e}^{2})=-a_2-a_3-a_4-a_5-a_6$
 & $1$ & \\
 & $l_6=(\tilde{e}^{1},-\tilde{e}^{0},-\tilde{e}^{2})=-a_1-a_2-a_3-a_4-a_5$
 & $2$ &\\
 & $l_7=(\tilde{e}^{0},-\tilde{e}^{1},\tilde{e}^{1})=
 -l_1-2l_2-3l_3-2l_4-l_5-2l_6$  & &\\
 & $~~~=a_1+a_2+2a_3+a_4+a_5+a_6$ 
 & $1$ &\\
 & & &\\
 \hline
 & & &\\
 & $k_1=(0,e_{1},0)=-a_1-a_2-2a_3-2a_4-a_5-a_6$ 
 &$1$ &\\
 & $k_2=(0,e_{2},0)=a_1+a_2+a_3+a_4$ 
 &$1$ &\\
 & $k_3=(0,e_{3},0)=a_3+a_4+a_5+a_6$ 
 & $1$ &\\
$i\partial X^{3}$ 
 & $k_1^{'}=(0,0,e_1)=a_1+a_2+a_3+a_6$ 
 &$1$  & $(0,0,-1)$ \\
 & $k_{2}^{'}=(0,0,e_2)=a_2+a_3+a_4+a_5$  
 &$1$ &\\
 & $k_{3}^{'}=(0,0,e_3)=-a_1-2a_2-2a_3-a_4-a_5-a_6$ 
 &$1$ & \\
 & & & \\
\end{tabular}
\caption{The terms that enter the expressions for the currents 
$i\partial X^{a}$ for $E1$ model. In the first column 
they are given in terms of their
quantum numbers under the Kac-Moody algebra ${\cal G}_{R}=E_6$ in the
$SU(3)^3$ basis. The weight vectors $\tilde{e}^1$, $\tilde{e}^2$ are 
defined by $\tilde{e}^{i} e_{j} = \delta^{i}_{j}$, and 
$\tilde{e}^{0}=\tilde{e}^{2}-\tilde{e}^{1}$. 
The co-marks of the roots are given in the second column.
The corresponding $H$-charges carried by the supercurrent 
are given in the last column.}
\label{E1super}
\end{table}

\begin{table}[t]
\begin{tabular}{|c|l|l|l|l|l|}
 $E2$ & Field & $SU(2) \otimes E_6 \otimes U(1)^{3}$
 & $Q^R$-charges in $SU(3)^3$ & $(H_1,H_2,H_3)_{-1}$ & $(H_1,H_2,H_3)_{-1/2}$
      \\ \hline
  & & & & & \\
  & $\Phi$ & $ ({\bf 1},{\bf 78})(0,0,0)_L$
  &$(0,0,0)$ 
  & $(0,0,+1)$ 
  & $(+{1\over 2}, +{1\over 2}, +{1\over 2})$ \\    
  $U$ & $U_0$ & $ ({\bf 1},{\bf 1})(0,+6,0)_L$
  &$(0,0,0)$ & $(0,0,+1)$ 
  &$(+{1\over 2}, +{1\over 2}, +{1\over 2})$ \\    
  & $D_{\pm}$ & $({\bf 2}, {\bf 1})(0,0,\pm 3)_L$ 
  &$(0,0,0)$ 
  & $(-1,0,0)$ 
  & $(-{1\over 2}, +{1\over 2}, -{1\over 2})$ \\    
  & ${\tilde U}_{\pm}$ & $({\bf 1},{\bf 1})(\pm 3,+{3},0)_L$
  &$(0,0,0)$ 
  & $(0,-1,0)$ 
  & $(+{1\over 2}, -{1\over 2}, -{1\over 2})$ \\
 & & & & & \\
  \hline
 & & & & & \\
 & $\chi_0$ & $ ({\bf 1},{\bf 27})(0,-{2},0)_L$
 & $ (0,-e_1/3,-e_1/3)$ 
 & $(0, -{2\over 3},+{1\over 3})$ 
 & $(+{1\over 2}, -{1\over 6},-{1\over 6})$\\ 
 $T3$ & ${\tilde \chi}_{+}$ & $({\bf 1},{\overline {\bf 27}}) (+1,-1,0)_L$
 & $(0,e_3/3,e_3/3)$ 
 & $(0, -{1\over 3},+{2\over 3})$ 
 & $(+{1\over 2}, +{1\over 6},+{1\over 6})$ \\
 & ${\tilde \chi}_{-}$ & $({\bf 1},{\overline {\bf 27}}) (-1,-1,0)_L$
 & $(0,e_2/3,e_2/3)$ 
 & $(0, -{1\over 3},+{2\over 3})$ 
 & $(+{1\over 2}, +{1\over 6},+{1\over 6})$ \\
 & & & & & \\ 
 \hline
 & & & & &\\
 $T6$ & ${\chi}_{++}$ &$({\bf  1}, {\bf 27})(0,+1,+1)_L$ 
 & $({\tilde e}^2/2,-e_3/3,-e_1/3)$ 
 & $(-{1\over 2},-{1\over 6},+{1\over 3})$ 
 & $(0,+{1\over 3},-{1\over 6})$\\

 & ${\chi}_{-+}$ &$({\bf  1}, {\bf 27})(0,+1,+1)_L$ 
 & $(-{\tilde e}^2/2,-e_1/3,-e_3/3)$ 
 & $(-{1\over 2},-{1\over 6},+{1\over 3})$ 
 & $(0,+{1\over 3},-{1\over 6})$\\

 & ${\chi}_{+-}$ &$({\bf  1}, {\bf 27})(0,+1,-1)_L$ 
 & $({\tilde e}^2/2,-e_1/3,-e_2/3)$ 
 & $(-{1\over 2},-{1\over 6},+{1\over 3})$ 
 & $(0,+{1\over 3},-{1\over 6})$\\

 & ${\chi}_{--}$ &$({\bf  1}, {\bf 27})(0,+1,-1)_L$ 
 & $(-{\tilde e}^2/2,-e_2/3,-e_1/3)$ 
 & $(-{1\over 2},-{1\over 6},+{1\over 3})$ 
 & $(0,+{1\over 3},-{1\over 6})$\\
  
 & & & & & \\
 \hline
 & & & & & \\ 
  $T2$   & ${U}_{+ \pm}$ &$({\bf 1},{\bf 1})(0,-{3},\pm 3)_L$
  & $(e_1/2,0,0)$ 
  & $(-{1\over 2},-{1\over 2},0)$ & $(0,0,-{1\over 2})$ \\
 
 & $U_{- \pm}$ &$({\bf 1},{\bf 1})(0,-{3},\pm 3)_L$
  & $(-e_1/2,0,0)$ 
  & $(-{1\over 2},-{1\over 2},0)$ & $(0,0,-{1\over 2})$ \\
  & & & & & \\
  \hline
 & & & & & \\
  $U(1)$ & & $(1/ \sqrt{6}, ~1/{3\sqrt{2}},  ~1/\sqrt{6})$
  & & $(1,1,1)$ & $(1,1,1)$
\end{tabular}
\caption{The $G$-, $Q^R$- and $H$-charges (in the $-1$ and $-1/2$ pictures) of 
the massless fields in the $E2$ model. The $U(1)$ normalization radii for 
the $G$- and $H$-charges are given at the bottom of the Table. 
The $Q^R$-charges (in the $-1$ picture which are the same as in the $-1/2$ 
picture) are written in $SU(3)^3$ basis.}
\label{E2charges}
\end{table}

\begin{table}[t]
\begin{tabular}{|c|l|l|c|}
 $E2$ & $Q^R$-charges in $SU(3)^3$ basis &co-marks  
      &$(H_1,H_2,H_3)$  \\
  \hline
 & & & \\
 & $\alpha_1=-(e_2,0,0)=-a_3$ 
 & $1$ & \\
 & $\alpha_2=(-\tilde{e}^1,\tilde{e}^0,-\tilde{e}^0)=a_1$ 
 &$1$  & \\
$i \partial X^1$
 & $\alpha_3=(-\tilde{e}^1,-\tilde{e}^2,\tilde{e}^2)=a_5$ 
 & $1$ & $(-1,0,0)$ \\
 & $\alpha_4=(-\tilde{e}^1,\tilde{e}^1,-\tilde{e}^1)
   =-a_1-2a_2-3a_3-2a_4-a_5-2a_6$ 
 & $1$ &\\
 & & & \\ 
 \hline
 & & &\\
 & $\beta_1=(\tilde{e}^{0},-\tilde{e}^{1},-\tilde{e}^{2})=-a_2$
 & $1$ & \\
 & $\beta_2=(-\tilde{e}^{0},\tilde{e}^{0},\tilde{e}^{2})
   =a_1+2a_2+2a_3+2a_4+a_5+a_6$
 & $1$ & \\
 & $\beta_3=(\tilde{e}^{0},-\tilde{e}^{0},\tilde{e}^{1})=-a_4$ 
 & $1$ & \\ 
$i\partial X^{2}$
 & $\beta_4=(-\tilde{e}^{0},-\tilde{e}^{2},-\tilde{e}^{1})=-a_1-a_2-a_3$
 & $1$ &$(0,-1,0)$ \\
 & $\beta_5=(\tilde{e}^{0},\tilde{e}^{2},\tilde{e}^{0})=-a_6$
 & $1$ & \\
 & $\beta_6=(-\tilde{e}^{0},\tilde{e}^{1},-\tilde{e}^{0})=-a_3-a_4-a_5$
 & $1$ & \\
 & & &\\
 \hline
 & & &\\
 & $k_1=(0,e_{1},0)=-a_1-a_2-2a_3-2a_4-a_5-a_6$ & 
 $1$ &\\
 & $k_2=(0,e_{2},0)=a_1+a_2+a_3+a_4$ & 
 $1$ &\\
 & $k_3=(0,e_{3},0)=a_3+a_4+a_5+a_6$ &
 $1$ &\\
$i\partial X^{3}$ 
 & $k_1^{'}=(0,0,e_1)=a_1+a_2+a_3+a_6$ & 
 $1$ & $(0,0,-1)$ \\
 & $k_{2}^{'}=(0,0,e_2)=a_2+a_3+a_4+a_5$ & 
 $1$ &\\
 & $k_{3}^{'}=(0,0,e_3)=-a_1-2a_2-2a_3-a_4-a_5-a_6$ &
 $1$ &\\
 & & &\\
\end{tabular}
\caption{The terms that enter the expressions for the currents 
$i\partial X^{a}$ for $E2$ model. In the first column 
they are given in terms of their
quantum numbers under the Kac-Moody algebra ${\cal G}_{R}=E_6$ in the
$SU(3)^3$ basis. The weight vectors $\tilde{e}^1$, $\tilde{e}^2$ are 
defined by $\tilde{e}^{i} e_{j} = \delta^{i}_{j}$, and 
$\tilde{e}^{0}=\tilde{e}^{2}-\tilde{e}^{1}$. 
The co-marks of the roots are given in the second column.
The corresponding $H$-charges carried by the supercurrent are
given in the last column.}
\label{E2super}
\end{table}

\begin{table}[t]
\begin{tabular}{|c|l|l|l|}
${\cal G}^{\prime}_R$ &
$SU(2)^3 \otimes U(1)^3$  & 
$SU(2)^4 \otimes U(1)^2$ \hspace{2cm} & $SU(3)^3$ \hspace{2cm}
 \\ \hline
&  &  & \\
& $({\bf 1},{\bf 1},{\bf 1})(+3,0,0)$ 
& $({\bf 1},{\bf 1},{\bf 1},{\bf 2}_+)(0,0,0)$ 
&$(e_1/2,0,0)$  \\
&$({\bf 1},{\bf 1},{\bf 1})(0,+12,0)$   
&$({\bf 1},{\bf 1},{\bf 1},{\bf 1})(+12,0)$ 
&$(0,e_1,e_1)$ \\
&$({\bf 1},{\bf 1},{\bf 1})(0,0,+4)$ 
&$({\bf 1},{\bf 1},{\bf 1},{\bf 1})(0,+4)$ 
&$(0,-\tilde{e}^2,-\tilde{e}^2)$ \\
&$({\bf 1},{\bf 1},{\bf 2}_+)(0,0,0)$ 
&$({\bf 1},{\bf 1},{\bf 2}_+,{\bf 1})(0,0,0)$
&$(\tilde{e}^2/2,\tilde{e}^0/2,-\tilde{e}^0/2)$ \\
&$({\bf 1},{\bf 2}_+,{\bf 1})(0,0,0)$
&$({\bf 1},{\bf 2}_+,{\bf 1},{\bf 1})(0,0,0)$
&$(\tilde{e}^2/2,\tilde{e}^1/2,-\tilde{e}^1/2)$ \\
&$({\bf 2}_+,{\bf 1},{\bf 1})(0,0,0)$ 
&$({\bf 2}_+,{\bf 1},{\bf 1},{\bf 1})(0,0,0)$
&$(\tilde{e}^2/2,-\tilde{e}^2/2,\tilde{e}^2/2)$ \\
& & & \\ \hline 
& & & \\
$U(1)$ &$(1/3 \sqrt{2},1/6,1/2 \sqrt{3})$ & $(1/6,1/2 \sqrt{3})$ & \\ 
\end{tabular}
\caption{The quantum numbers under the Kac-Moody algebra 
${\cal G}^{\prime}_R$
in the $SU(3)^3$ basis. For $E1$ model, 
${\cal G}^{\prime}_R=SU(2)^3 \otimes U(1)^3$. For $E2$ model,
${\cal G}^{\prime}_R=SU(2)^4 \otimes U(1)^2$.
The $U(1)$ normalization radii are given at the bottom of the Table.
Here 
${\bf 2}_+$ and ${\bf 2}_-$ stand for the upper and lower components of an 
$SU(2)$ doublet. In the $SU(2)\supset U(1)$ basis 
we have ${\bf 2}_\pm =
(\pm 1)$.}
\label{convert}
\end{table}

\begin{table}[t]
\begin{tabular}{|c|l|c|l|c|}
Field & $SU(2) \otimes E_6 \otimes U(1)^3$
& Field & $SU(2) \otimes SO(10)\otimes U(1)^4$ 
& $D$-charge \\ \hline
 & & & & \\
$\Phi$ & $ ({\bf 1},{\bf 78})(0,0,0)_L$
& $\Phi$ & $ ({\bf 1},{\bf 45})(0,0,0,0)_L$
& $2$ \\
& & $\phi$ & $ ({\bf 1},{\bf 1})(0,0,0,0)_L$
& $2$ \\
& & $Q$ & $({\bf 1},{\bf 16})(0,0,0,3)_L$
& $1$ \\
& & $\overline{Q}$ & $({\bf 1},{\overline{\bf 16}})(0,0,0,-3)_L$
& $0$ \\ 
 & & & & \\ \hline
 & & & & \\
$\chi_0$ & $ ({\bf 1},{\bf 27})(0,-{2},0)_L$
&$Q_0$ & $ ({\bf 1},{\bf  16})(0,-{2},0,-1)_L$
& $1$ \\
& & $H_0$ & $ ({\bf 1},{\bf  10})(0,-{2},0,+2)_L$
& $0$ \\
& & $S_0$ & $ ({\bf 1},{\bf  1})(0,-{2},0,-4)_L$
& $2$ \\
& & & & \\ \hline
 & & & & \\
$\chi_{+\pm}$ & $({\bf  1}, {\bf 27})(0,+1,\pm1)_L$
& $Q_{+ \pm}$ & $({\bf  1}, {\bf  16})(0,+1,\pm1,-1)_L$
& $2$ \\
& & $H_{+ \pm}$ & $({\bf  1}, {\bf 10})(0,+1,\pm1,+2)_L$
& $1$ \\
& & $S_{+ \pm}$ & $({\bf  1}, {\bf 1})(0,+1,\pm1,-4)_L$
& $0$ \\
 & & & & \\ \hline
 & & & & \\
$\chi_{-\pm}$ & $({\bf  1}, {\bf 27})(0,+1,\pm1)_L$
& $Q_{- \pm}$ & $({\bf  1}, {\bf  16})(0,+1,\pm1,-1)_L$
& $2$ \\
& & $H_{- \pm}$ & $({\bf  1}, {\bf 10})(0,+1,\pm1,+2)_L$
& $1$ \\
& & $S_{- \pm}$ & $({\bf  1}, {\bf 1})(0,+1,\pm1,-4)_L$
& $0$ \\ 
 & & & & \\ \hline
 & & & & \\
$\tilde{\chi}_{\pm}$ & $ ({\bf 1},{\overline {\bf 27}}) (\pm 1,-1,0)_L$ 
& $\tilde{Q}_{\pm}$ & $ ({\bf 1},{\overline {\bf 16}}) (\pm 1,-1,0,+1)_L$
& $1$ \\
& & $\tilde{H}_{\pm}$ &  $ ({\bf 1},{{\bf 10}}) (\pm 1,-1,0,-2)_L$
& $2$ \\
& & $\tilde{S}_{\pm}$ & $ ({\bf 1},{{\bf 1}}) (\pm 1,-1,0,+4)_L$
& $0$ \\
 & & & & \\
\end{tabular}
\caption{The discrete $D$-charges for $E1$ model. 
The second column gives the gauge quantum
numbers of the fields in $E1$ model.
The fourth column gives the gauge quantum numbers of the fields 
in the
branching $E_6 \supset SO(10) \otimes U(1)$. 
The last column gives the discrete $D$-charges. This $D$-charge
comes from a ${\bf Z}_3$ symmetry and must be conserved modulus $3$ in
the scattering amplitude.
Fields in the $E1$ model
that have $D=0$ for the entire $E_6$ multiplet are not shown.}

\label{discrete}
\end{table}

\begin{table}[t]
\begin{tabular}{|c|l|l||l|l|}
 &$T1(1,1)$ & &$T2(1,1)$ & \\
 M & $SU(2) \otimes SO(10)\otimes U(1)^4$ & Field& 
$SU(2) \otimes SO(10) \otimes U(1)^4$
  & Field
      \\ \hline
 & & & & \\
   & $ ({\bf 1},{\bf 45})(0,0,0,0)_L$ & $\Phi$ & 
$ ({\bf 1},{\bf  45})(0,0,0,0)_L$ & $\Phi$\\
   & $ ({\bf 1},{\bf 1})(0,0,0,0)_L$ & $\phi$ & 
$ ({\bf 1},{\bf  1})(0,0,0,0)_L$ & $\phi$\\ 
  $U$ & $2 ({\bf 1},{\bf 1})(0,-{3},\pm 3,0)_L$ & $U_{+\pm},U_{-\pm}$& 
     $ ({\bf 1},{\bf 1})(\pm 3,+{3},0,0)_L$ & ${\tilde U}_\pm$\\
   & $ ({\bf 1},{\bf 1})(0,+6,0,0)_L$ & $U_0$ &
$ ({\bf 1},{\bf 1})(0,+6,0,0)_L$ & $U_0$\\
    & & & $({\bf 2}, {\bf 1})(0,0,\pm 3,0)_L$ &$D_{\pm}$\\
  & & & & \\
  \hline
 & &  & & \\
   &  $ ({\bf 1},{\bf  16})(0,-{2},0,-1)_L$ & $Q_0$ &
$ ({\bf 1},{\bf  16})(0,-{2},0, -1)_L$ &$Q_0$\\
   &   $ ({\bf 1},{\bf  10})(0,-{2},0,+2)_L$ & $H_0$ &  
$ ({\bf 1},{\bf  10})(0,-{2},0,+2)_L$ & $H_0$\\
   &   $ ({\bf 1},{\bf  1})(0,-{2},0,-4)_L$ & $S_0$ & 
$ ({\bf 1},{\bf 1})(0,-{2},0,-4)_L$ &$S_0$ \\
 $T3$  
   & $2({\bf  1}, {\bf  16})(0,+1,\pm1,-1)_L$ & $Q_{+\pm},Q_{-\pm}$
  &  $ ({\bf 1},{\overline {\bf 16}}) (\pm 1,-1,0,+1)_L$ &${\tilde Q}_\pm$\\
   & $2({\bf  1}, {\bf 10})(0,+1,\pm1,+2)_L$ & $H_{+\pm},H_{-\pm}$& 
  $ ({\bf 1},{ {\bf 10}}) (\pm 1,-1,0,-2)_L$ &${\tilde H}_\pm$\\
   & $2({\bf  1}, {\bf 1})(0,+1,\pm1,-4)_L$ & $S_{+\pm},S_{-\pm}$&  
 $ ({\bf 1},{ {\bf 1}}) (\pm 1,-1,0,+4)_L$ &${\tilde S}_\pm$\\
 & & & & \\
 \hline
  & &  & &\\
  $T6$  &  $ ({\bf 1},{\overline {\bf 16}}) (\pm 1,-1,0,+1)_L$ &
${\tilde Q}_\pm$& 2({\bf  1}, 
 ${\bf 16})(0,+1,\pm1,-1)_L$  & $Q_{+\pm},Q_{-\pm}$\\
      &  $ ({\bf 1},{{\bf 10}}) (\pm 1,-1,0,-2)_L$ &${\tilde H}_\pm$
& 2({\bf  1}, 
 ${\bf 10})(0,+1,\pm1,+2)_L$  & $H_{+\pm},H_{-\pm}$\\
     &  $ ({\bf 1},{{\bf 1}}) (\pm 1,-1,0,+4)_L$ &${\tilde S}_\pm$& 
$2({\bf  1}, 
 {\bf 1})(0,+1,\pm1,-4)_L$ & $S_{+\pm},S_{-\pm}$ \\

    & &  & &\\
 \hline
 & &  & & \\
   $T2$ & $({\bf 2},{\bf 1})(0,0,{\pm 3},0)_L$ & $D_\pm$ & 
$2 ({\bf 1},{\bf 1})(0,-{3},\pm 3,0)_L$ 
  & $U_{+\pm},U_{-\pm}$ \\
    &  $ ({\bf 1},{\bf 1})(\pm {3},+{3},0,0)_L$ & ${\tilde U}_\pm$& &\\
  & & & &\\
   \hline
 & &  &&\\
 $U(1)$ & $(1/ \sqrt{6}, ~1/3\sqrt{2}, ~1/\sqrt{6},~1/6)$ & &$(1/ \sqrt{6}, 
  ~1/{3\sqrt{2}},    ~1/\sqrt{6},~1/6)$ &\\
\end{tabular}
\caption{The massless spectra of the two $SO(10)$ models $T1(1,1)$ and 
$T2(1,1)$ both with gauge symmetry $SU(2)_1 \otimes SO(10)_3\otimes U(1)^4$.
The $U(1)$ normalization
radii are given at the bottom of the Table.
The gravity, dilaton and gauge supermultiplets are not shown.}

\label{SO(10)spectra}
\end{table}
 
\begin{table}[t]
\begin{tabular}{|c|l|l|l|l|l|}
 $S1$ & Field & $SU(3)   \otimes SU(6) \otimes U(1)^3$
 & $Q^R$-charges in $SU(3)^3$ & $(H_1,H_2,H_3)_{-1}$ & $(H_1,H_2,H_3)_{-1/2}$
      \\ \hline
  & & & & &\\
  & $\Phi$ & $({\bf 1},{\bf 35})(0,0,0)_L$
  &$(0,0,0)$
  & $(0,0,+1)$ & $(-{1\over 2}, -{1\over 2}, +{1\over 2})$ \\    
  & $\phi$ & $({\bf 1},{\bf 1})(0,0,0)_L$
  &$(0,0,0)$
  & $(0,0,+1)$ & $(-{1\over 2}, -{1\over 2}, +{1\over 2})$ \\  
  & $U_0$ & $ ({\bf 1},{\bf 1}) (+6,0,0)_L$
  &$(0,0,0)$
  & $(0,0,+1)$ & $(-{1\over 2}, -{1\over 2}, +{1\over 2})$ \\
$U$
  & $T_0$  & $({\bf 3},{\bf 1})(0,0,-4)_L$
  &$(0,0,0)$
  & $(0,0,+1)$ & $(-{1\over 2}, -{1\over 2}, +{1\over 2})$ \\    
  & $U_{+ \pm}$ & $({\bf 1},{\bf 1})(-3, {\pm 3},{\pm 3})_L$
  & $(0,0,0)$
  & $(+1,0,0)$ & $(+{1\over 2}, -{1\over 2}, -{1\over 2})$ \\
  & ${\tilde T}_{+}$ & $({\overline {\bf 3}},{\bf 1})(+3,-3,+1)_L$
  &$(0,0,0)$
  & $(+1,0,0)$ & $(+{1\over 2}, -{1\over 2}, -{1\over 2})$ \\
  & $U_{- \pm}$ &$({\bf 1},{\bf 1})(-3, {\pm 3},{\pm 3})_L$
  &$(0,0,0)$
  & $(0,+1,0)$ & $(-{1\over 2}, +{1\over 2}, -{1\over 2})$ \\
  & ${\tilde T}_{-}$ & $({\overline {\bf 3}},{\bf 1})(+3,-3,+1)_L$
  &$(0,0,0)$
  & $(0,+1,0)$ & $(-{1\over 2}, +{1\over 2}, -{1\over 2})$ \\
 & & & & &\\
  \hline
 & & & & &\\
 & ${\tilde S}^1_0$ & $({\bf 1},{\overline {\bf 6}})(+1,0,+2)_L$ 
 & $ -{1\over 3}(e_1,e_1,e_1)$ 
 & $(+{1\over 3}, +{1\over 3},+{1\over 3})$ 
 & $(-{1\over 6}, -{1\over 6},-{1\over 6})$\\ 
 & ${\tilde S}^2_0$ & $({\bf 1},{\overline {\bf 6}})(+1,0,+2)_L$ 
 & $ -{1\over 3}(e_1,e_2,e_2)$ 
 & $(+{1\over 3}, +{1\over 3},+{1\over 3})$ 
 & $(-{1\over 6}, -{1\over 6},-{1\over 6})$\\ 
 & ${\tilde S}^3_0$ & $({\bf 1},{\overline {\bf 6}})(+1,0,+2)_L$ 
 & $ -{1\over 3}(e_1,e_3,e_3)$ 
 & $(+{1\over 3}, +{1\over 3},+{1\over 3})$ 
 & $(-{1\over 6}, -{1\over 6},-{1\over 6})$\\ 
 & $F^1_\pm$ & $({\bf 1},{\bf 15})(+1,-1,-{1})_L$
 & $-{1\over 3}(e_3,e_2,e_3),-{1\over 3}(e_2,e_3,e_2)$
 & $(+{1\over 3}, +{1\over 3},+{1\over 3})$ 
 & $(-{1\over 6}, -{1\over 6},-{1\over 6})$\\ 
$T3$
 & ${\tilde S}^1_\pm$ & $({\bf 1},{\overline {\bf 6}})(-2,+1,-{1})_L$
 & $-{1\over 3}(e_3,e_2,e_3),-{1\over 3}(e_2,e_3,e_2)$
 & $(+{1\over 3}, +{1\over 3},+{1\over 3})$ 
 & $(-{1\over 6}, -{1\over 6},-{1\over 6})$\\ 
 & $F^2_\pm$ & $({\bf 1},{\bf 15})(+1,-1,-{1})_L$
 & $-{1\over 3}(e_3,e_3,e_1),-{1\over 3}(e_2,e_1,e_3)$
 & $(+{1\over 3}, +{1\over 3},+{1\over 3})$ 
 & $(-{1\over 6}, -{1\over 6},-{1\over 6})$\\ 
 & ${\tilde S}^2_\pm$ & $({\bf 1},{\overline {\bf 6}})(-2,+1,-{1})_L$
 & $-{1\over 3}(e_3,e_3,e_1),-{1\over 3}(e_2,e_1,e_3)$
 & $(+{1\over 3}, +{1\over 3},+{1\over 3})$ 
 & $(-{1\over 6}, -{1\over 6},-{1\over 6})$\\
 & $F^3_\pm$ & $({\bf 1},{\bf 15})(+1,-1,-{1})_L$
 & $-{1\over 3}(e_3,e_1,e_2),-{1\over 3}(e_2,e_2,e_1)$
 & $(+{1\over 3}, +{1\over 3},+{1\over 3})$ 
 & $(-{1\over 6}, -{1\over 6},-{1\over 6})$\\
 & ${\tilde S}^3_\pm$ & $({\bf 1},{\overline {\bf 6}})(-2,+1,-{1})_L$
 & $-{1\over 3}(e_3,e_1,e_2),-{1\over 3}(e_2,e_2,e_1)$
 & $(+{1\over 3}, +{1\over 3},+{1\over 3})$ 
 & $(-{1\over 6}, -{1\over 6},-{1\over 6})$\\
  & & & & &\\ 
 \hline
 & & & & &\\
 & $S^1$ & $({\bf 1},{\bf 6})(+2,+1,+1)_L$
 & $(-e_1/6,e_1/3,e_1/3)$
 &  $(+{1\over 6},+{1\over 6},+{2\over 3})$
 & $(-{1\over 3},-{1\over 3},+{1\over 6})$\\ 
 & ${\tilde F}^1$ & $({\bf 1},{\overline {\bf 15}}) ( -1,-{1},+{1})_L$
 & $(-e_1/6,e_1/3,e_1/3)$
 &  $(+{1\over 6},+{1\over 6},+{2\over 3})$
 & $(-{1\over 3},-{1\over 3},+{1\over 6})$\\ 
 & $S^2$ & $({\bf 1},{\bf 6})(+2,+1,+1)_L$
 & $(-e_1/6,e_2/3,e_2/3)$
 &  $(+{1\over 6},+{1\over 6},+{2\over 3})$
 & $(-{1\over 3},-{1\over 3},+{1\over 6})$\\ 
$T6$
 & ${\tilde F}^2$ & $({\bf 1},{\overline {\bf 15}}) ( -1,-{1},+{1})_L$
 & $(-e_1/6,e_2/3,e_2/3)$
 &  $(+{1\over 6},+{1\over 6},+{2\over 3})$
 & $(-{1\over 3},-{1\over 3},+{1\over 6})$\\ 
 & $S^3$ & $({\bf 1},{\bf 6})(+2,+1,+1)_L$
 & $(-e_1/6,e_3/3,e_3/3)$
 & $(+{1\over 6},+{1\over 6},+{2\over 3})$
 & $(-{1\over 3},-{1\over 3},+{1\over 6})$\\ 
 & ${\tilde F}^3$ & $({\bf 1},{\overline {\bf 15}}) ( -1,-{1},+{1})_L$
 & $(-e_1/6,e_3/3,e_3/3)$
 & $(+{1\over 6},+{1\over 6},+{2\over 3})$
 & $(-{1\over 3},-{1\over 3},+{1\over 6})$\\ 
 & & & & &\\
 \hline
 & & & & &\\ 
 & $T_{+}$ & $({\bf 3},{\bf 1})(+3,-3,-1)_L$
 & $(e_1/2,0,0)$
 & $(+{1\over 2},+{1\over 2},0)$
 & $(0,0,-{1\over 2})$ \\
$T2$
 & ${\tilde T}_0$ & $({\overline {\bf 3}},{\bf 1})(-3,+3,+1)_L$
 & $(e_1/2,0,0)$
 & $(+{1\over 2},+{1\over 2},0)$
 & $(0,0,-{1\over 2})$ \\
 & $T_{-}$ & $({\bf 3},{\bf 1})(-3,-3,-1)_L$
 & $(-e_1/2,0,0)$ 
 & $(+{1\over 2},+{1\over 2},0)$ 
 & $(0,0,-{1\over 2})$ \\
 & $\tilde{U}_{\pm}$ &$ ({\bf 1},{\bf 1})(+3,\pm 3,\mp 3)_L$
 & $(-e_1/2,0,0)$ 
 & $(+{1\over 2},+{1\over 2},0)$ 
 & $(0,0,-{1\over 2})$ \\
  & & & & &\\
  \hline
 & & & & &\\
  $U(1)$ &
  & $({1\over{3\sqrt{2}}},~{1\over{2\sqrt{3}}},~{1\over {2\sqrt{3}}})$ 
  & & $(1,1,1)$ & $(1,1,1)$
\end{tabular}
\caption{The $G$-, $Q^R$- and $H$-charges (in the $-1$ and $-1/2$ pictures) for
the massless fields of the $S1$ model. The $U(1)$ normalization 
radii for the $G$- and $H$-charges are given at the bottom of the 
Table.} 
\label{S1charges}
\end{table}

\begin{table}[t]
\begin{tabular}{|c|l|l|l|l|l|}
 $S2$ & Field & $SU(2)^{2} \otimes SU(6) \otimes U(1)^3$
 & $Q^R$-charges in $SU(3)^3$ & $(H_1,H_2,H_3)_{-1}$ & $(H_1,H_2,H_3)_{-1/2}$
      \\ \hline
  & & & & &\\
  & $\Phi$ & $ ({\bf 1},{\bf 1},{\bf 35})(0,0,0)_L$
  &$(0,0,0)$
  & $(0,0,+1)$ & $(-{1\over 2}, -{1\over 2}, +{1\over 2})$ \\    
  & $\phi$ & $ ({\bf 1},{\bf 1},{\bf 1})(0,0,0)_L$
  &$(0,0,0)$
  & $(0,0,+1)$ & $(-{1\over 2}, -{1\over 2}, +{1\over 2})$ \\  
$U$
  & $U_0$ & $ ({\bf 1},{\bf 1},{\bf 1}) (0,0,-6)_L$
  &$(0,0,0)$
  & $(0,0,+1)$ & $(-{1\over 2}, -{1\over 2}, +{1\over 2})$ \\
  & $D_{\pm}$ & $  ({\bf 2},{\bf 1}, {\bf 1})({\pm 2},0,+3)_L$
  &$(0,0,0)$
  & $(0,0,+1)$ & $(-{1\over 2}, -{1\over 2}, +{1\over 2})$ \\
  & $d_{+ \pm}$ & $({\bf 1},{\bf 2},{\bf 1})({\pm 1},{\mp 3},+3)_L$
  &$(0,0,0)$
  & $(+1,0,0)$ & $(+{1\over 2}, -{1\over 2}, -{1\over 2})$ \\
  & $d_{- \pm}$ & $({\bf 1},{\bf 2},{\bf 1})({\pm 1},{\mp 3},+3)_L$
  & $(0,0,0)$
  & $(0,+1,0)$ & $(-{1\over 2}, +{1\over 2}, -{1\over 2})$ \\
 & & & & &\\
  \hline
 & & & & &\\
 & ${F}^1$ & $({\bf 1},{\bf 1},{\bf 15})(0,0,+{2})_L$
 & $-(e_1/3,e_1/3,e_1/3)$ 
 & $(+{1\over 3}, +{1\over 3},+{1\over 3})$ 
 & $(-{1\over 6}, -{1\over 6},-{1\over 6})$\\ 
 & ${F}^2$ & $({\bf 1},{\bf 1},{\bf 15})(0,0,+{2})_L$
 & $-(e_1/3,e_2/3,e_2/3)$ 
 & $(+{1\over 3}, +{1\over 3},+{1\over 3})$ 
 & $(-{1\over 6}, -{1\over 6},-{1\over 6})$\\ 
 & ${F}^3$ & $({\bf 1},{\bf 1},{\bf 15})(0,0,+{2})_L$
 & $-(e_1/3,e_3/3,e_3/3)$ 
 & $(+{1\over 3}, +{1\over 3},+{1\over 3})$ 
 & $(-{1\over 6}, -{1\over 6},-{1\over 6})$\\ 
 & $\tilde{S}^1_{+ \pm}$  
 & $({\bf 1},{\bf 1},{\overline {\bf 6}})(\pm 1,\mp {1},-1)_L$
 & $-(e_3/3,e_2/3,e_3/3)$ 
 & $(+{1\over 3}, +{1\over 3},+{1\over 3})$ 
 & $(-{1\over 6}, -{1\over 6},-{1\over 6})$\\ 
$T3$
 & $\tilde{S}^1_{- \pm}$ 
 & $({\bf 1},{\bf 1},{\overline {\bf 6}})(\pm 1,\mp {1},-1)_L$
 & $-(e_2/3,e_3/3,e_2/3)$ 
 & $(+{1\over 3}, +{1\over 3},+{1\over 3})$ 
 & $(-{1\over 6}, -{1\over 6},-{1\over 6})$\\ 
 & $\tilde{S}^2_{+ \pm}$
 & $({\bf 1},{\bf 1},{\overline {\bf 6}})(\pm 1,\mp {1},-1)_L$
 & $-(e_3/3,e_3/3,e_1/3)$
 & $(+{1\over 3}, +{1\over 3},+{1\over 3})$ 
 & $(-{1\over 6}, -{1\over 6},-{1\over 6})$\\ 
 & $\tilde{S}^2_{- \pm}$ 
 & $({\bf 1},{\bf 1},{\overline {\bf 6}})(\pm 1,\mp {1},-1)_L$
 & $-(e_2/3,e_1/3,e_3/3)$ 
 & $(+{1\over 3}, +{1\over 3},+{1\over 3})$ 
 & $(-{1\over 6}, -{1\over 6},-{1\over 6})$\\ 
 & $\tilde{S}^3_{+ \pm}$
 & $({\bf 1},{\bf 1},{\overline {\bf 6}})(\pm 1,\mp {1},-1)_L$
 & $-(e_3/3,e_1/3,e_2/3)$ 
 & $(+{1\over 3}, +{1\over 3},+{1\over 3})$ 
 & $(-{1\over 6}, -{1\over 6},-{1\over 6})$\\
 & $\tilde{S}^3_{- \pm}$ 
 & $({\bf 1},{\bf 1},{\overline {\bf 6}})(\pm 1,\mp {1},-1)_L$
 & $-(e_2/3,e_2/3,e_1/3)$ 
 & $(+{1\over 3}, +{1\over 3},+{1\over 3})$ 
 & $(-{1\over 6}, -{1\over 6},-{1\over 6})$\\
   & & & & &\\ 
 \hline
 & & & & &\\
 & $S^1_{\pm}$ & $({\bf 1},{\bf 1}, {\bf 6})(\pm 1,\pm {1},+1)_L$
 & $(-e_1/6,e_1/3,e_1/3)$
 &  $(+{1\over 6},+{1\over 6},+{2\over 3})$
 & $(-{1\over 3},-{1\over 3},+{1\over 6})$\\ 
$T6$
 & $S^2_{\pm}$ & $({\bf 1},{\bf 1}, {\bf 6})(\pm 1,\pm {1},+1)_L$
 & $(-e_1/6,e_2/3,e_2/3)$
 &  $(+{1\over 6},+{1\over 6},+{2\over 3})$
 & $(-{1\over 3},-{1\over 3},+{1\over 6})$\\ 
 & $S^3_{\pm}$ & $({\bf 1},{\bf 1}, {\bf 6})(\pm 1,\pm {1},+1)_L$
 & $(-e_1/6,e_3/3,e_3/3)$ &  
 $(+{1\over 6},+{1\over 6},+{2\over 3})$
 & $(-{1\over 3},-{1\over 3},+{1\over 6})$\\ 
 & & & & &\\
 \hline
 & & & & &\\ 
$T2$
 & $\Delta_{\pm}$ & $({\bf 2},{\bf 2},{\bf 1})(\pm 1,\mp 3,0)_L$
 & $(e_1/2,0,0)$
 & $(+{1\over 2},+{1\over 2},0)$
 & $(0,0,-{1\over 2})$ \\
 & ${\tilde d}_{\pm}$ & $({\bf 1},{\bf 2},{\bf 1})(\pm 1,\pm 3, -3)_L$
 & $(-e_1/2,0,0)$ 
 & $(+{1\over 2},+{1\over 2},0)$ 
 & $(0,0,-{1\over 2})$ \\
  & & & & &\\
  \hline
 & & & & &\\
  $U(1)$ &
  & $({1\over 2},~{1\over {2\sqrt{3}}},~{1\over {3\sqrt{2}}})$
  & & $(1,1,1)$ & $(1,1,1)$
\end{tabular}
\caption{The $G$-, $Q^R$- and $H$-charges (in the $-1$ and $-1/2$ pictures) for
the massless fields of the $S2$ model. The $U(1)$ normalization 
radii for the $G$- and $H$-charges are given at the bottom of the 
Table.}
\label{S2charges}
\end{table}

\begin{table}[t]
\begin{tabular}{|c|l|l|l|l|l|}
$S3$ & Field & $SU(3)   \otimes SU(6) \otimes U(1)^3$
 & $Q^R$-charges in $SU(3)^3$ & $(H_1,H_2,H_3)_{-1}$ & 
$(H_1,H_2,H_3)_{-1/2}$
      \\ \hline
  & & & & &\\
  & $\Phi$ & $({\bf 1},{\bf 35})(0,0,0)_L$
  & $(0,0,0)$ 
  & $(0,0,+1)$ & $(+{1\over 2}, +{1\over 2}, +{1\over 2})$ \\ 
  & $\phi$ & $({\bf 1},{\bf 1})(0,0,0)_L$
  & $(0,0,0)$ 
  & $(0,0,+1)$ & $(+{1\over 2}, +{1\over 2}, +{1\over 2})$ \\ 
  & $U_0$ & $ ({\bf 1},{\bf 1}) (+6,0,0)_L$
  & $(0,0,0)$ 
  & $(0,0,+1)$ & $(+{1\over 2}, +{1\over 2}, +{1\over 2})$ \\ 
$U$
  & $T_0$ & $({\bf 3},{\bf 1})(0,0,-4)_L$
  & $(0,0,0)$ 
  & $(0,0,+1)$ & $(+{1\over 2}, +{1\over 2}, +{1\over 2})$ \\ 
  & $T_+$ & $({\bf 3},{\bf 1})(+3,-3,-1)_L$
  & $(0,0,0)$ 
  & $(-1,0,0)$ & $(-{1\over 2}, +{1\over 2}, -{1\over 2})$ \\  
  & ${\tilde T}_0$ & $({\overline {\bf 3}},{\bf 1})(-3,+3,+1)_L$
  & $(0,0,0)$ 
  & $(-1,0,0)$ & $(-{1\over 2}, +{1\over 2}, -{1\over 2})$ \\  
  & $T_-$ & $({\bf 3},{\bf 1})(-3,-3,-1)_L$
  &$(0,0,0)$ 
  & $(0,-1,0)$ & $(+{1\over 2}, -{1\over 2}, -{1\over 2})$ \\
  & ${\tilde U}_{\pm}$ & $ ({\bf 1},{\bf 1})(+3,\pm 3,\mp 3)_L$
  &$(0,0,0)$ 
  & $(0,-1,0)$ & $(+{1\over 2}, -{1\over 2}, -{1\over 2})$ \\
 & & & & &\\
  \hline
 & & & & &\\
 & ${\tilde S}^1_0$ &$({\bf 1},{\overline {\bf 6}})(+1,0,+2)_L$
 & $ (0,-e_1/3,-e_1/3)$ 
 & $(0, -{2\over 3},+{1\over 3})$ & $(+{1\over 2}, -{1\over 6},-{1\over 6})$\\ 
 & ${\tilde S}^2_0$ &$({\bf 1},{\overline {\bf 6}})(+1,0,+2)_L$
 & $ (0,-e_2/3,-e_2/3)$ 
 & $(0, -{2\over 3},+{1\over 3})$ & $(+{1\over 2}, -{1\over 6},-{1\over 6})$\\ 
 & ${\tilde S}^3_0$ &$({\bf 1},{\overline {\bf 6}})(+1,0,+2)_L$
 & $ (0,-e_3/3,-e_3/3)$ 
 & $(0, -{2\over 3},+{1\over 3})$ & $(+{1\over 2}, -{1\over 6},-{1\over 6})$\\
 & ${\tilde F}^1$ &$({\bf 1},{\overline {\bf 15}}) ( -1,-{1},+{1})_L$
 & $ (0,e_1/3,e_1/3)$ 
 & $(0, -{1\over 3},+{2\over 3})$ & $(+{1\over 2}, +{1\over 6},+{1\over 6})$\\ 
$T3$
 & $S^1$ &$({\bf 1},{\bf 6})(+2,+1,+1)_L$
 &$ (0,e_1/3,e_1/3)$ 
 & $(0, -{1\over 3},+{2\over 3})$ & $(+{1\over 2}, +{1\over 6},+{1\over 6})$\\
 & ${\tilde F}^2$ &$({\bf 1},{\overline {\bf 15}}) ( -1,-{1},+{1})_L$
 & $ (0,e_2/3,e_2/3)$ 
 & $(0, -{1\over 3},+{2\over 3})$ 
 & $(+{1\over 2}, +{1\over 6},+{1\over 6})$\\ 
 & $S^2$ &$({\bf 1},{\bf 6})(+2,+1,+1)_L$
 & $ (0,e_2/3,e_2/3)$ 
 & $(0, -{1\over 3},+{2\over 3})$ 
 & $(+{1\over 2}, +{1\over 6},+{1\over 6})$\\ 
 & ${\tilde F}^3$ &$({\bf 1},{\overline {\bf 15}}) ( -1,-{1},+{1})_L$
 & $ (0,e_3/3,e_3/3)$ 
 & $(0, -{1\over 3},+{2\over 3})$ 
 & $(+{1\over 2}, +{1\over 6},+{1\over 6})$\\  
 & $S^3$ &$({\bf 1},{\bf 6})(+2,+1,+1)_L$
 & $ (0,e_3/3,e_3/3)$ 
 & $(0, -{1\over 3},+{2\over 3})$ 
 & $(+{1\over 2}, +{1\over 6},+{1\over 6})$\\  
 & & & & &\\ 
 \hline
 & & & & &\\
 & ${\tilde S}^1_{\pm}$ & $({\bf 1},{\overline {\bf 6}})(-2,+1,-{1})_L$
 & $({\tilde{e}^2 \over 2},-{e_2 \over 3},-{e_3 \over 3}),
   -({\tilde{e}^2 \over 2},{e_3 \over 3},{e_2 \over 3})$
 & $(-{1\over 2},-{1\over 6},+{1\over 3})$ & $(0,+{1\over 3},-{1\over 6})$\\
 & $F^1_{\pm}$ &$({\bf 1},{\bf 15})(+1,-1,-{1})_L$
 & $({\tilde{e}^2 \over 2},-{e_2 \over 3},-{e_3 \over 3}),
   -({\tilde{e}^2 \over 2},{e_3 \over 3},{e_2 \over 3})$
 & $(-{1\over 2},-{1\over 6},+{1\over 3})$ & $(0,+{1\over 3},-{1\over 6})$\\
$T6$
 & ${\tilde S}^2_{\pm}$ & $({\bf 1},{\overline {\bf 6}})(-2,+1,-{1})_L$
 & $({\tilde{e}^2 \over 2},-{e_3 \over 3},-{e_1 \over 3}),
   -({\tilde{e}^2 \over 2},{e_1 \over 3},{e_3 \over 3})$
 & $(-{1\over 2},-{1\over 6},+{1\over 3})$ & $(0,+{1\over 3},-{1\over 6})$\\
 & $F^2_{\pm}$ &$({\bf 1},{\bf 15})(+1,-1,-{1})_L$
 & $({\tilde{e}^2 \over 2},-{e_3 \over 3},-{e_1 \over 3}),
   -({\tilde{e}^2 \over 2},{e_1 \over 3},{e_3 \over 3})$
 & $(-{1\over 2},-{1\over 6},+{1\over 3})$ & $(0,+{1\over 3},-{1\over 6})$\\
 & ${\tilde S}^3_{\pm}$ & $({\bf 1},{\overline {\bf 6}})(-2,+1,-{1})_L$
 & $({\tilde{e}^2 \over 2},-{e_1 \over 3},-{e_2 \over 3}),
   -({\tilde{e}^2 \over 2},{e_2 \over 3},{e_1 \over 3})$ 
 & $(-{1\over 2},-{1\over 6},+{1\over 3})$ & $(0,+{1\over 3},-{1\over 6})$\\
 & $F^3_{\pm}$ &$({\bf 1},{\bf 15})(+1,-1,-{1})_L$
 & $({\tilde{e}^2 \over 2},-{e_1 \over 3},-{e_2 \over 3}),
   -({\tilde{e}^2 \over 2},{e_2 \over 3},{e_1 \over 3})$ 
 & $(-{1\over 2},-{1\over 6},+{1\over 3})$ & $(0,+{1\over 3},-{1\over 6})$\\
 & & & & &\\
 \hline
 & & & & &\\ 
 & ${U}_{\pm+}$ &$({\bf 1},{\bf 1})(-3, +{3},+{3})_L$
 & $(e_1/2,0,0),(-e_1/2,0,0)$ 
 & $(-{1\over 2},-{1\over 2},0)$ & $(0,0,-{1\over 2})$ \\
$T2$
 & $U_{\pm -}$ &$({\bf 1},{\bf 1})(-3, -{3},-{3})_L$
 & $(e_1/2,0,0),(-e_1/2,0,0)$
 & $(-{1\over 2},-{1\over 2},0)$ & $(0,0,-{1\over 2})$ \\
 & ${\tilde T}_\pm$ & $({\overline {\bf 3}},{\bf 1})(+3,-3,+1)_L$
 & $(e_1/2,0,0),(-e_1/2,0,0)$ 
 & $(-{1\over 2},-{1\over 2},0)$ & $(0,0,-{1\over 2})$ \\
  & & & & &\\
  \hline
 & & & & &\\
  $U(1)$ & 
& $({1\over{3\sqrt{2}}},~{1\over{2\sqrt{3}}},~{1\over {2\sqrt{3}}})$
& & $(1,1,1)$ & $(1,1,1)$
\end{tabular}
\caption{The $G$-, $Q^R$- and $H$-charges (in the $-1$ and $-1/2$ pictures) 
for 
the massless fields of the $S3$ model. The $U(1)$ normalization radii for 
the $G$- and $H$-charges are given at the bottom of the Table.} 
\label{S3charges}
\end{table}

\begin{table}[t]
\begin{tabular}{|c|l|l|l|l|l|}
$S4$ & Field & $SU(2)^2 \otimes SU(6) \otimes U(1)^3$
     & $Q^R$-charges in $SU(3)^3$ & $(H_1,H_2,H_3)_{-1}$ & 
$(H_1,H_2,H_3)_{-1/2}$
      \\ \hline
  & & & & &\\
  & $\Phi$ & $ ({\bf 1},{\bf 1},{\bf 35})(0,0,0)_L$
  & $(0,0,0)$ 
  & $(0,0,+1)$ & $(+{1\over 2}, +{1\over 2}, +{1\over 2})$ \\ 
  & $\phi$ & $ ({\bf 1},{\bf 1},{\bf 1})(0,0,0)_L$
  & $(0,0,0)$ 
  & $(0,0,+1)$ & $(+{1\over 2}, +{1\over 2}, +{1\over 2})$ \\ 
  & $U_0$ & $ ({\bf 1},{\bf 1},{\bf 1})(0,0,-6)_L$
  & $(0,0,0)$ 
  & $(0,0,+1)$ & $(+{1\over 2}, +{1\over 2}, +{1\over 2})$ \\ 
$U$
  & $D_{\pm}$ &$  ({\bf 2},{\bf 1}, {\bf 1})({\pm 2},0,+3)_L$
  & $(0,0,0)$ 
  & $(0,0,+1)$ & $(+{1\over 2}, +{1\over 2}, +{1\over 2})$ \\ 
  & $\Delta_{\pm}$ & $({\bf 2},{\bf 2},{\bf 1})(\pm 1,\mp 3,0)_L$
  & $(0,0,0)$ 
  & $(-1,0,0)$ & $(-{1\over 2}, +{1\over 2}, -{1\over 2})$ \\  
  & $\tilde{d}_{\pm}$ &$({\bf 1},{\bf 2},{\bf 1})(\pm 1,\pm 3, -3)_L$
  &$(0,0,0)$ 
  & $(0,-1,0)$ & $(+{1\over 2}, -{1\over 2}, -{1\over 2})$ \\
 & & & & &\\
  \hline
 & & & & &\\
 & ${F}^1$ & $({\bf 1},{\bf 1},{\bf 15})(0,0,+2)_L$
 & $ (0,-e_1/3,-e_1/3)$ 
 & $(0, -{2\over 3},+{1\over 3})$ & $(+{1\over 2}, -{1\over 6},-{1\over 6})$\\ 
 & ${F}^2$ & $({\bf 1},{\bf 1},{\bf 15})(0,0,+2)_L$
 & $ (0,-e_2/3,-e_2/3)$ 
 & $(0, -{2\over 3},+{1\over 3})$ & $(+{1\over 2}, -{1\over 6},-{1\over 6})$\\ 
$T3$
 & ${F}^3$ & $({\bf 1},{\bf 1},{\bf 15})(0,0,+2)_L$
 & $ (0,-e_3/3,-e_3/3)$ 
 & $(0, -{2\over 3},+{1\over 3})$ & $(+{1\over 2}, -{1\over 6},-{1\over 6})$\\
 & ${S}^1_{\pm}$ &$({\bf 1},{\bf 1}, {\bf 6})(\pm 1,\pm {1},+1)_L$
 & $ (0,e_1/3,e_1/3)$ 
 & $(0, -{1\over 3},+{2\over 3})$ & $(+{1\over 2}, +{1\over 6},+{1\over 6})$\\ 
 & ${S}^2_{\pm}$ &$({\bf 1},{\bf 1}, {\bf 6})(\pm 1,\pm {1},+1)_L$
 & $ (0,e_2/3,e_2/3)$ 
 & $(0, -{1\over 3},+{2\over 3})$ 
 & $(+{1\over 2}, +{1\over 6},+{1\over 6})$\\ 
 & ${S}^3_{\pm}$ &$({\bf 1},{\bf 1}, {\bf 6})(\pm 1,\pm {1},+1)_L$
 & $ (0,e_3/3,e_3/3)$ 
 & $(0, -{1\over 3},+{2\over 3})$ 
 & $(+{1\over 2}, +{1\over 6},+{1\over 6})$\\  
 & & & & &\\ 
 \hline
 & & & & &\\
 & ${\tilde S}^{1}_{+ \pm}$
 &$({\bf 1},{\bf 1},{\overline {\bf 6}})(\pm 1,\mp {1},-1)_L$
 & $(\tilde{e}^2/2,-e_2/3,-e_3/3)$
 & $(-{1\over 2},-{1\over 6},+{1\over 3})$ & $(0,+{1\over 3},-{1\over 6})$\\
 & ${\tilde S}^{1}_{- \pm}$ 
 &$({\bf 1},{\bf 1},{\overline {\bf 6}})(\pm 1,\mp {1},-1)_L$
 & $(-\tilde{e}^2/2,-e_3/3,-e_2/3)$
 & $(-{1\over 2},-{1\over 6},+{1\over 3})$ & $(0,+{1\over 3},-{1\over 6})$\\
$T6$  
 & ${\tilde S}^2_{+ \pm}$ 
 &$({\bf 1},{\bf 1},{\overline {\bf 6}})(\pm 1,\mp {1},-1)_L$
 & $(\tilde{e}^2/2,-e_3/3,-e_1/3)$
 & $(-{1\over 2},-{1\over 6},+{1\over 3})$ & $(0,+{1\over 3},-{1\over 6})$\\
 & ${\tilde S}^{1}_{- \pm}$ 
 &$({\bf 1},{\bf 1},{\overline {\bf 6}})(\pm 1,\mp {1},-1)_L$
 & $(-\tilde{e}^2/2,-e_1/3,-e_3/3)$
 & $(-{1\over 2},-{1\over 6},+{1\over 3})$ & $(0,+{1\over 3},-{1\over 6})$\\
 & ${\tilde S}^3_{+ \pm}$
 &$({\bf 1},{\bf 1},{\overline {\bf 6}})(\pm 1,\mp {1},-1)_L$
 & $(\tilde{e}^2/2,-e_1/3,-e_2/3)$
 & $(-{1\over 2},-{1\over 6},+{1\over 3})$ & $(0,+{1\over 3},-{1\over 6})$\\
 & ${\tilde S}^{1}_{- \pm}$ 
 &$({\bf 1},{\bf 1},{\overline {\bf 6}})(\pm 1,\mp {1},-1)_L$
 & $(-\tilde{e}^2/2,-e_2/3,-e_1/3)$ 
 & $(-{1\over 2},-{1\over 6},+{1\over 3})$ & $(0,+{1\over 3},-{1\over 6})$\\
 & & & & &\\
 \hline
 & & & & &\\ 
$T2$
 & ${d}_{+ \pm}$ 
 &$({\bf 1},{\bf 2},{\bf 1})({\pm 1},{\mp 3},+3)_L$
 & $(e_1/2,0,0)$ 
 & $(-{1\over 2},-{1\over 2},0)$ & $(0,0,-{1\over 2})$ \\
 & ${d}_{- \pm}$ 
 &$({\bf 1},{\bf 2},{\bf 1})({\pm 1},{\mp 3},+3)_L$
 & $(-e_1/2,0,0)$ 
 & $(-{1\over 2},-{1\over 2},0)$ & $(0,0,-{1\over 2})$ \\
  & & & & &\\
  \hline
 & & & & &\\
  $U(1)$ & 
& $({1\over 2},~{1\over {2\sqrt{3}}},~{1\over {3\sqrt{2}}})$
& & $(1,1,1)$ & $(1,1,1)$
\end{tabular}
\caption{The $G$-, $Q^R$- and $H$-charges (in the $-1$ and $-1/2$ pictures) 
for 
the massless fields of the $S4$ model. The $U(1)$ normalization radii for 
the $G$- and $H$-charges are given at the bottom of the Table.} 
\label{S4charges}
\end{table}


\begin{references}

\bibitem{DHVW}L. Dixon, J. Harvey, C. Vafa and E. Witten, Nucl. Phys.
{\bf B261} (1985) 678; {\bf B274} (1986) 285; \\
L.E. Ib\'a\~nez, H.P. Nilles and F. Quevedo, Phys. Lett. {\bf 187B} (1987) 25.

\bibitem{NSV} K.S. Narain, M.H. Sarmadi and C. Vafa, Nucl. Phys. {\bf B288} 
(1987) 551.

\bibitem{kt} Z. Kakushadze and S.-H.H. Tye, Phys. Rev. Lett. {\bf 77} 
(1996) 2612, hep-th/9605221; Phys. Rev. {\bf D54} (1996) 7520, hep-th/9607138;
Phys. Lett. {\bf B392} (1996) 335, hep-th/9609027; Phys. Rev. {\bf D55}
(1997) 7878, hep-th/9610106; 
Phys. Rev. {\bf D55} (1997) 7896, hep-th/9701057.

\bibitem{DFMS} L. Dixon, D. Friedan, E. Martinec and S. Shenker, Nucl. Phys.
{\bf B282} (1987) 13; \\
S. Hamidi and C. Vafa, Nucl. Phys. {\bf B279} (1987) 465.

\bibitem{LMN} See, {\em e.g.},\\
M. Cvetic, Phys. Rev. Lett. {\bf 59} (1987) 1795; \\
M. Cvetic, J. Molera and B.A. Ovrut, Phys. Rev. {\bf D40} (1989) 1140; \\
A. Font, L.E. Ibanez, H.P. Nilles and F. Quevedo,
Phys. Lett. {\bf 210B} (1988) 101; Nucl. Phys. {\bf 307} (1988) 109; \\
A. Font, L.E. Ibanez, F. Quevedo and A. Sierra,
Nucl. Phys. {\bf B331} (1990) 421; \\ 
A. Font, L.E. Ibanez and F. Quevedo, Nucl. Phys. {\bf B345} (1990) 389; \\
J.A. Casas and C. Munoz, Phys. Lett. {\bf 212B} (1988) 343; \\
V. Kaplunovsky, Nucl. Phys. {\bf B307} (1988) 145; {\bf 382} 
(1992) 436; \\
L. Dixon, V. Kaplunovsky and J. Louis,
Nucl. Phys. {\bf B329} (1990) 27; {\bf B355} (1991) 649; \\
J. Lauer, J. Mas and H.P. Nilles, Nucl. Phys. {\bf B351} (1991) 353; \\
S. Stieberger , D. Jungnickel, J. Lauer and M. Spalinski, Mod. Phys. Lett. 
{\bf A7} (1992) 3059;\\
J. Erler, D. Jungnickel, M. Spalinski and S. Stieberger, Nucl. Phys. B397 
(1993) 379; \\
J. Erler and M. Spalinski, Int. J. Mod. Phys. {\bf A9} (1994) 4407; \\
S. Kalara, J.L. Lopez and D.V. Nanopoulos, Phys. Lett. {\bf B245} (1990)
421; Nucl. Phys. {\bf B353} (1991) 650;\\
A. Faraggi, Phys. Lett. {\bf B278} (1992) 131.

\bibitem{bsc} H. Kawai, D.C. Lewellen and S.-H.H. Tye, Phys. Lett. {\bf 191B}
(1987) 63; Int. J. Mod. Phys. {\bf A3} (1988) 279; \\
W. Lerche, B.E.W. Nilsson and A.N. Schellekens, Nucl. Phys. {\bf B294} (1987) 
136.

\bibitem{sw} A.N. Schellekens and N.P. Warner, Nucl. Phys. {\bf B308} (1988)
397; \\
S. Chaudhuri, H. Kawai and S.-H.H. Tye, Nucl. Phys. {\bf B322} (1989) 373; \\
W. Lerche, A.N. Schellekens and N.P. Warner, Phys. Rep. {\bf 177} (1989) 1.

\bibitem{KLT} H. Kawai, D.C. Lewellen and S.-H.H. Tye, Phys. Rev. Lett.
{\bf 57} (1986) 1832; {\bf 58} (1987) 492E; Nucl. Phys. {\bf B288} (1987) 1; \\
I. Antoniadis, C. Bachas and C. Kounnas, Nucl. Phys. {\bf B289} (1987) 87.

\bibitem{narain} K.S. Narain, Phys. Lett. {\bf B169} (1986) 41; \\
K.S. Narain, M.H. Sarmadi and E. Witten, Nucl. Phys. {\bf B279} (1987) 369.

\bibitem{FMS} D. Friedan, E.Martinec and S. Shenker, Phys. Lett. 
{\bf 160B} (1985) 55; Nucl. Phys. {\bf B271} (1986) 93.

\bibitem{global} M. Dine and N. Seiberg, Nucl. Phys. {\bf B306} (1988) 137; \\
T. Banks and L. Dixon, Nucl. Phys. {\bf B307} (1988) 93; \\
L.M. Krauss and F. Wilczek, Phys. Rev. Lett. {\bf 62} (1989) 1221. 

\bibitem{cocycle} H. Kawai, D.C. Lewellen and S.-H.H. Tye, Nucl. Phys. 
{\bf B269} (1986) 1.

\bibitem{KST} D.S. Freed and C. Vafa, Comm. Math. Phys. 
{\bf 110} (1987) 349;\\
P. Ginsparg, Nucl. Phys. {\bf B295} (1988) 153;\\
R. Dijkgraaf, C. Vafa, E. Verlinde and H. Verlinde,
Comm. Math. Phys. {\bf 123} (1989) 485;\\
Z.S. Li and C.S. Lam, Int. J. Mod. Phys. {\bf A7} 
(1992) 5739;\\
Z. Kakushadze, G. Shiu and S.-H.H. Tye, Phys. Rev. {\bf D54}
(1996) 7545, hep-th/9607137.

\end{references}
\end{document}